\documentclass{aastex63}
\usepackage{amsmath, amssymb, mathrsfs, mathtools}
\usepackage{graphics}
\usepackage{color, enumerate}

\allowdisplaybreaks

\newcommand{\bfv}{\boldsymbol{v}}

\newcommand{\bfk}{\boldsymbol{k}}
\newcommand{\bfb}{\boldsymbol{B}}
\newcommand{\bfg}{\boldsymbol{g}}
\newcommand{\bfA}{\boldsymbol{A}}
\newcommand{\bfkappa}{\boldsymbol{\kappa}}
\newcommand{\HL}{\mathscr{L}}
\newcommand{\bfeb}{\boldsymbol{e}_B}
\newcommand{\eps}{\varepsilon}

\newcommand{\bey}{\boldsymbol{u}_2}
\newcommand{\bez}{\boldsymbol{u}_3}
\newcommand{\ben}{\boldsymbol{\hat{e}}_n}
\newcommand{\bfX}{\boldsymbol{X}}
\newcommand{\texp}{\text{e}}
\newcommand\T{\rule{0pt}{3ex}}			
\newcommand\B{\rule[-1.5ex]{0pt}{0pt}}	
\newcommand{\ts}{\textsuperscript}

\begin{document}

\title{Legolas: a modern tool for magnetohydrodynamic spectroscopy}

\author[0000-0002-8720-9119]{Niels Claes}
\affiliation{Centre for mathematical Plasma-Astrophysics, Celestijnenlaan 200B, 3001 Leuven, KU Leuven, Belgium}

\author[0000-0003-2443-3903]{Jordi De Jonghe}
\affiliation{Centre for mathematical Plasma-Astrophysics, Celestijnenlaan 200B, 3001 Leuven, KU Leuven, Belgium}

\author[0000-0003-3544-2733]{Rony Keppens}
\affiliation{Centre for mathematical Plasma-Astrophysics, Celestijnenlaan 200B, 3001 Leuven, KU Leuven, Belgium}

\begin{abstract}
Magnetohydrodynamic (MHD) spectroscopy is central to many astrophysical disciplines, ranging from helio- to asteroseismology, over solar coronal (loop) seismology, to the study of waves and instabilities in jets, accretion disks, or solar/stellar atmospheres. MHD spectroscopy quantifies all linear (standing or travelling) wave modes, including overstable (i.e. growing) or damped modes, for a given configuration that achieves force and thermodynamic balance. Here, we present \texttt{Legolas}\footnote{The Legolas code is available on GitHub: \url{https://github.com/n-claes/legolas}}, a novel, open-source numerical code to calculate the full MHD spectrum of one-dimensional equilibria with flow, that balance pressure gradients, Lorentz forces, centrifugal effects and gravity, enriched with non-adiabatic aspects like radiative losses, thermal conduction and resistivity. The governing equations use Fourier representations in the ignorable coordinates, and the set of linearised equations are discretised using Finite Elements in the important height or radial variation, handling Cartesian and cylindrical geometries using the same implementation. A weak Galerkin formulation results in a generalised (non-Hermitian) matrix eigenvalue problem, and linear algebraic algorithms calculate all eigenvalues and corresponding eigenvectors. We showcase a plethora of well-established results, ranging from p- and g-modes in magnetised, stratified atmospheres, over modes relevant for coronal loop seismology, thermal instabilities and discrete overstable Alfv\'en modes related to solar prominences, to stability studies for astrophysical jet flows. We encounter (quasi-)Parker, (quasi-)interchange, current-driven and Kelvin-Helmholtz instabilities, as well as non-ideal quasi-modes, resistive tearing modes, up to magneto-thermal instabilities. The use of high resolution sheds new light on previously calculated spectra, revealing interesting spectral regions that have yet to be investigated.
\end{abstract}

\section{Introduction} \label{sec:intro}
The study of stability for plasmas and fluids alike has been a major topic of research over the last century. Understanding how and why a given medium reacts to a linear perturbation is of central importance to many astrophysical phenomena. In incompressible or compressible fluids, governed by hydrodynamic equations, notable instabilities are the Kelvin-Helmholtz instability (KHI), which arises due to a velocity shear at the interface of two fluids, and the Rayleigh-Taylor instability (RTI) where gravitational stratification can lead to an unstable configuration of layered fluids of different density~\citep{book_chandrasekhar, book_choudhuri}. In plasmas, governed by the magnetohydrodynamic (MHD) equations, the study of waves and instabilities becomes much richer due to the inclusion of magnetic fields, with a modern overview provided in~\cite{book_MHD}. Magnetic fields modify the two aforementioned instabilities in various ways, and the combination of flow, magnetic fields and pressure gradients introduces many new modes, e.g. the magnetorotational instability~\citep{balbus1991} relevant for (weakly magnetised) accretion disks or the Trans-Slow-Alfv\'en Continuum modes in disks of arbitrary magnetisation~\citep{goedbloed2004}. In the highly magnetised solar corona, observed coronal loop oscillations (periods and damping times) are routinely used to infer loop parameters like their field strength~\citep{nakariakov2001}. Embedded in the hot solar corona, we find stable and long-lived quiescent prominences, with internal dynamics due to KHI \citep{hillier2018_KHI} and RTI \citep{hillier2018}. The formation of prominences is due to the thermal instability (TI), as demonstrated in direct observations by e.g.~\citet{berger2012} or in simulations by~\citet{xia2016, claes2020}. Together with categorising all instabilities, knowing the stable eigenoscillations such as p-modes or g-modes in stratified atmospheres or stellar interiors, is of prime importance to link theoretical understanding with observed periodic phenomena. In all of these cases, we need to compute eigenoscillations and corresponding eigenfunctions from the linearised set of governing equations. Linear MHD spectroscopy, which encompasses the entirety of helio- and asteroseismology, but incorporates laboratory fusion plasma MHD spectroscopy~\citep{goedbloed1993}, MHD spectroscopy of accretion disks \citep{keppens2002} and jets, as well as solar coronal seismology \citep{book_roberts}, is thus a powerful tool for studying many astrophysical processes.

Since the advent of more powerful computational resources, the main focus of computational astrophysical research has gradually shifted towards solving the fully non-linear MHD equations, where many non-adiabatic/non-ideal effects are incorporated, depending on the application at hand. While this approach successfully reproduced many physical phenomena, especially for realistic solar setups (e.g. sunspots~\citep{rempel2012}, flares~\citep{ruan2019}, or prominences~\citep{xia2016}), it usually fails to answer which specific perturbation produces the complex evolution as witnessed. At the same time, theoretical insight showed that MHD spectral theory actually governs the stability of flowing, (self-)gravitating single fluid evolutions of nonlinear, time-dependent plasmas, and this at any time during their nonlinear evolution~\citep{demaerel2016}. Hence, in order to predict the reaction of a certain physical state to perturbations, we should really quantify all its waves and instabilities using linear theory. This has been recognised fully in laboratory fusion plasmas, where MHD spectroscopy is very successful for identifying waves and stability aspects of a given toroidal Grad-Shafranov equilibrium. That this can meaningfully be done for states that include important non-adiabatic effects, like optically thin radiative losses, is important for investigations into prominences and their intriguing fine structure, as revealed by means of direct observations \citep{engvold1998, mackay2010, ballester2006} or through numerical simulations \citep{xia2016, xia2017, claes2020}. In that context, early analytical work by \citet{vanderlinden1991} based on linear MHD suggests the hypothesis that finite perpendicular thermal conduction induces fine structure in unstable linear eigenmodes. Since this pioneering work of \citet{vanderlinden1991}, not much research has been done regarding the full MHD spectrum when non-adiabatic effects are at play, for the simple reason that to date, there existed no numerical tool to solve the full system of linearised MHD equations with all the physical effects included. This is why we developed the new and open-source \texttt{Legolas} solver.

\texttt{Legolas} builds on the heritage of early numerical codes, most notably LEDA~\citep{kerner1985}, which allowed studies of the ideal or resistive MHD spectrum for laboratory plasmas, approximated by a diffuse cylindrical plasma column (or flux tube) and CASTOR~\citep{kerner1998}, which applied to resistive spectra of general tokamak configurations. The latter has follow-up codes such as FINESSE \citep{belien2002} and PHOENIX \citep{blokland2007phoenix}, extending it to stationary and axisymmetric truly 2D configurations. LEDA was later extended in \citet{vanderlinden1992}, where non-adiabatic effects like anisotropic thermal conduction and optically thin radiative losses were added to the equations, using a simple analytic function to treat radiative cooling effects. A different branch of LEDA, called LEDAFLOW \citep{nijboer1997}, was developed to investigate the resistive MHD spectrum, augmented with gravitational and flow effects, but omitting those non-adiabatic terms. Since these codes were developed decades ago and focus shifted away from linear MHD, their further development was stalled, although in laboratory fusion context, tools to compute multidimensional equilibria and their linear modes are very important for diagnosing experiments. The original codes, like LEDA(FLOW), were not flexible, in the sense that adding different equilibria or accounting for additional terms in the equations would be a major undertaking, as parts were hard-coded to (limited) computational resources of that time. Furthermore, programming languages and numerical tools like LAPACK~\citep{lapack} to solve eigenvalue problems have come a long way. This prompted us to develop a brand new, modern MHD spectral code which we named \texttt{Legolas}, short for ``Large Eigensystem Generator for One-dimensional pLASmas". The \texttt{Legolas} code is able to handle both Cartesian and cylindrical geometries, and introduces many new features, e.g. selecting between modern cooling curves that treat optically thin radiative cooling effects. Furthermore, every aspect of the code is modularised, making it ready to be extended with additional physics or modern algorithmic requirements (such as mesh refinement). The main goal of this paper is to present the new code in terms of its implementation details and to validate it against a plethora of test cases that ensure a correct treatment of the governing equations.

These tests include eigenmode quantifications of ideal, static MHD configurations under adiabatic conditions, where the static (that is, no equilibrium flow) and adiabatic linear MHD equations make the problem self-adjoint. When performing a standard Fourier analysis in the ignorable directions, the resulting eigenvalue problem is then Hermitian, meaning that all eigenfrequencies will be either fully real (stable waves) or fully complex (pure damped or unstable modes), hence they are found on the real or imaginary axis of the complex eigenfrequency plane, and the full MHD spectrum will be both left-right and up-down symmetric. However, in nature, physical conditions may be far from ideal. The inclusion of non-ideal effects like resistivity or thermal conduction lifts the self-adjointness of the eigenvalue problem, allowing the eigenmodes to move away from the axes into the complex plane, and the up-down symmetry gets broken. As long as the equilibrium configuration is static, all (adiabatic or non-adiabatic) modes will still have a complementary mode that lies mirrored around the imaginary axis, making the entire spectrum left-right symmetric. This is related to the forward and backward propagating mode symmetry, or the equivalent statement on the parity-time (PT) symmetry. However, for typical astrophysical plasmas, the conditions are far from static: tokamak plasmas, astrophysical jets, solar coronal loops, accretion discs\ldots~all have equilibrium flows. The inclusion of a background flow breaks the left-right symmetry of the MHD spectrum, resulting in an even more complicated structure. However, the study of the ideal, linear MHD spectrum of flowing plasmas is still governed by a pair of self-adjoint operators~\citep{goedbloed2011, book_MHD}, and it leaves the up-down symmetry of the spectrum intact (where every overstable mode has an equivalent damped counterpart at the same frequency). The combination of flow and non-adiabatic effects, where both left-right and up-down eigenfrequency symmetries are broken, has never been explored in earnest. All of the above makes it clear that a numerical approach becomes essential, especially when the equilibrium is no longer homogeneous. Since in reality, virtually no astrophysical configuration is spatially homogeneous, we need a flexible numerical tool to explore the spectrum systematically.

\texttt{Legolas} solves the linearised MHD equations including various non-adiabatic effects, resistivity and gravity, assuming a one-dimensional (1D) equilibrium profile with the possibility of background flow. A standard Fourier analysis for the perturbations is combined with a Finite Element representation of the eigenfunctions in the important coordinate. We transform the original system into an eigensystem for the complex eigenfrequencies $\omega$. We use a general formalism to include two kinds of geometries, a plane Cartesian stratified slab or a (possibly also stratified) cylinder, through the inclusion of a scale factor originating from the divergence, gradient and curl operators. \texttt{Legolas} can handle the hydrodynamic limit (where all equilibrium magnetic field components are set to zero), enabling us to investigate stability of hydrodynamic static and stationary equilibria. The resulting system of equations is solved in weak form, transforming the original system into a non-Hermitian complex eigenvalue problem, which is solved using the QR algorithm. This results in a calculation of all eigenfrequencies and corresponding eigenfunctions of the system, such that a detailed analysis of mode stability, but also the entire overview on all supported linear wave modes, becomes possible.

Section \ref{sect: model_equations} introduces the system of equations, along with the linearisation procedure and Fourier mode representation. The treatment of the final system using the finite element method is given in Appendix \ref{sect: numerical_approach} where we explain the basic mathematical formalism behind the FEM, with a complete treatment of the finite element matrix assembly process and the boundary conditions. \texttt{Legolas} is tested against a large amount of spectra found in the literature in Section \ref{sect: results}, which is subdivided into multiple categories. First we treat ideal MHD with only a gravitational term included, which has as main advantage that solutions can be obtained analytically. The first case handles gravito-MHD waves in a Cartesian slab, after which we quantify quasi-Parker instabilities in stratified atmospheres. Results obtained with \texttt{Legolas} are compared with results found in various modern textbooks, including results for cylindrical geometries, discussing modes for ideal flux tubes and tokamak current equilibria where we show liability to interchanges.
For the inclusion of flow into the equations, we consider a non-trivial case related to astrophysical jet stability, looking at Kelvin-Helmholtz and current-driven instabilities. We demonstrate that we can compute Suydam cluster modes, originating from a surface where the wave vector is locally perpendicular to the magnetic field. We then transit to the inclusion of non-ideal effects such as resistivity, looking first at the resistive spectrum for a homogeneous plasma. We then extend to recover the resistive quasi-mode in a non-homogeneous case. This is further complemented by adding an inhomogeneous medium and background flow, in such a way as to give rise to resistive tearing modes, all of which in a Cartesian geometry. Additionally, by allowing for resistivity and current variation, we show that we can also compute resistive rippling modes. Lastly, we treat non-adiabatic effects, including optically thin radiative losses and thermal conduction into the equations, revisiting some pioneering results of discrete Alfv\'en waves and magnetothermal instabilities. Along the way, we find interesting extensions to the original published works, due to the much higher resolutions employed here.
Since \texttt{Legolas} is the first, modern linear MHD code to investigate realistic astrophysical plasmas, this opens the door to further in-depth studies of non-ideal equilibrium configurations at high resolutions, ranging from loops, jets or accretion disks.

\newpage

\section{Problem description and model equations}	\label{sect: model_equations}
The MHD equations with the inclusion of non-adiabatic effects, resistivity and gravity can be written in a (normalised) Eulerian representation as
\begin{align}
	\frac{\partial \rho}{\partial t} &= -\nabla \cdot (\rho\bfv),	\label{eq: continuity} \\
	\rho\frac{\partial \bfv}{\partial t} &= -\nabla p - \rho\bfv \cdot \nabla \bfv + (\nabla \times \bfb) \times \bfb + \rho\bfg,	\label{eq: momentum}	\\
	\rho\frac{\partial T}{\partial t} &= -\rho\bfv \cdot \nabla T - (\gamma - 1)p\nabla\cdot\bfv - (\gamma - 1)\rho\HL + (\gamma - 1)\nabla\cdot(\bfkappa \cdot \nabla T) + (\gamma - 1)\eta(\nabla \times \bfb)^2, \label{eq: energy}\\
	\frac{\partial \bfb}{\partial t} &= \nabla \times (\bfv \times \bfb) - \nabla \times (\eta\nabla \times \bfb),	\label{eq: induction}
\end{align}
where $\rho$ is the plasma density, $\bfv$ is the velocity, $T$ is the temperature, $p$ is the pressure, $\bfb$ denotes the magnetic field (satisfying $\nabla\cdot\bfb = 0$), $\eta$ is the resistivity and $\bfg$ the gravitational acceleration. To close the system the (normalised) ideal gas law $p = \rho T$ is used, while $\gamma$ denotes the ratio of specific heats, taken to be $5/3$. The symbol $\HL$ in Eq. \eqref{eq: energy} represents the heat-loss function, defined as energy losses minus energy gains due to optically thin radiative cooling effects \citep{parker1953} and is given by
\begin{equation}	\label{eq: radiative_cooling}
	\HL = \rho\Lambda(T) - \mathcal{H},
\end{equation}
where $\mathcal{H}$ represents the total energy gains. In general, $\mathcal{H}$ can be anything (for example heating through dissipative Alfv\'en waves \citep{vanderholst2014}), but since there is still no well-defined parametrisation for coronal heating to date this term is assumed to be as convenient as possible, that is, constant in time but possibly varying in space to ensure thermodynamic balance. Note that $\mathcal{H}$ should always be consistent with a given equilibrium profile as to exactly balance out the radiative losses (and possibly the thermal conduction effects and/or ohmic heating effects), reaching a thermal equilibrium state. This indirectly implies that the heating term is not necessarily independent of location, but dependent on the connection between the radiative losses and equilibrium temperature profile, which, in general, are both spatially dependent. The first term in \eqref{eq: radiative_cooling} denotes the radiative losses, dependent on the cooling curve $\Lambda(T)$. These curves are tabulated sets resulting from detailed calculations, and can hence be interpolated to high temperature resolutions. \texttt{Legolas} has multiple cooling curves implemented, most notably those by \citet{colgan2008} and \cite{spex2009}, where the latter one is extended to the low-temperature limit using \cite{dalgarno1972}. In addition we also implemented a piecewise power law as described by \cite{rosner1978}, which is an explicit (piecewise) function over the entire temperature domain. In the solar and astrophysical literature, these cooling curves collect detailed knowledge on radiative processes, which are all assumed to be in the optically thin regime. It is worth noting that the inclusion of time-dependent background heating or flows can influence the spectrum in itself, as shown in for example
\citet{barbulescu2019, hillier2019}, but due to the time-dependence of those effects the assumption of a stationary background state is no longer applicable. A possibility however may be varying the background heating or flow in subsequent runs, provided the background state is known at every snapshot.

Thermal conduction in magnetised plasmas is highly anisotropic, as the effect is a few orders of magnitude stronger along the field lines than across. We hence use a tensor representation to model this anisotropy, denoting the thermal conductivity tensor $\bfkappa$ by
\begin{equation}	\label{eq: conduction_tensor}
	\bfkappa = \kappa_\parallel\bfeb\bfeb + \kappa_\bot(\boldsymbol{I} - \bfeb\bfeb),
\end{equation}
where $\boldsymbol{I}$ denotes the unit tensor and $\bfeb = \bfb / B$ is a unit vector along the magnetic field. The coefficients $\kappa_\parallel$ and $\kappa_\bot$ denote the conductivity coefficients parallel and perpendicular to the local direction of the magnetic field. For typical astrophysical applications the Spitzer conductivity is used, given by
\begin{equation}	\label{eq: conduction_coeffs}
	\begin{aligned}
		\kappa_\parallel & \approx 8 \times 10^{-7}T^{5/2}~\text{erg cm$^{-1}$s$^{-1}$K$^{-1}$},	&\qquad\qquad \kappa_\bot \approx 4 \times 10^{-10} n^2B^{-2}T^{-3}\kappa_\parallel,	\\
		\kappa_\parallel & \approx 8 \times 10^{-12}T^{5/2}~\text{W m$^{-1}$K$^{-1}$},				&\qquad\qquad \kappa_\bot \approx 4 \times 10^{-30} n^2B^{-2}T^{-3}\kappa_\parallel,
	\end{aligned}
\end{equation}
where $n$ denotes the number density, given by $\rho = n m_p$ with $m_p$ the proton mass. The first and second row in Eq. \eqref{eq: conduction_coeffs} give the thermal conduction coefficients in cgs and mks units, respectively. For the solar corona we typically find that $\kappa_\parallel$ is about $12$ orders of magnitude larger than $\kappa_\bot$ \citep{book_priest}, so perpendicular thermal conduction is usually ignored. Nevertheless, in \texttt{Legolas} both parallel and perpendicular thermal conduction are implemented. For the resistivity $\eta$ one can in principle take any profile. We implemented the Spitzer resistivity,
\begin{equation}	\label{eq: spitzer_eta}
	\eta = \frac{4\sqrt{2\pi}}{3}\frac{Z_\text{ion}e^2\sqrt{m_e}\ln(\lambda)}{(4\pi\epsilon_0)^2(k_BT)^{3/2}},
\end{equation}
where $Z_\text{ion}$ denotes the ionisation taken to be unity, $e$ and $m_e$ denote the electron charge and mass, respectively, and $\epsilon_0$ and $k_B$ are the electrical permittivity and Boltzmann constant. The Coulomb logarithm is given by $\ln(\lambda)$ and is approximately equal to 22 for solar coronal conditions \citep{book_MHD}. It is important to emphasise that, since we will further linearise the governing non-linear equations, we can adopt fully realistic values for all the non-ideal coefficients, such as the resistivity or thermal conduction coefficients. This is in contrast to fully nonlinear computations, which are severely restrained in reaching magnetic Reynolds numbers beyond $10^4-10^5$.

\subsection{Equilibrium conditions}
\begin{figure}[t]
	\centering
	\includegraphics[width=0.9\textwidth]{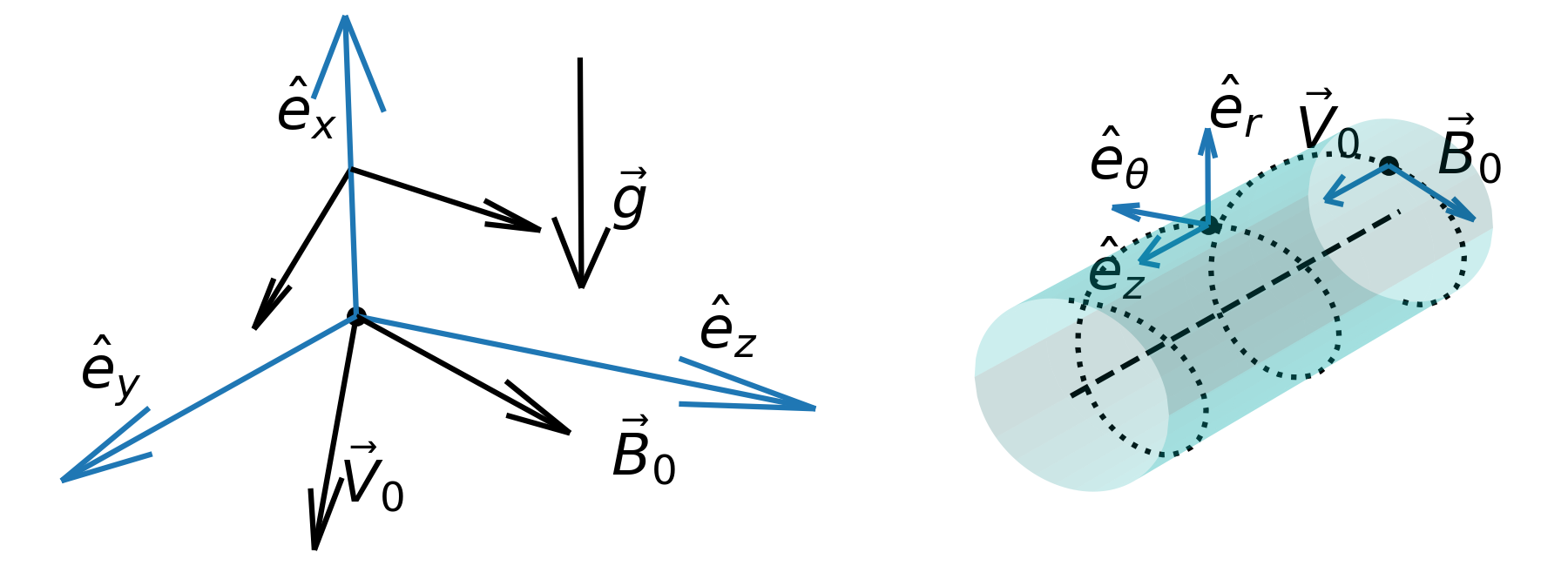}
	\caption{Unit vectors and examples of $\bfb_0$ and $\bfv_0$ for the Cartesian (left) and cylindrical case (right).}
	\label{fig: axes}
\end{figure}
We consider a general coordinate system denoted by $(u_1, u_2, u_3)$, corresponding to three orthogonal basis vectors. The main advantage of this approach is that it allows us to include two different geometries with only one basic formalism (and implementation). First we consider a standard plane slab geometry in Cartesian coordinates, that is, a plasma which is confined in height, and considered to be bounded by two horizontal, perfectly conducting walls at a fixed distance apart, extending outwards to infinity in the other two ignorable coordinates. This case also approximates the limit of a fully infinite free space when the walls are moved off to infinity.
In Cartesian geometry the coordinate system can be written as $(x, y, z)$ and the vectors $\{ \mathbf{u}_1, \mathbf{u}_2, \mathbf{u}_3 \}$ are the standard Cartesian triad $\{ \hat{\mathbf{e}}_x, \hat{\mathbf{e}}_y, \hat{\mathbf{e}}_z \}$ along the axes. This makes it quite convenient to include for example gravitational effects which will induce an equilibrium stratification in the $u_1$ coordinate. The second geometry is that of an infinitely long plasma cylinder encased by a solid wall at a certain distance away from the cylinder axis, for which the coordinate system can be defined as $(r, \theta, z)$. At each point the vectors $\{ \mathbf{u}_1, \mathbf{u}_2, \mathbf{u}_3 \}$ are defined as the triad of tangent vectors, $\{ \hat{\mathbf{e}}_r, \hat{\mathbf{e}}_\theta, \hat{\mathbf{e}}_z \}$, with $\hat{\mathbf{e}}_r$ along the radial direction, $\hat{\mathbf{e}}_z$ in the direction of the cylinder axis and $\hat{\mathbf{e}}_\theta = \hat{\mathbf{e}}_z \times \hat{\mathbf{e}}_r$ tangent to the cylinder. A detailed view of both geometries and their corresponding coordinate systems is shown in Figure \ref{fig: axes}. The basic operators present in Equations \eqref{eq: continuity}-\eqref{eq: induction}, that is, the divergence, gradient and curl, introduce a scale factor $\eps = r$ for cylindrical geometries, which is reduced to $\eps = 1$ for a Cartesian coordinate system. Hence exploiting this scale factor in the mathematical formalism allows for one implementation, where one can conveniently switch between both cases.
We note that the cylindrical setup is also applicable to the so-called cylindrical accretion disk limit, as for example exploited to study MHD instabilities in disks by \citet{blokland2007}.

Linearisation involves splitting variables into two parts: a time-independent part, usually denoted with subscript $0$, and a perturbed part, denoted by a subscript $1$. \texttt{Legolas} handles one-dimensional equilibria which depend only on $u_1$, or, more specifically, time-independent equilibria of the form
\begin{equation}	\label{eq: equilibrium}
	\begin{aligned}
		\rho_0 &= \rho_0(u_1),		\\
		p_0 &= p_0(u_1),			\\
		T_0 &= T_0(u_1),			\\
	\end{aligned}
	\qquad\qquad\qquad
	\begin{aligned}
		\bfv_0 &= v_{02}(u_1)\bey + v_{03}(u_1)\bez,	\\
		\bfb_0 &= B_{02}(u_1)\bey + B_{03}(u_1)\bez. \\
		&
	\end{aligned}
\end{equation}
In general we have $\bfg = -g(u_1)\mathbf{u}_1$, where the Cartesian case is for a stratified atmosphere or layer, and the cylindrical case can also allow for gravitational stratification of an accretion disk situated for $u_1 = r \in [1, R]$.
In the case of a cylinder where $u_1 = r \in [0, R]$, this gravitational term is absent. Using these equations in combination with Eqs. \eqref{eq: continuity}-\eqref{eq: induction} yields two conditions for the time-independent parts, given by
\begin{align}
	\left(p_0 + \frac{1}{2}\bfb_0^2\right)' + \rho_0 g - \frac{\eps'}{\eps}\left(\rho_0v_{02}^2 - B_{02}^2\right) &= 0,	\label{eq: eq_condition_1} \\
	\frac{1}{\eps}\left(\eps \kappa_\bot T_0'\right)' - \rho_0\HL_0 &= 0,	\label{eq: eq_condition_2}
\end{align}
where the first condition originates from the momentum equation \eqref{eq: momentum} and should always be satisfied as it expresses a force-balanced state. The second condition originates from the non-adiabatic terms in the energy equation and should be accounted for if these terms are included, the prime denotes the derivative with respect to $u_1$. It should be noted that resistive terms are not considered here, which is justified by considering that the time scales on which the magnetic fields decay due to resistivity is much, much larger than the time scales of resistive modes. This is a consequence of large magnetic Reynolds numbers $R_m$ in typical astrophysical cases, yielding magnetic decay time scales of $\tau \sim R_m \tau_A$ (with $\tau_A$ the typical Alfv\'en time in ideal MHD) compared to much faster resistive mode time scales of $\tau \sim R_m^\alpha$ (where typically $0 < \alpha < 1$). We can hence consider the equilibrium itself to be independent of resistivity, which removes some stringent extra conditions on the energy and induction equations. Also note that the third term in Eq. \eqref{eq: eq_condition_1} is only included for a cylinder, since $\eps' = 0$ in a Cartesian geometry. This translates to the well-known fact that the centrifugal and tensional part of the Lorentz force are absent for a Cartesian slab. Furthermore a cylindrical equilibrium profile should satisfy on-axis regularity conditions, meaning that $v_{02}, v_{03}', B_{02}$ and $B_{03}'$ all have to be equal to zero at $r = 0$. When considering an accretion disk in the cylindrical limit, the inner edge of the disk is at $r = 1$, so lengths are then expressed in this inner disk radius and no regularity conditions apply then.

\subsection{Linearised equations}
Now we linearise Eqs. \eqref{eq: continuity}-\eqref{eq: induction} around the equilibrium specified in \eqref{eq: equilibrium}, where the unperturbed time-independent parts are denoted with a subscript $0$ and the perturbed time-dependent parts are denoted with a subscript $1$. It follows from the adopted equilibrium configuration that $\nabla \cdot \bfv_0 = 0$ and $\nabla \cdot \bfb_0 = 0$, such that the divergence-free condition on the magnetic field is fulfilled and that the equilibrium flow field is incompressible. However, the perturbed quantities can represent both incompressible or compressible eigenoscillations. The no-monopole condition should also be taken into account for the perturbed magnetic field. Therefore, we adopt a vector potential to write $\bfb_1 = \nabla \times \bfA_1$ such that $\nabla \cdot \bfb_1 = 0$ is automatically satisfied. The system of linearised equations is thus given by
\begin{align}
	\frac{\partial \rho_1}{\partial t} &= -\rho_0 \nabla \cdot \bfv_1 - \bfv_0 \cdot \nabla\rho_1,			\label{eq: linearised_continuity}	\\
	\rho_0 \frac{\partial \bfv_1}{\partial t} &=
		\begin{alignedat}[t]{1}
			&-\rho_0\bfv_0 \cdot \nabla \bfv_1 - \rho_0\bfv_1 \cdot \nabla \bfv_0 - \rho_1\bfv_0\cdot\nabla\bfv_0 - \nabla\left(\rho_1T_0 + \rho_0T_1\right)	\\
			&+ (\nabla \times \bfb_0) \times (\nabla \times \bfA_1) + \left[\nabla \times (\nabla \times \bfA_1)\right] \times \bfb_0 + \rho_1 \bfg,	\\
		\end{alignedat}	\label{eq: linearised_momentum}	\\
	\rho_0\frac{\partial T_1}{\partial t} &=
		\begin{alignedat}[t]{1}
			&-\rho_0\bfv_1 \cdot \nabla T_0 - \rho_0\bfv_0\cdot \nabla T_1 - (\gamma - 1)\rho_0T_0\nabla \cdot \bfv_1 - (\gamma - 1)\rho_1\HL_0 - (\gamma - 1)\rho_0\left(\HL_T T_1 + \HL_\rho \rho_1\right)	\\
			&+ (\gamma - 1)\nabla \cdot (\bfkappa_0 \cdot \nabla T_1) + (\gamma - 1)\nabla \cdot (\bfkappa_1 \cdot \nabla T_0)	\\
			&+ 2(\gamma - 1)\eta(\nabla \times \bfb_0) \cdot \left[\nabla \times (\nabla \times \bfA_1)\right] + (\gamma - 1)\frac{d\eta}{dT}T_1(\nabla \times \bfb_0)^2,
		\end{alignedat}	\label{eq: linearised_energy} \\
	\frac{\partial \bfA_1}{\partial t} &= \bfv_1 \times \bfb_0 + \bfv_0 \times (\nabla \times \bfA_1) - \eta\nabla \times (\nabla \times \bfA_1) - \frac{d\eta}{dT}T_1(\nabla \times \bfb_0),	\label{eq: linearised_induction}
\end{align}
where $p_1$ is replaced by $\rho_1T_0 + \rho_0T_1$ resulting from the linearised ideal gas law. The perturbation $\bfkappa_1$ of the thermal conduction tensor is obtained by linearising the expressions \eqref{eq: conduction_tensor}-\eqref{eq: conduction_coeffs},
while the derivatives of the heat-loss function with respect to density and temperature are given by
\begin{equation}
	\HL_\rho = \left.\frac{\partial \HL}{\partial \rho}\right|_T = \Lambda(T_0),	\qquad\qquad 	\HL_T = \left.\frac{\partial \HL}{\partial T}\right|_\rho = \rho_0\frac{d\Lambda(T_0)}{dT},
\end{equation}
which should be evaluated using the equilibrium quantities. The terms containing $\partial \eta / \partial T$ follow from a linearisation of the resistivity parameter, which can be written in terms of the variable $T_1$ using the temperature dependence originating from the assumed Spitzer resistivity in Eq. \eqref{eq: spitzer_eta}, that is, $\eta_1 = T_1 \partial \eta / \partial T$.
In addition, \texttt{Legolas} allows for an anomalous resistivity prescription in which we typically have $\eta(\mathbf{u}_1, \mathbf{j}(\mathbf{u}_1, t))$, that is, a resistivity profile that is spatio-temporal in general, but for a fixed time, depends on position and current profile. This in turn implies that a total derivative should be used for the resistivity, given by $\eta' = \partial \eta / \partial u_1 + T_0'\partial \eta/ \partial T$.
In most use cases a resistivity profile $\eta(T(\mathbf{u}_1))$ is sufficient, which is only temperature dependent (and hence indirectly spatially varying as well for inhomogeneous temperature profiles).
Also note that these linear equations (as also the nonlinear set above) assumed an external gravitational field, so we did not need to linearise the gravity term $\bfg$ (this is the so-called Cowling approximation). In the future, we can extend the set of equations with the Poisson equation and also allow for self-gravity driven Jeans instabilities.

Next, we perform a Fourier analysis with an exponential time dependence, imposing standard Fourier modes on the $u_2$ and $u_3$ coordinates of a perturbed quantity $f_1$, given by
\begin{equation}	\label{eq: fourier}
	f_1 = \widehat{f_1}(u_1)\exp{\left[i(k_2u_2 + k_3u_3 - \omega t)\right]}.
\end{equation}
In this form the wave numbers $k_2$ and $k_3$ correspond to $k_y$ and $k_z$ in Cartesian geometry, and to $m$ and $k$ in a cylindrical geometry, respectively. Note that $m$ is quantified to integer values, since the $\theta$-direction is periodic.
Additionally, we apply the following transformation to the perturbed quantities
\begin{equation}	\label{eq: transformation}
	\begin{alignedat}{4}
		\eps\widehat{\rho}_1 &= \widetilde{\rho}_1,	\qquad\qquad	i\eps\widehat{v}_1 &= \widetilde{v}_1,	\qquad\qquad	\widehat{v}_2 &= \widetilde{v}_2,		\qquad\qquad	\eps\widehat{v}_3 &= \widetilde{v}_3,	\\
		\eps\widehat{T}_1 &= \widetilde{T}_1,		\qquad\qquad	i\widehat{A}_1 &= \widetilde{a}_1,		\qquad\qquad	\eps\widehat{A}_2 &= \widetilde{a}_2,	\qquad\qquad	\widehat{A}_3 &= \widetilde{a}_3,
	\end{alignedat}
\end{equation}
This particular transformation simplifies the resulting set of equations and has as additional effect that all terms are real except for the non-adiabatic and resistive contributions, such that we are only dealing with imaginary terms when these physical effects are included. This is in analogy to the fact that the purely adiabatic case is governed by self-adjoint operators: one for the case without flow and two for the case with flow included \citep{goedbloed2018web1, goedbloed2018web2}. The final set of Fourier-analysed linearised equations is given below, where the tilde notation in Eqs. \eqref{eq: transformation} is dropped for the sake of simplicity. From now on, tildes will no longer be written explicitly since there is no confusion possible.

\begin{equation}
	\omega \frac{1}{\eps}\rho_1	=
			-\frac{1}{\eps}\rho_0'v_1 - \frac{1}{\eps}\rho_0\left(v_1' - k_2v_2 - k_3v_3\right) + \frac{1}{\eps}\left(\frac{1}{\eps}k_2v_{02} + k_3v_{03}\right)\rho_1,	\label{eq: linearised_rho_trans}
\end{equation}
\begin{align}
	\omega \rho_0\frac{1}{\eps}v_1 &=
		\begin{alignedat}[t]{1}
			&\left(\frac{\rho_1T_0 + \rho_0T_1}{\eps}\right)' + \frac{1}{\eps}B_{03}'\left(a_2' - k_2a_1\right) - \frac{(\eps B_{02})'}{\eps}\left(a_3' - k_3a_1\right) \\
			&-B_{03}\left\{\frac{k_3^2}{\eps}a_2 - \frac{k_2k_3}{\eps}a_3 - \left[\frac{1}{\eps}\left(a_2' - k_2a_1\right)\right]'\right\} + \frac{1}{\eps}\left(\frac{k_2v_{02}}{\eps} + k_3v_{03}\right)\rho_0 v_1	\\
			&+B_{02}\left\{\frac{k_2^2}{\eps^2}a_3 - \frac{k_2k_3}{\eps^2}a_2 - \frac{1}{\eps}\Bigl[\eps\left(a_3' - k_3a_1\right)\Bigr]'\right\} - 2\frac{\eps'}{\eps}\rho_0v_{02}v_2 - \frac{\eps'}{\eps^2}v_{02}^2\rho_1 + \frac{1}{\eps}g\rho_1,
		\end{alignedat}	\label{eq: linearised_v1_trans}
\end{align}
\begin{align}
	\omega\rho_0 v_2 &=
		\begin{alignedat}[t]{1}
			&\frac{1}{\eps^2}\left(\rho_1T_0 + \rho_0 T_1\right)k_2 + \rho_0\left(\frac{k_2v_{02}}{\eps} + k_3v_{03}\right)v_2 + \frac{1}{\eps^2}(\eps B_{02})'(k_3a_2 - k_2a_3)		\\
			&+ B_{03}\left[-\left(k_3^2 + \frac{k_2^2}{\eps^2}\right)a_1 + \frac{k_2}{\eps^2}a_2' + k_3a_3'\right] - \frac{1}{\eps^2}\rho_0(\eps v_{02})'v_1,
		\end{alignedat}	\label{eq: linearised_v2_trans}
\end{align}
\begin{align}
	\omega\rho_0\frac{1}{\eps}v_3 &=
		\begin{alignedat}[t]{1}
			&\frac{1}{\eps}(\rho_1T_0 + \rho_0 T_1)k_3 + \frac{1}{\eps}\rho_0\left(\frac{k_2v_{02}}{\eps} + k_3v_{03}\right)v_3 + \frac{1}{\eps}B_{03}'(k_3a_2 - k_2a_3)	\\
			&- B_{02}\left[-\left(k_3^2 + \frac{k_2^2}{\eps^2}\right)a_1 + \frac{k_2}{\eps^2}a_2' + k_3a_3'\right] - \frac{1}{\eps}\rho_0v_{03}'v_1,
		\end{alignedat}	\label{eq: linearised_v3_trans}
\end{align}
\begin{align}
	\omega\rho_0\frac{1}{\eps}T_1 &=
		\begin{alignedat}[t]{1}
			&-\frac{1}{\eps}\rho_0T_0'v_1 + \frac{1}{\eps}\rho_0\left(\frac{k_2v_{02}}{\eps} + k_3v_{03}\right)T_1 - (\gamma - 1)\frac{1}{\eps}\rho_0T_0\left(v_1' - k_2v_2 - k_3v_3\right)		\\
			&- i(\gamma - 1)\frac{\left(\kappa_{\parallel, 0} - \kappa_{\bot, 0}\right)}{\eps}\frac{1}{B^2}\left(\frac{k_2B_{02}}{\eps} + k_3B_{03}\right)^2T_1 + i(\gamma - 1)\frac{1}{\eps}\left[\eps\kappa_{\bot, 0}\left(\frac{T_1}{\eps}\right)'\right]'	\\
			&- i(\gamma - 1)\frac{\kappa_{\bot, 0}}{\eps}\left(\frac{k_2^2}{\eps^2} + k_3^2\right)T_1 + i(\gamma - 1)\frac{1}{\eps}\left(\eps\kappa_{\bot, 1}T_0'\right)'	\\
			&+ i(\gamma - 1)\frac{\left(\kappa_{\parallel, 0} - \kappa_{\bot, 0}\right)}{\eps}\frac{1}{B^2}T_0'\left[\left(\frac{k_2k_3}{\eps}B_{02} + k_3^2B_{03}\right)a_2 - \left(\frac{k_2^2}{\eps}B_{02} + k_2k_3B_{03}\right)a_3\right]	\\
			&- i(\gamma - 1)\HL_0\frac{1}{\eps}\rho_1 - i(\gamma - 1)\frac{1}{\eps}\rho_0\left(\HL_T T_1 + \HL_\rho \rho_1\right) 	\\
			&+ 2i(\gamma - 1)\eta\Biggl\{ \Biggr.
					\begin{alignedat}[t]{2}
						&B_{03}'\biggl[\Bigl(\frac{1}{\eps}(a_2' - k_2a_1)\Bigr)' + \frac{k_2k_3}{\eps}a_3 - \frac{k_3^2}{\eps}a_2\biggr] \\
						&-\frac{(\eps B_{02})'}{\eps^2}\left[\Bigl(\eps(a_3' - k_3a_1)\Bigr)' + \frac{k_2k_3}{\eps}a_2 - \frac{k_2^2}{\eps}a_3\right] \Biggl. \Biggr\}
					\end{alignedat} \\
			&+ i(\gamma - 1)\frac{d\eta}{dT}\frac{1}{\eps}\left[\bigl(B_{02}'\bigr)^2 + \bigl(B_{03}'\bigr)^2 + 2\frac{\eps'}{\eps}B_{02}B_{02}' + \left(\frac{\eps'}{\eps}B_{02}\right)^2\right] T_1,
		\end{alignedat}	\label{eq: linearised_T1_trans}
\end{align}
\begin{equation}
	\omega a_1 =
			- B_{03}v_2 + \frac{1}{\eps}B_{02}v_3 - \frac{1}{\eps}v_{02}a_2' - v_{03}a_3' + \left(\frac{k_2v_{02}}{\eps} + k_3v_{03}\right)a_1
			+ i\eta\left[-\left(\frac{k_2^2}{\eps^2} + k_3^2\right)a_1 + \frac{k_2}{\eps^2}a_2' + k_3a_3'\right], 	\label{eq: linearised_a1_trans}
\end{equation}
\begin{equation}
	\omega\frac{1}{\eps}a_2 =
			- \frac{1}{\eps}B_{03}v_1 + \frac{k_3v_{03}}{\eps}a_2 - \frac{k_2v_{03}}{\eps}a_3
			+ i\eta\left[\left(\frac{1}{\eps}\left(a_2' - k_2a_1\right)\right)' + \frac{k_2k_3}{\eps}a_3 - \frac{k_3^2}{\eps}a_2\right] + i\frac{d\eta}{dT}\frac{1}{\eps}B_{03}'T_1,		\label{eq: linearised_a2_trans}
\end{equation}
\begin{equation}
	\omega a_3 =
			\frac{1}{\eps}B_{02}v_1 - \frac{k_3v_{02}}{\eps}a_2 + \frac{k_2v_{02}}{\eps}a_3
			+ i\frac{1}{\eps}\eta\left[\Bigl(\eps\left(a_3' - k_3a_1\right)\Bigr)' + \frac{k_2k_3}{\eps}a_2 - \frac{k_2^2}{\eps}a_3\right] - i\frac{d\eta}{dT}\frac{1}{\eps^2}\left(\eps B_{02}\right)'T_1.		\label{eq: linearised_a3_trans}
\end{equation}
The perturbed thermal conductivity tensor $\kappa_{\bot, 1}$ in Eq. \eqref{eq: linearised_T1_trans} written in terms of the perturbed variables is given by
\begin{equation}	\label{eq: kappa_perp_perturbed}
	\eps\kappa_{\bot, 1} = \frac{\partial \kappa_{\bot, 0}}{\partial T}T_1 + \frac{\partial \kappa_{\bot, 0}}{\partial \rho}\rho_1
							- 2\eps B_{02}\bigl(a_3' - k_3a_1\bigr)\frac{\partial \kappa_{\bot, 0}}{\partial \left(B^2\right)} + 2B_{03}\bigl(a_2' - k_2a_1\bigr)\frac{\partial \kappa_{\bot, 0}}{\partial \left(B^2\right)}.
\end{equation}
An interesting side note is that $\kappa_{\parallel, 1}$ does not appear in the equations, which is due to the fact that this term is accompanied by a $\bfb_0 \cdot \nabla T_0$ contribution, and this is zero due to the equilibrium profile in Eq. \eqref{eq: equilibrium} \citep{vanderlinden1991, vanderlinden1991TI}. We now have a system of eight ordinary differential equations in $u_1$ for the perturbed quantities $\rho_1, v_1, v_2, v_3, T_1, a_1, a_2$ and $a_3$.

\subsection{Boundary conditions}	\label{subsect: boundary_conditions}
The above system of differential equations \eqref{eq: linearised_rho_trans}-\eqref{eq: linearised_a3_trans} has to be complemented by a set of boundary conditions on both sides of the domain.
For a Cartesian geometry we look at a domain enclosed by two conducting walls. Clearly, the velocity component perpendicular to the walls has to be zero since there can not be any propagation into a solid boundary.
Mathematically, this translates into $\bfv \cdot \ben = 0$, where $\ben$ represents the normal vector to the wall. Following the same reasoning we also require that $\bfb \cdot \ben = 0$, so in terms of a vector potential this implies $\ben \times \bfA = 0$.
Hence, applying this to the set of linearised equations this means that for the Cartesian case $v_1$, $a_2$ and $a_3$ all have to be zero on the boundaries. Furthermore, since we are dealing with a perfectly conducting wall one has to take care when thermal conduction is included. In that case the rigid wall directly influences the temperature, since it acts as an energy reservoir essentially eliminating the temperature perturbation. Hence, if and only if perpendicular thermal conduction is taken into account we have to supplement the boundary conditions by the additional condition $T_1 = 0$ at the boundary. In theory there is a second possibility, which is treating the wall as a perfect insulator instead of a perfect conductor. In that case there is no heat flux, which translates to the boundary condition $T_1' = 0$ instead of $T_1 = 0$. For now we only consider the latter condition, that is, the one corresponding to a perfectly conducting wall.

In a cylindrical geometry we have the exact same boundary conditions as for the Cartesian case at the outer wall $r = R$, or at the outer edge of the accretion disk at $R$. The same is true at the inner disk edge, but for a flux tube extending to $r=0$ we have to take the regularity conditions into account when treating the cylinder axis $r = 0$, which comes down to the fact that $rv_r$ should go to zero when approaching $r = 0$. Looking back at the transformations \eqref{eq: transformation} we applied, it follows that this condition is equivalent to $v_1 = 0$.
Analogously, the same holds true for $a_2$ and $a_3$ such that these conditions are identical to the ones we applied for the Cartesian case, which is convenient implementation-wise. We again have to consider an additional condition if perpendicular thermal conduction is taken into account, since then $r\widehat{T}_1 = 0$ should also hold on the cylinder axis, which, similarly as for $v_1$, translates into $T_1 = 0$ at $r = 0$.

In the case of confinement by a perfectly conducting wall we thus have straightforward boundary conditions, that is, $v_1 = 0, a_2 = 0, a_3 = 0$ and $T_1 = 0$ for both the Cartesian and cylindrical geometries on both sides. This latter boundary condition should only be taken into account if and only if perpendicular thermal conduction is included.

\subsection{Solving the equations}
The system of equations \eqref{eq: linearised_rho_trans}-\eqref{eq: linearised_a3_trans} is solved through usage of a Finite Element discretisation. Applying a weak Galerkin formalism turns this system of equations into a generalised matrix-eigenvalue problem.
A detailed explanation on how this is done can be found in Appendix \ref{sect: numerical_approach}, where we describe the structure of the finite element approach and the matrix assembly process, along with a detailed treatment of how the boundary conditions are handled.

\section{Results}	\label{sect: results}
As is common practice when developing a new numerical code we tested \texttt{Legolas} against various results previously obtained in the literature. We divided this section into four subsections, each of which handles different physical effects.
To begin with, we discuss results for adiabatic equilibria where only gravity is included in Subsection \ref{subsect: adiabatic_cartesian_results}. In this case we can compare numerical spectra obtained through \texttt{Legolas} with analytical solutions acquired by solving dispersion relations, here we focus on stratified atmospheres containing $p$- and $g$-modes. We then move on to cylindrical geometries in \ref{subsect: adiabatic_cylindrical_results} where we first look at adiabatic flux tubes, followed by the inclusion of flow effects by considering equilibria with Kelvin-Helmholtz instabilities and Suydam cluster modes. Next the focus shifts to non-adiabatic effects in \ref{subsect: resistive_results} by looking at a resistive MHD computation for a case without gravity, where a quasi-mode is known analytically. Resistive tearing modes are also discussed, combining the effects of flow and resistivity. The final subsection \ref{subsect: nonadiabatic_results} treats the inclusion of thermal conduction and optically thin radiative cooling effects, where we look at non-adiabatic discrete Alfv\'en waves and magnetothermal modes.

\subsection{Cartesian cases: waves in stratified atmospheres}		\label{subsect: adiabatic_cartesian_results}
First of all we discuss multiple theoretical results for adiabatic equilibria in a Cartesian geometry, where only gravity is included. We consider $p$- and $g$-modes in stratified layers, and pay special attention to specific unstable branches.

\subsubsection{Gravito-MHD waves}	\label{subsect: gravito-mhd}
\begin{figure}[t]
	\centering
	\includegraphics[width=0.8\textwidth]{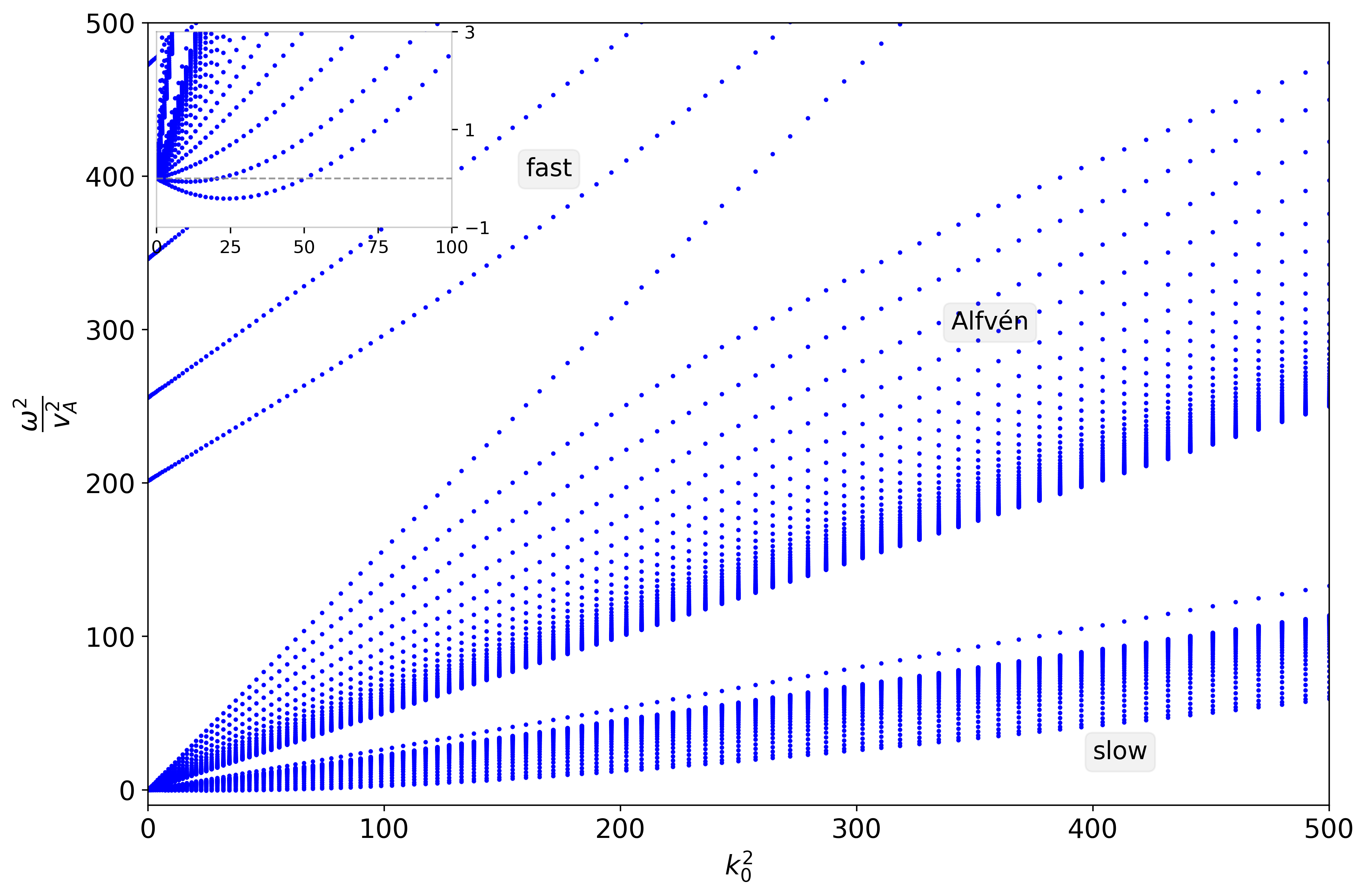}
	\caption{Spectrum of gravito-MHD modes, obtained through 100 \texttt{Legolas} runs of 351 gridpoints each. The fast (top), Alfv\'en (middle) and slow (bottom) branches of the MHD spectrum are clearly visible.
		The inset shows unstable slow modes at low frequencies.}
	\label{fig: gravito-mhd}
\end{figure}
The first test case covers gravito-MHD waves as discussed in \citet[fig. 7.9]{book_MHD}, which handles an exponentially stratified atmosphere with constant sound and Alfv\'en speeds. This magnetised atmosphere contains the generalisation of the $p$- and $g$-modes of an unmagnetised layer, and the constancy of the sound and Alfv\'en speed renders it analytically tractable, since the slow and Alfv\'en continua collapse to points.
The geometry is Cartesian, with $x \in [0, 1]$ and an equilibrium configuration given by
\begin{equation}	\label{eq: gravito-mhd_eq}
	\rho_0 = \rho_c\texp^{-\alpha x},	\qquad 		p_0 = p_c\texp^{-\alpha x},		\qquad 		\bfb_0 = B_c\texp^{-\frac{1}{2}\alpha x}\hat{\mathbf{e}}_z,		\qquad 		\alpha = \frac{\rho_c g}{p_c + \frac{1}{2}B_c},
\end{equation}
where $p_c$ and $B_c$ are taken to be 0.5 and 1, respectively, as to yield a plasma beta equal to unity. The parameter $\alpha$ is taken to be 20, which, together with $g = 20$, is used to constrain the value for the constant $\rho_c$.
These four equations completely determine the equilibrium configuration, since the temperature is $T_0 = p_0/\rho_0$, following the ideal gas law. The spectrum discussed in \citet{book_MHD} is actually the solution to the analytic dispersion relation for gravito-MHD waves, which shows the squared eigenvalue as a function of wave number for a fixed angle $\theta = \pi/4$ between the wave vector $\mathbf{k_0}$ and the magnetic field $\bfb_0$. However, the spectrum as calculated by \texttt{Legolas} corresponds to one single equilibrium configuration, meaning one value for $k_y$ and $k_z$. In order to reproduce figure 7.9 from \citet{book_MHD} and compare the results, we performed 100 different runs where the equilibrium parameters in Eq. \eqref{eq: gravito-mhd_eq} remained unchanged, but $k_y$ and $k_z$ took on 100 different values between $0$ and $\sqrt{250}$ as to yield a wave number range for $k_0^2$ between 0 and 500. Since the magnetic field is purely aligned with the $z$-axis we can write $k_\parallel = k_z = k_0\cos(\theta)$ and $k_\bot = k_y = k_0\sin(\theta)$. All runs were performed using 351 gridpoints, yielding a matrix size of $5616 \times 5616$.

Our results are shown in Figure \ref{fig: gravito-mhd}, where every vertical collection of points at the same $k_0^2$ value represents one single \texttt{Legolas} run. Since we are in an MHD regime with $\beta = 1$, the three MHD subspectra can be clearly distinguished, showing the fast $p$-modes (top-left branches), Alfv\'en $g$-modes (middle branches) and slow $g$-modes (bottom branches). The inset shows a zoom-in near the marginal frequency of the spectrum, showing unstable ($\omega^2 < 0$) slow MHD modes.
These long-wavelength unstable modes are related to the Parker instabilities, due to magnetic buoyancy, as we will show in Section \ref{subsect: quasi-parker}. Note that since this case is adiabatic and fully self-adjoint, every individual MHD spectrum is left-right and up-down symmetric in the complex eigenfrequency plane, but this aspect is hidden from the $\omega^2-k_0^2$ view shown here.

\subsubsection{Quasi-Parker instabilities}	\label{subsect: quasi-parker}
Next we discuss a modified case of the gravito-MHD waves, namely a spectrum showing quasi-Parker instabilities as done in \citet[fig. 12.2]{book_MHD}. The difference with the previous case is that a fully analytic description is no longer possible, since the introduction of magnetic shear leads to continuous ranges in the MHD spectrum. Instead of showing the spectrum for one single value for $\theta$, we now vary the direction of the wave vector
$\mathbf{k_0}$ between 0 and $\pi$. The equilibrium configuration is similar to the one in Section \ref{subsect: gravito-mhd}, given in Cartesian geometry by
\begin{equation}	\label{eq: quasi-parker_eq}
	\begin{aligned}
		\rho_0 &= \rho_c\texp^{-\alpha x},	\qquad\qquad 		p_0 = p_c\texp^{-\alpha x},		\qquad\qquad 		\alpha = \frac{\rho_c g}{p_c + \frac{1}{2}B_c},	\\
		B_{02} &= B_c\texp^{-\frac{1}{2}\alpha x}\sin(\lambda x),			\qquad\qquad 		B_{03} = B_c\texp^{-\frac{1}{2}\alpha x}\cos(\lambda x),
	\end{aligned}
\end{equation}
where magnetic shear was introduced through the parameter $\lambda$.
\begin{figure}[t]
	\centering
	\includegraphics[width=\textwidth]{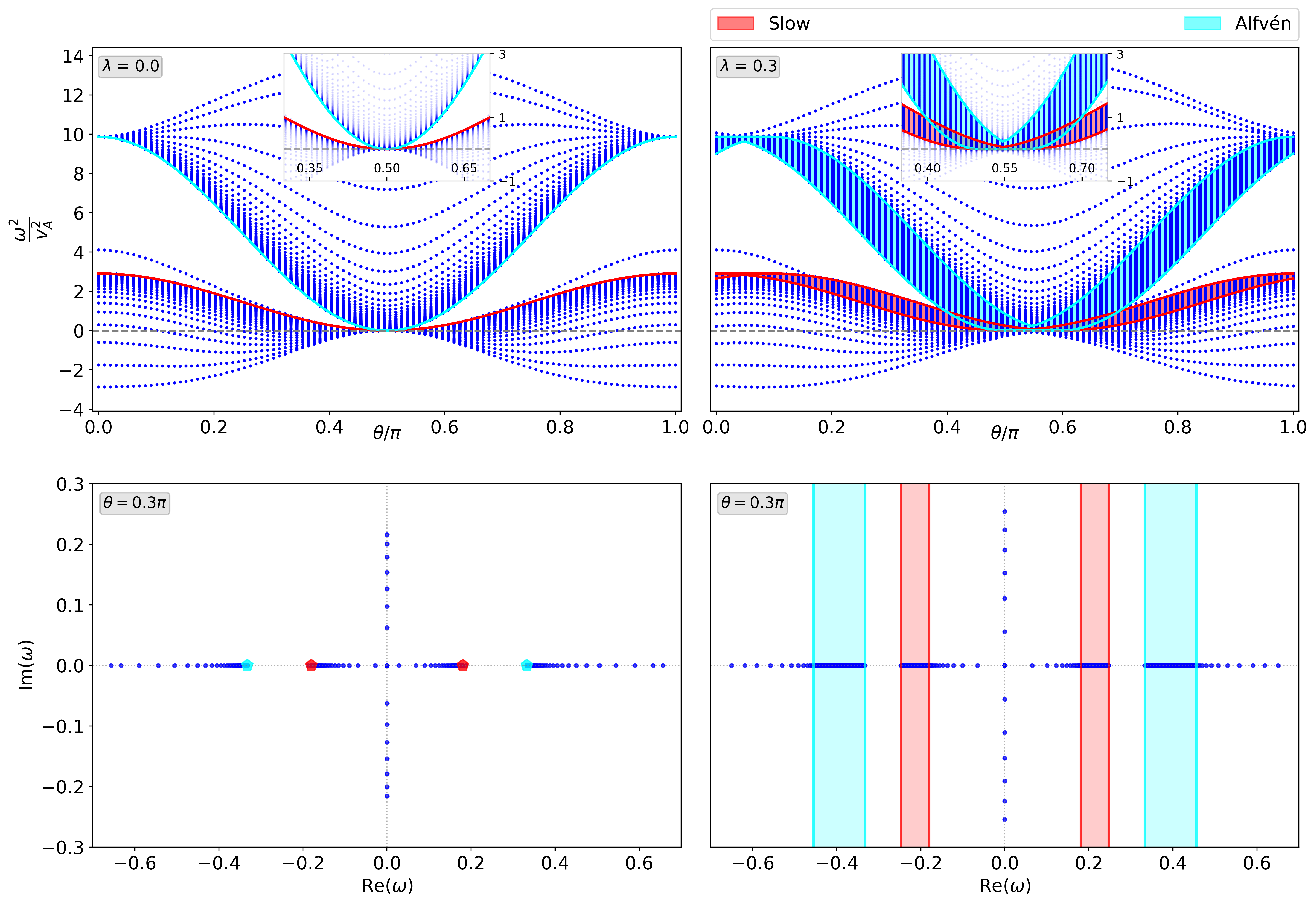}
	\caption{Spectrum showing Parker and quasi-Parker modes without (left) and with (right) magnetic shear. The slow and Alfv\'en continua are shown in red and cyan, respectively, where the insets zoom into the region of quasi-interchange modes.
			 The bottom row of panels show the eigenfrequency view for the single case $\theta = 0.3\pi$. The continua are again annotated on the figures, visualising the collapsed single point values (left) as well as the genuine continuum ranges (right).
		 	 The grey dashed line in the top two panels denotes $\omega = 0$.}
	\label{fig: quasi-parker}
\end{figure}
The quantities $\alpha$ and $B_c$ are assigned the same values as in Eq. \eqref{eq: gravito-mhd_eq}, except that $g = 0.5$ and $p_c = 0.25$ which yields a plasma beta $\beta = 0.5$.
The wave vectors are given by $k_y = \pi\sin(\theta)$ and $k_z = \pi\cos(\theta)$, such that $k_0^2 \approx 10$. The angle $\theta$ was varied between 0 and $\pi$ for a total of 100 runs at 351 gridpoints each, shown in Figure \ref{fig: quasi-parker}.

The left panels handle the case without magnetic shear, that is, $\lambda = 0$, which basically reduces to the one from the previous subsection. In this case the slow and Alfv\'en continua collapse into single point values, denoted in red and cyan, respectively. The right panels show the same configuration where $\lambda = 0.3$ was taken, introducing magnetic shear, which introduces genuine continua seen as bands. These continua affect the overall stability, and organise the entire MHD spectrum: all discrete modes are fully aware of the essential spectrum formed by these (slow and Alfv\'en) continua and the (fast) accumulation points at infinite frequency. All features of the original figure in \citet{book_MHD} are reproduced. The inset zooms into the region where both continua overlap, showing quasi-interchange and interchange instabilities. Once more, each run shown here collectively in Figure \ref{fig: quasi-parker} actually has a spectrum that is left-right and up-down symmetric in the eigenfrequency plane. This is depicted on the bottom two panels, which show the eigenfrequency view for one single case ($\theta = 0.3\pi$). The continuum ranges separate nicely: the collapsed single point values are denoted by cyan (Alfv\'en) and red (slow) points on the left panel, the genuine continua are shown with cyan and red bands on the right panel. The instabilities themselves are situated on the (positive) imaginary axis, due to the self-adjointness of the eigenvalue problem mentioned earlier.

As explained in \citet{book_MHD}, we see from this eigenmode computation that the Parker instability, which is there for $\mathbf{k_0}$ parallel to $\bfb_0$, becomes a quasi-Parker instability away from perfect alignment, and connects smoothly to well-known quasi-interchange instabilities that occur here (marginally) away from perpendicular orientation. Quantifying how the equilibrium parameters influence the growth rates of these unstable branches can only be done numerically, e.g. with \texttt{Legolas}.

\newpage
\subsection{Adiabatic, cylindrical cases}	\label{subsect: adiabatic_cylindrical_results}
Next we move on to cylindrical configurations, which provide tests for the scale factor $\eps$ in the equations. Analytical results from the literature are again well reproduced.
Furthermore we look at different spectra previously obtained by the LEDA code, discussed in various papers, and compare those with the new spectra from \texttt{Legolas}.
\subsubsection{Magnetic flux tubes}
The first case that we describe in this subsection is a magnetic flux tube embedded in a uniform magnetic environment, discussed in \citet{book_roberts}. The equilibrium configuration is simple, in the sense that we have a uniform magnetic field aligned with the $z$-axis both inside and outside of the flux tube, with a similar structure for the other equilibrium parameters:
\begin{equation}
	B_0(r), \rho_0(r), p_0(r), T_0(r) =
	\begin{cases}
		B_0, \rho_0, p_0, T_0,	&\qquad r < a	\\
		B_e, \rho_e, p_e, T_e,	&\qquad r > a
	\end{cases}
\end{equation}
where the subscripts $0$ and $e$ refer to values inside the tube and for the environment, respectively. The outer radius of the tube is denoted by $a$ and hence represents a discontinuous interface between the tube itself and the environment. Since total pressure balance should be preserved across the boundary, which is something that follows from Eq. \eqref{eq: eq_condition_1}, this yields a relation between pressures and magnetic field components inside and outside of the tube, which in turn implies a connection between the plasma densities, sound speeds and Alfv\'en speeds across the boundary:
\begin{equation}	\label{eq: fluxtube_relations}
	p_0 + \frac{1}{2}B_0^2 = p_e + \frac{1}{2}B_e^2,		\qquad\qquad 	\frac{\rho_e}{\rho_0} = \frac{c_s^2 + \frac{1}{2}\gamma c_A^2}{c_{se}^2 + \frac{1}{2}\gamma c_{Ae}^2},
\end{equation}
where $c_s^2 = \gamma p_0/\rho_0$ and $c_A^2 = B_0^2/\rho_0$ denote the sound speed and Alfv\'en speed, respectively, in which the values outside of the flux tube are used if there is a subscript $e$ present.

It should be noted that this extremely simple equilibrium configuration is the standard case used in many solar coronal loop seismology efforts. Since it simply has two uniform media (one inside the tube and one in its exterior), it has no continuous spectra (they reduce to point values), but the interface makes it possible to again have surface modes that would be affected by true radial variation. Also note that these flux tubes have only stable waves, but we can distinguish between body and surface waves, depending on the variation of the eigenfunctions within the flux tube. In the exterior of the flux tube all eigenfunctions are exponentially varying.

We should also clarify here that the original dispersion relation as given in \citet{book_roberts} assumes a flux tube embedded in an environment extending towards infinity, while \texttt{Legolas} on the other hand assumes a fixed wall boundary at the outer edge of the domain. Hence, we assume here that the domain is situated in $r \in [0, 10]$ with the inner flux tube wall at $r = 1$, in order to minimise the outer wall influence. However, this introduces an additional computational challenge, in the sense that we are (mainly) interested in the behaviour of the inner modes, since we know that the outer modes all have exponentially varying eigenfunctions which decay to infinity (or towards our far-away outer wall). Hence, in order to resolve those inner waves huge resolutions are needed due to the $1:10$ ratio. In order to circumvent this issue we used a simple prescription for mesh refinement, that is, a $60-30-10$ division of the initial nodes. This means that $60\%$ of the gridpoints are used for the inner tube region $r \in [0, 0.95]$, $30\%$ of the gridpoints are located near the transition region $r \in [0.95, 1.05]$, and the remaining $10\%$ are used for the environment $r \in [1.05, 10]$.

\begin{figure}[t]
	\centering
	\includegraphics[width=\textwidth]{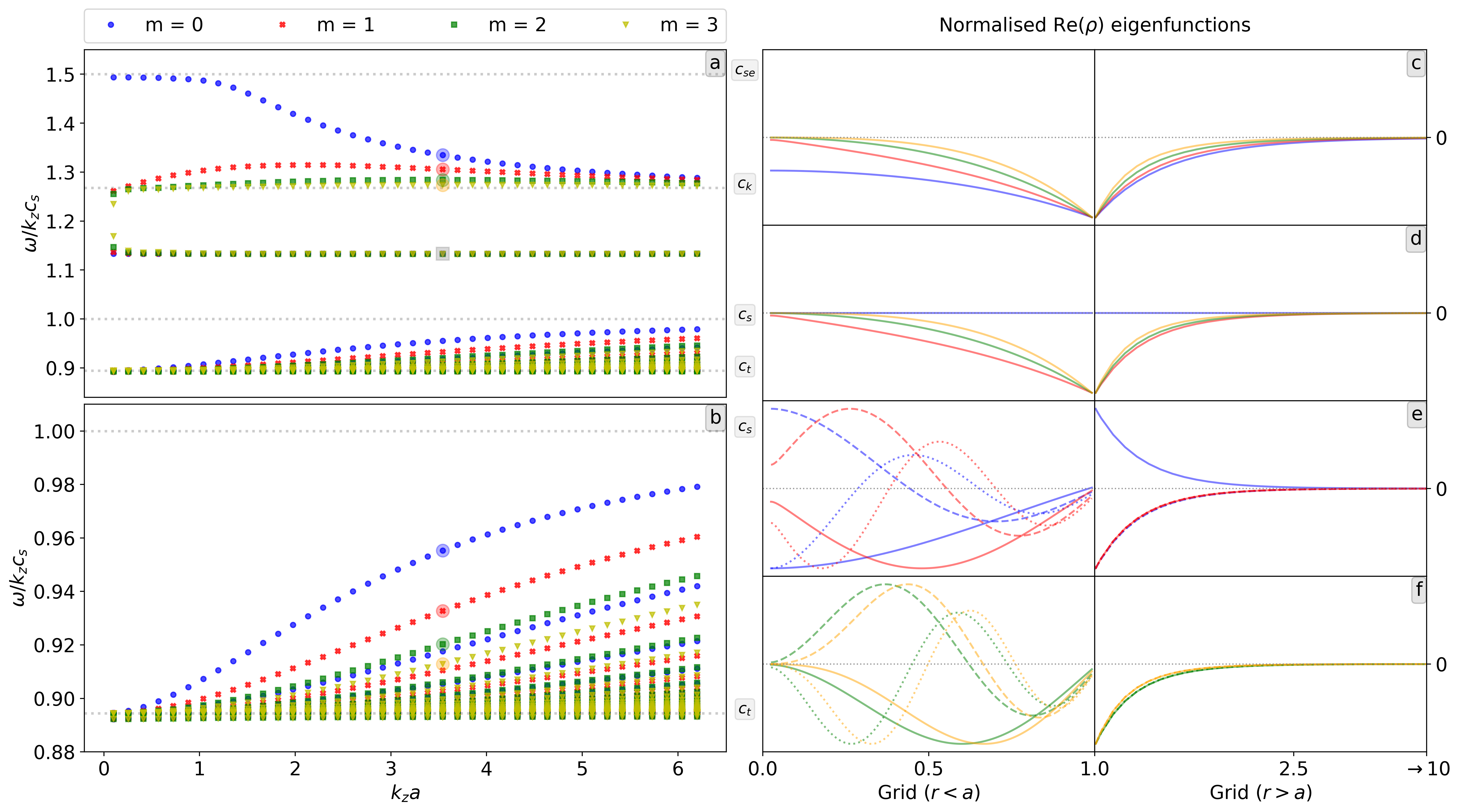}
	\caption{Panel $a$: spectrum showing the dispersion relation for a flux tube under photospheric conditions. The dimensionless phase speed $\omega/k_zc_s$ is displayed as a function of the dimensionless wave number $k_za$,
			 for four values of the azimuthal wave number $m = 0$ (blue dots), $m = 1$ (red crosses), $m = 2$ (green squares) and $m = 3$ (yellow triangles). Both sound speeds, the kink speed and the tube speed are denoted using dashed grey horizontal lines, all
			 normalised to $c_s$. Panel $b$: zoom-in of panel $a$ between the tube speed $c_t$ and internal sound speed $c_s$.
			 Panels $c$ and $d$: eigenfunctions of the modes annotated with circles ($c$) and squares ($d$) on the top-left panel $a$. Panel $e$: first three eigenfunctions of the $m=0$ and $m=1$ body wave sequence on the bottom-left panel $b$.
		 	 Panel $f$: first three eigenfunctions of the $m=2$ and $m=3$ body wave sequence. For both panels $e$ and $f$ the $n=1$, $n=2$ and $n=3$ modes are shown in solid, dashed and dotted lines, respectively, only the first mode is annotated on the bottom-left panel. Every eigenfunction on panels $c$ through $f$ is normalised to its maximum absolute value in that particular grid interval.}
	\label{fig: photospheric_fluxtube}
\end{figure}

\textbf{a) Photospheric flux tube.} First we look at a flux tube under photospheric conditions, that is, an equilibrium for which $c_{Ae} < c_s < c_{se} < c_A$. More specifically, we take $c_{Ae} = c_s/2$, $c_{se} = 3c_s/2$ and $c_A = 2c_s$ following
\citet[fig. 6.5]{book_roberts}. The relations between the inner and outer regions of the flux tube follow straightforward from Eq. \eqref{eq: fluxtube_relations}, and hence we only have two degrees of freedom, namely $\rho_0$ and $p_0$ which are both taken to be unity, with $\gamma = 5/3$. This results in $\rho_0 \approx 0.567~\rho_e$, $c_t \approx 0.89~c_s$, $c_k \approx 0.63~c_A \approx 1.27~c_s$. Here we also introduced the tube and kink speeds, given by
\begin{equation}
	c_t^2 = \frac{c_s^2 c_A^2}{c_s^2 + c_A^2},	\qquad\qquad\qquad	 c_k^2 = \frac{\rho_0 c_A^2 + \rho_e c_{Ae}^2}{\rho_0 + \rho_e},
\end{equation}
using the same notation as in Eqs. \eqref{eq: fluxtube_relations}. As described before we take $r \in [0, 10]$ and place the flux tube boundary at $a = 1$. Next we perform 40 runs at 300 gridpoints each for four azimuthal wave numbers $m = 0$ to $m = 3$. For the wave number $k_3 = k_z$ we take 40 values in such a way that the dimensionless wave number $k_za$ has values in $[0, 6.2]$. The spectrum showing the dispersion relation, where the dimensionless phase speed $\omega/k_zc_s$ is plotted as a function of $k_z a$, is depicted in Figure \ref{fig: photospheric_fluxtube}. All speeds indicated on the figure are normalised to the internal sound speed $c_s$. Panel $a$ clearly shows the fast surface waves, including the sausage ($m = 0$), kink ($m = 1$) and first two fluting modes ($m=2$ and $m=3$). The eigenfunctions in the top-right panel $c$ correspond to the modes annotated with a transparent circle on panel $a$, with colours indicating the mode number $m$. The left and right side of panels $c$ through $f$ show the eigenfunctions for the inner and outer regions of the flux tube, respectively. All eigenfunctions are normalised to their maximum value, and all eigenvalues having a normalised phase speed larger than 1.5 are not shown. It should be noted that the first three runs, that is, the first three dots according to the $k_z a$ axis, were done using 1001 gridpoints.
The reason for this is that when we divide the eigenvalues $\omega$ by $k_z$ small errors are increased selectively for small $k_z$, explaining why those modes seem slightly scattered, and hence why we have to employ such high resolutions in order to minimise said error.

The panel $b$ of Figure \ref{fig: photospheric_fluxtube} zooms in between the tube speed ($c_t$) and internal sound speed ($c_s$), showing a clear representation of the various body waves that accumulate to the tube speed at long wavelengths. Panel $e$ shows the first three modes of the $m=0$ and $m=1$ sequences in solid, dashed and dotted lines, respectively; that is, the first mode in the sequence (solid line) corresponds to the mode annotated on panel $b$. The next two modes in that sequence are the next two blue dots moving vertically downwards (same $k_z a$ value), these are not annotated to avoid cluttering the figure. Analogously, panel $f$ shows the first three modes of the $m=2$ and $m=3$ sequences, with everything colour-coded according to the legend. As indicated before all eigenfunctions in the outer region are exponentially varying. The panels $a$ and $b$ in Figure \ref{fig: photospheric_fluxtube} reproduce the analytical results in \citet[fig. 6.5]{book_roberts}.

One mode has not yet been discussed, and that is the horizontal line of modes between the internal sound speed and kink speed. The eigenfunctions are shown in panel $d$ and correspond to the annotated squares on the top-left panel $a$.
These modes are not present in the original work, and it is not a priori clear how to interpret these. However, what we do know is that their phase speed is approximately $0.5(c_k + c_s)$ and that these modes are degenerate, meaning that their position does not change when the mode number $m$ or wave number $k_z$ changes. The position of the outer wall also does not seem to have any influence on their value. This, together with the fact that the eigenfunctions seem to indicate that these are surface waves, strengthens the belief that these solutions could be actual waves and not some numerical remnant.

\begin{figure}[t]
	\centering
	\includegraphics[width=\textwidth]{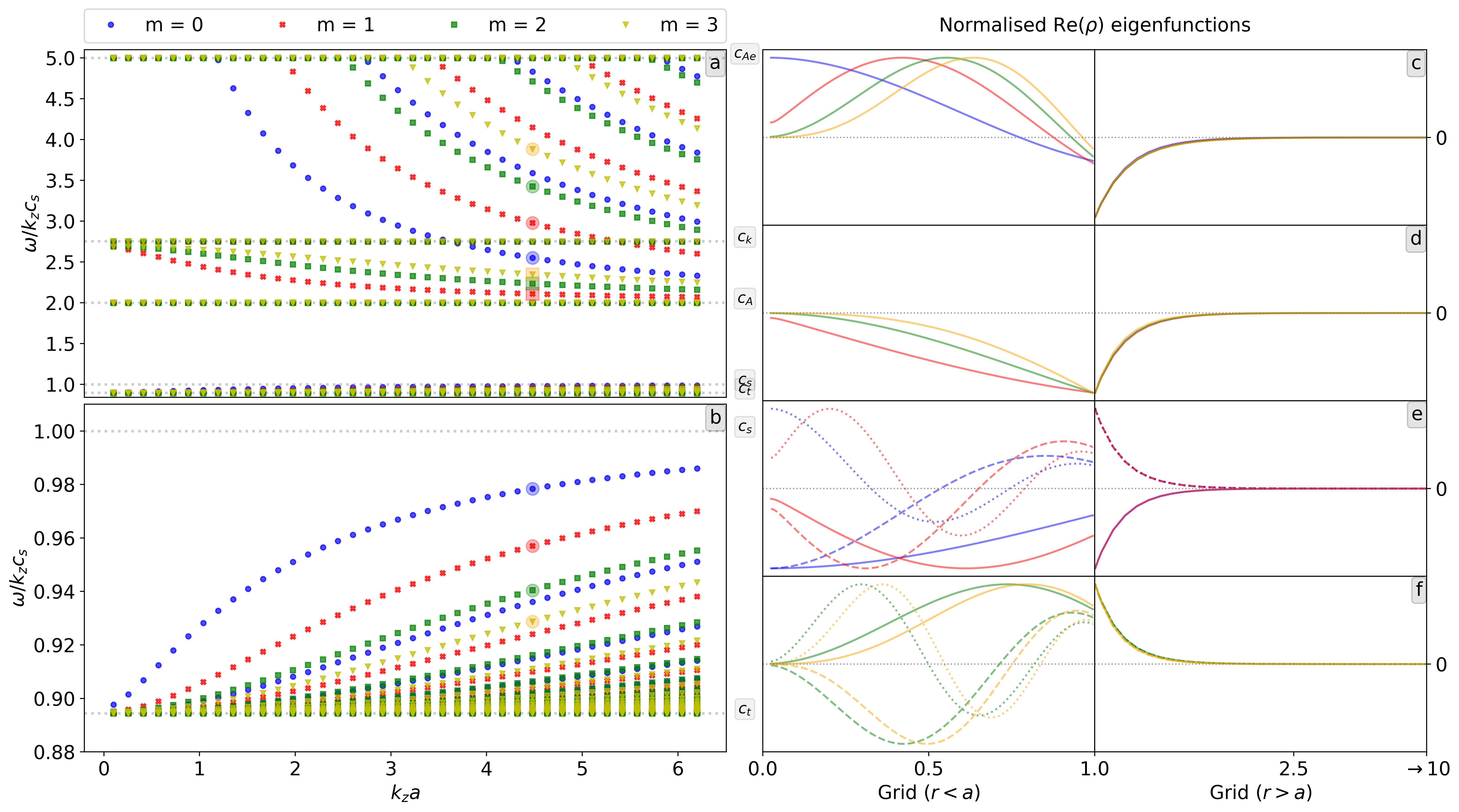}
	\caption{Panel $a$: spectrum showing the dispersion relation for a flux tube under coronal conditions. The dimensionless phase speed $\omega/k_zc_s$ is displayed as a function of the dimensionless wave number $k_za$,
		for four values of the azimuthal wave number $m = 0$ (blue dots), $m = 1$ (red crosses), $m = 2$ (green squares) and $m = 3$ (yellow triangles). Both Alfv\'en speeds, together with the kink, tube and sound speeds are denoted by grey horizontal lines and are all normalised to $c_s$. Panel $b$ zooms in on the tube speed-related body mode sequences. Panel $b$: zoom-in of panel $a$ between the tube speed $c_t$ and internal sound speed $c_s$.
		Panels $c$ and $d$: eigenfunctions of the modes annotated with circles ($c$) and squares ($d$) on the top-left panel $a$.
		Panel $e$: first three eigenfunctions of the $m=0$ and $m=1$ body wave sequence on the bottom-left panel $b$.
		Panel $f$: first three eigenfunctions of the $m=2$ and $m=3$ body wave sequence. For both panels $e$ and $f$ the $n=1$, $n=2$ and $n=3$ modes are shown in solid, dashed and dotted lines, respectively,
		only the first mode is annotated on the bottom-left panel $b$.
		Every eigenfunction on panels $c$ through $f$ is normalised to its maximum absolute value in that particular grid interval.}
	\label{fig: coronal_fluxtube}
\end{figure}

\textbf{b) Coronal flux tube.} The second application of the magnetic flux tube is one under coronal conditions, that is, an equilibrium for which $c_{se} < c_s < c_A < c_{Ae}$. More specifically, we take $c_{Ae} = 5c_s$, $c_{se} = c_s/2$ and $c_A = 2c_s$. Analogous to the previous case, the relations between the equilibrium values inside and outside of the flux tube follow from Eq. \eqref{eq: fluxtube_relations}, where we again take $p_0$ and $\rho_0$ equal to one. This results in $\rho_0 \approx 4.86~\rho_e$, $c_t \approx 0.89~c_s$ and $c_k \approx 1.38~c_A \approx 0.55~c_{Ae}$, and we use the same values as for the photospheric case for $k_z$ and the flux tube and outer wall boundaries. Similar to case \textbf{a)} we perform 40 runs at 300 gridpoints each for four values of $k_2 = m$ (where again the first three runs have 1000 gridpoints) and plot the spectrum showing the dispersion relation in Figure \ref{fig: coronal_fluxtube}. Again all speeds indicated on the figure are normalised to the sound speed $c_s$. The top-left panel $a$ focuses on the fast body waves, the bottom-left panel $b$ on the slow body waves. Panel $c$ shows the eigenfunctions corresponding to the four body waves indicated with circles on panel $a$. Panel $d$ depicts eigenfunctions of the modes annotated with squares between the internal Alfv\'en ($c_A$) and kink ($c_k$) speeds, representing the kink $m=1$ and first two fluting ($m=2$ and $m=3$) modes.

Similar to Figure \ref{fig: photospheric_fluxtube} panels $e$ and $f$ represent the first three body modes in the $m=0$ and $m=1$ sequences (panel $e$), of which the first mode is indicated on panel $b$.
Panel $f$ shows the first three modes of the $m=2$ and $m=3$ sequences, with the first, second and third mode indicated with a solid, dashed and dotted line, respectively. The colours of panels $c$ through $f$ are consistent with the legend.
Figure \ref{fig: coronal_fluxtube} reproduces the analytical results in \citet[fig. 6.7]{book_roberts}, which are based on the analytic dispersion relation containing Bessel functions.

\subsubsection{Tokamak constant current}
Next we discuss an example initially given in \citet{kerner1985}, which shows the ideal MHD spectrum in the presence of an unstable $m = -2$ interchange mode in a cylindrical geometry. We start from a so-called tokamak current profile, in which an axial current density of the form $\boldsymbol{j_0} = (0, 0, j_0\left(1 - r^2\right)^\nu)$ is assumed, with $j_0$ a given constant. This yields a twisted magnetic profile in which the longitudinal component $B_{0z}$ is uniform and equals one, while the poloidal component $B_{0\theta}$ is given by
\begin{equation}	\label{eq: Bnu_profile}
	B_{0\theta} = \frac{j_0}{2r(\nu + 1)}\left[1 - \left(1 - r^2\right)^{\nu + 1}\right],
\end{equation}
for a given value of $\nu$. For the equilibrium considered here we take $\nu = 0$, making the current profile constant over the flux tube. This means that $B_{0\theta}$ has a linear profile in $r$ such that the magnetic field lines have a constant pitch and the current is distributed equally in the plasma. An expression for the pressure (and hence temperature) can be found by integrating \eqref{eq: eq_condition_1} and assuming that, for example, the pressure vanishes at the outer boundary, resulting in a parabolic pressure profile. Hence, for a cylindrical geometry in which $r \in [0, 1]$ this yields the following equilibrium configuration:
\begin{equation}	\label{eq: kerner-tokamak}
	\rho_0 = 1,		\qquad\qquad 	p_0 = \frac{1}{4}j_0^2\left(1 - r^2\right),		\qquad\qquad 	B_{02} = \frac{1}{2}j_0 r,		\qquad\qquad 		B_{03} = 1,
\end{equation}
where we assumed a uniform density. We introduce an additional parameter $q$, called the safety factor, given by
\begin{equation}
	q(r) = \frac{r k_z B_{0z}}{B_{0\theta}} = \frac{2 k_z}{j_0}.
\end{equation}
\begin{figure}[t]
	\centering
	\includegraphics[width=0.9\textwidth]{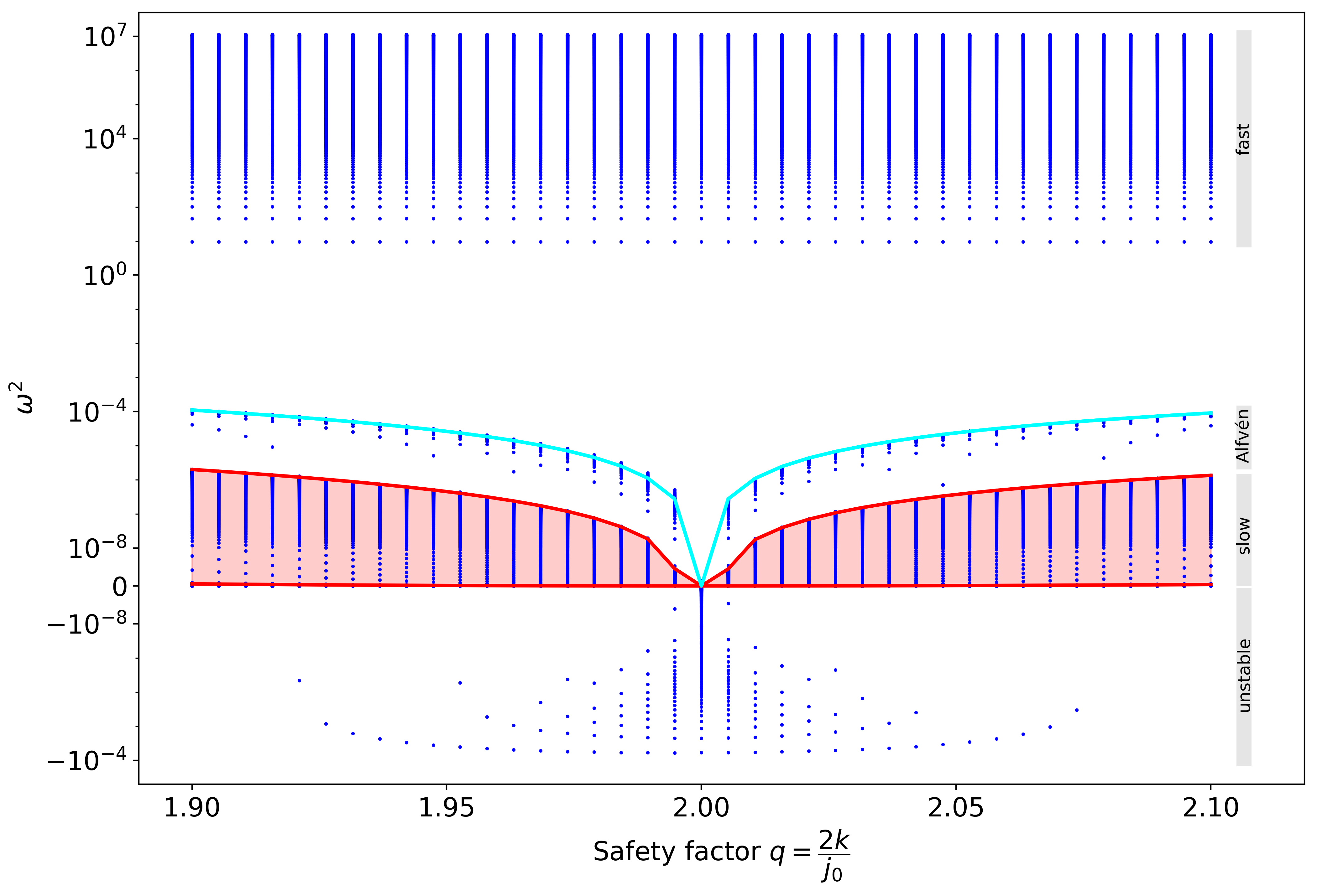}
	\caption{Complete MHD spectrum for a tokamak current profile in the presence of $m = -2$ interchange modes. The squared eigenvalues are plotted against the safety factor $q$, instabilities correspond to $\omega^2 < 0$.
		The various branches are indicated on the right side of the figure.}
	\label{fig: kerner-tokamak}
\end{figure}
We performed 39 runs, varying the q-factor between 1.9 and 2.1 in order to probe the regime containing the unstable $m = -2$ interchange mode, which is associated with the vanishing of the factor $m + kq$, that is, the $\bfk \cdot \bfb$ product. This implicitly constrains the value for $j_0$, and we assigned $k_2 = m = -2$ and $k_3 = k = 0.2$ for all runs. It should be noted that this particular equilibrium configuration requires a high resolution near $q = 2$ to correctly resolve the unstable modes. We thus used 501 gridpoints for all runs. The complete spectrum is shown in Figure \ref{fig: kerner-tokamak}, where the squared eigenvalues are plotted as a function of the safety factor. The 3 main branches, that is, fast, Alfv\'en and slow, are denoted on the right side of the figure, as well as the region where the modes become unstable ($\omega^2 < 0$). We see that in the region near the $m = -2$ interchange instabilities the slow and Alfv\'en modes collapse to zero, which is due to the vanishing of the combination $F = mB_{\theta}/r + kB_z$ in the ideal MHD equations \citep{book_MHD}. The slow and Alfv\'en continua are annotated on the figure in red and cyan, respectively. The Alfv\'en continuum is collapsed to a single point for this equilibrium configuration, while the slow continuum covers a range in frequency. Note in particular how the full spectrum ranges over many orders of magnitude in the $\omega^2$ view shown here: an intrinsic property and challenge posed by MHD spectral theory.

\subsubsection{KH and CD instabilities}	\label{subsect: kh_cd}
As a first test for the inclusion of flow into the equations, we look at the interaction between Kelvin-Helmholtz (KH) and current-driven (CD) instabilities in a magnetised astrophysical jet, following \citet{baty2002}. This model uses a cylindrical jet with a supersonic background flow aligned with the axis and sheared in the radial direction. The equilibrium configuration is taken such that KH surface modes can develop, and is generally given by
\begin{equation}	\label{eq: kh_cd}
	\begin{aligned}
		\rho_0 &= 1,											\qquad\qquad&		v_{02} &= 0,		\qquad\qquad& 	v_{03} &= \frac{1}{2}V\tanh\left(\frac{r_j - r}{a}\right),		\\
		B_{02} &= B_{\theta 0} \frac{r r_c}{r_c^2 + r^2},		\qquad\qquad&		B_{03} &= B_{z0},	\qquad\qquad&	T_0 &= \frac{p_0}{\rho_0} - \frac{B_{\theta 0}^2}{2\rho_0}\left(1 - \frac{r_c^4}{\left(r_c^2 + r^2\right)^2}\right).
	\end{aligned}
\end{equation}
Here $r_j$ denotes the jet radius and $r_c$ quantifies the radial variation, and they are taken to be $r_j = 1$ and $r_c = 0.5$, with $r \in [0, 2]$. The parameter $V$ represents the amplitude of the velocity, given by $V = 1.63$, while the chosen velocity profile ensures that the shear layer is situated at the jet radius with a radial width given by $a = 0.1r_j$. Both the density $\rho_0$ and the pressure on-axis $p_0$ are chosen to be equal to unity. The parameters $B_{\theta 0}$ and $B_{z0}$ control the amplitude and twist of the magnetic field, respectively, as explained in \citet{baty2002}. The original work discusses three different magnetic field configurations, we choose the profile with
\begin{equation}
	B_{\theta 0} = 0.4\left(\frac{r_c^2 + r_j^2}{r_jr_c}\right),	 \qquad\qquad 	B_{z0} = 0.25,
\end{equation}
such that $B_{02} = 0.4$ at the jet radius. Furthermore, the azimuthal and longitudinal wave numbers are taken to be $k_2 = m = -1$ and $k_3 = k = \pi$.

\begin{figure}[t]
	\centering
	\includegraphics[width=\textwidth]{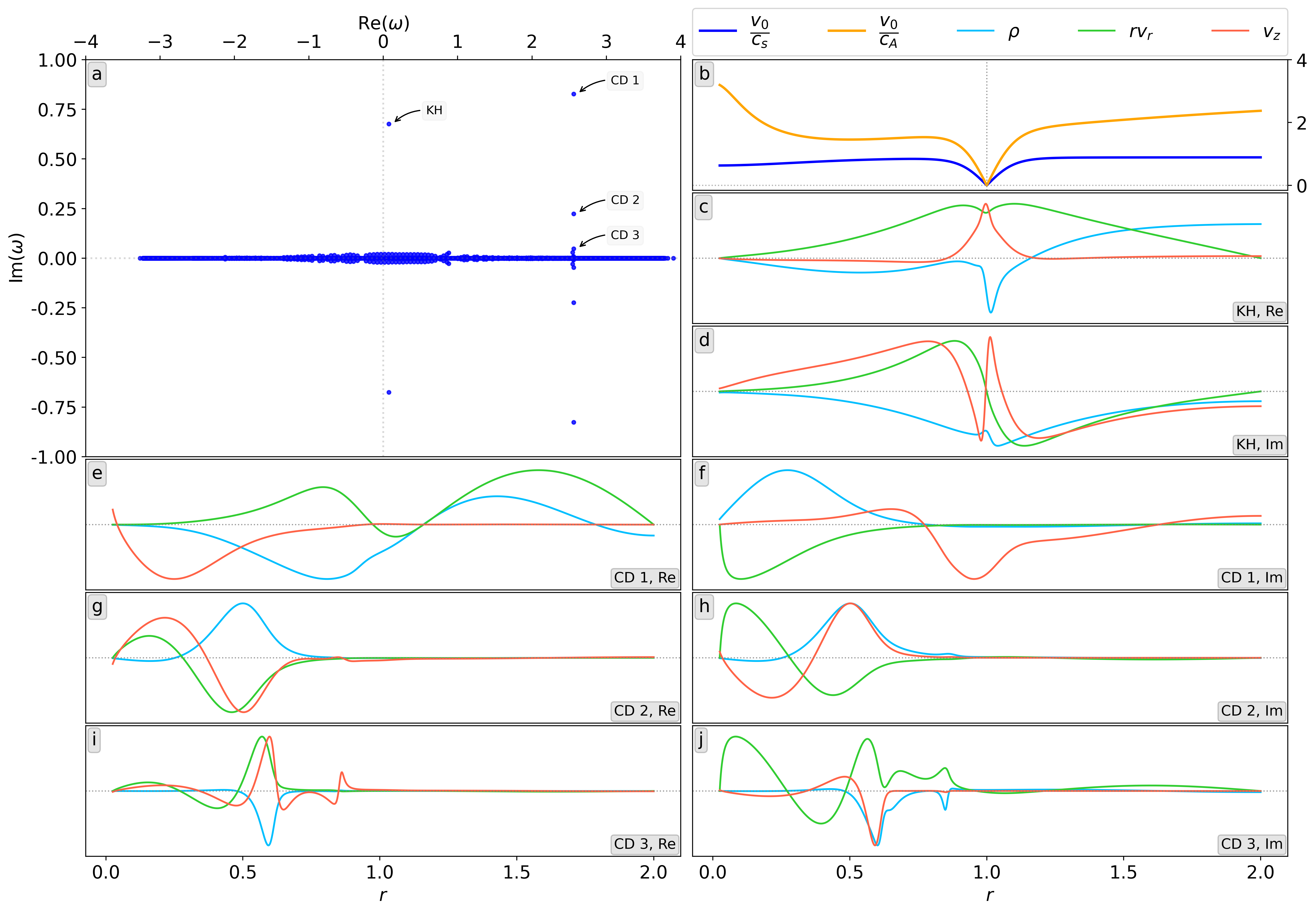}
	\caption{Panel $a$: MHD spectrum for the equilibrium configuration given in \eqref{eq: kh_cd}. The KH and CD instabilities are indicated on the panel. Panel $b$: sonic and Alfv\'enic Mach numbers.
			 The other panels show the real and imaginary parts of the $\rho$ (blue), $rv_r$ (green) and $v_z$ (red) eigenfunctions, for the KH mode ($c-d$), first CD mode ($e-f$), second CD mode ($g-h$) and third CD mode ($i-j$), respectively.}
	\label{fig: kh_cd}
\end{figure}

The entire spectrum is calculated at high resolution using 501 gridpoints and shown in Figure \ref{fig: kh_cd}. Panel $a$ depicts the full slow and Alfv\'en spectra with the KH and first three CD unstable modes ($\omega_I > 0)$ denoted on the figure itself.
The real and imaginary parts of the $\rho$, $rv_r$ and $v_z$ eigenfunctions for each of these modes are shown on the subsequent panels. Those for the KH mode (panels $c-d$) are localised around the jet radius $r = 1$, which is also the point where the sonic and Alfv\'enic Mach numbers drop to zero (panel $b$). Panels $e$ through $j$ depict the eigenfunctions of the first three CD modes, with the first, second and third shown on the first, second and third row of the bottom panels, respectively. The left column shows the real part, the right column the imaginary part of the eigenfunction. The CD modes have an increasing number of nodes on $r \in [0, 1]$, which is most clearly visible by looking at the $rv_r$ eigenfunction (green): no nodes for the first CD mode (panels $e-f$), one node
for the second CD mode (panels $g-h$) and two nodes for the third CD mode (panels $i-j$).

Note that here, we are still adiabatic such that the up-down symmetry (relating to time reversal) is still present in the eigenfrequency plane, but the introduction of equilibrium flow caused left-right symmetry breaking between forwards and backwards propagating modes. As pointed out in \citet{goedbloed2018web1}, the study of MHD spectra of stationary (with flow) equilibria is still governed by two self-adjoint operators, but as seen in Figure \ref{fig: kh_cd}, modes can enter the complex plane at various locations (identified by the spectral web \citep{goedbloed2018web2}). The correspondence with the original figure in \citet{baty2002} is one-to-one for the KH and CD modes, but here we have a lot more detail near the axes due to the higher resolution. We will discuss this resolution aspect in Section \ref{sect: convergence}.

\subsubsection{Suydam cluster modes}
Next we look at Suydam cluster modes in a cylindrical geometry, which arise from the presence of a Suydam surface in the equilibrium configuration, that is, a location where $\bfk \cdot \bfb_0 = 0$. Shear flow effects are included, and the equilibrium is given by
\begin{equation}	\label{eq: suydam}
	\begin{aligned}
		\rho_0 &= 1,							\qquad\qquad& 		v_{02} &= 0,								\qquad\qquad&		v_{03} &= v_{z0}\left(1 - r^2\right),			\\
		B_{02} &= J_1(\alpha r),				\qquad\qquad&		B_{03} &= \sqrt{1 - P_1}J_0(\alpha r),		\qquad\qquad& 		p_0 &= P_0 + \dfrac{1}{2}P_1J_0^2(\alpha r),
	\end{aligned}
\end{equation}
where $\alpha = 2$, $P_0 = 0.05$, $P_1 = 0.1$ and $v_{z0} = 0.14$, the functions $J_0$ and $J_1$ denote the Bessel functions of the first kind. The wave numbers were chosen to be $k_2 = m = 1$ and $k_3 = k = -1.2$, ensuring a Suydam surface at $r \approx 0.483$.

\begin{figure}[t]
	\centering
	\includegraphics[width=\textwidth]{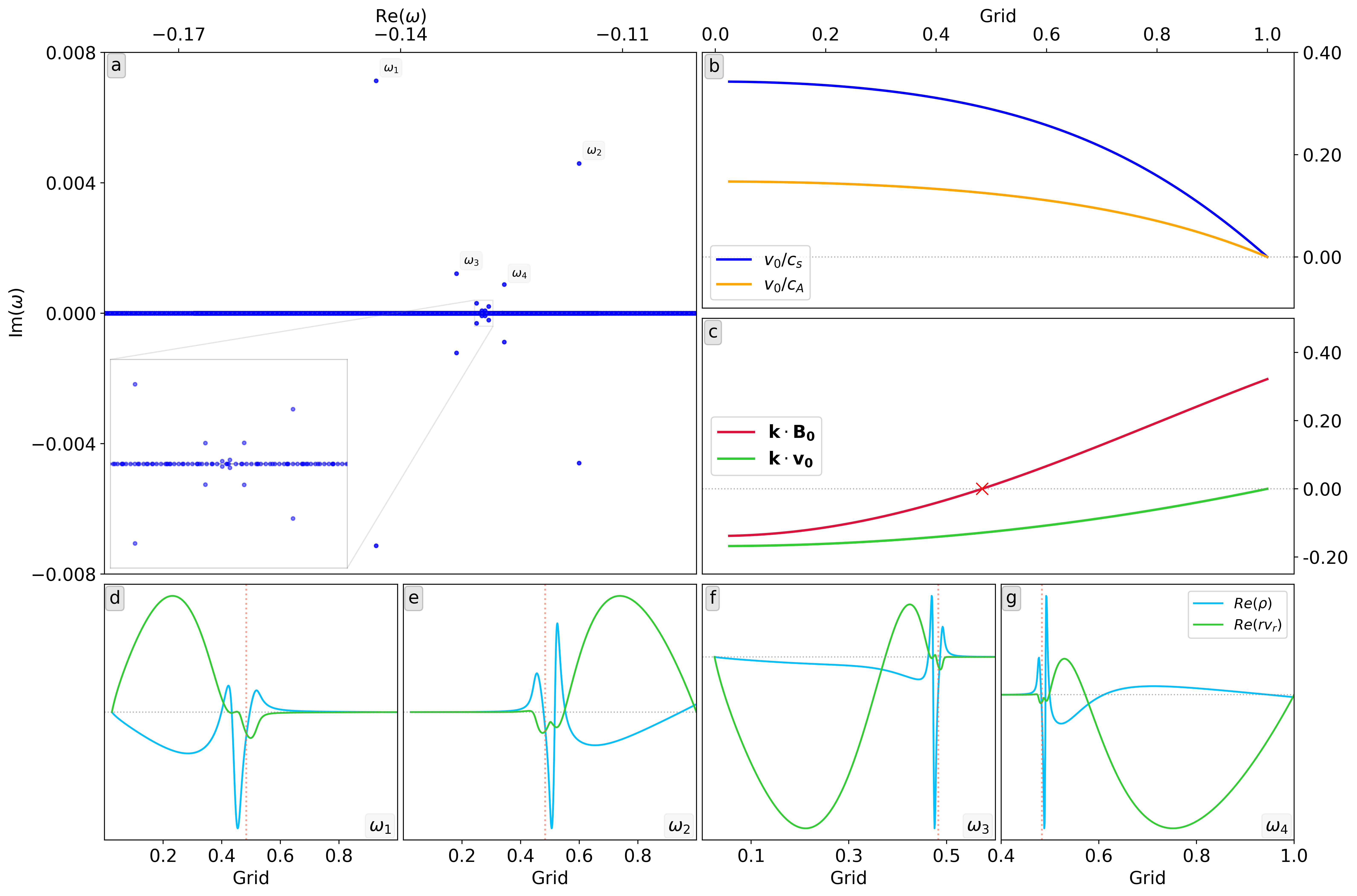}
	\caption{Panel $a$: Suydam cluster spectrum for the equilibrium given in \eqref{eq: suydam}. Inset: zoom-in near the Suydam surface, revealing more detail. Panel $b$ depicts the sonic and Alfv\'enic Mach numbers as a function of radius,
		panel $c$ shows the $\bfk \cdot \bfb_0$ and $\bfk \cdot \bfv_0$ curves where the red cross denotes the Suydam surface. The bottom row of panels shows the real part of the $\rho$ and $rv_r$ eigenfunctions, for the four modes
		in the Suydam sequence annotated on panel $a$. The dotted red line denotes the location of the Suydam surface.}
	\label{fig: suydam}
\end{figure}

The spectrum is calculated using 501 gridpoints for $r \in [0, 1]$ and is shown in Figure \ref{fig: suydam}. The resulting locations of the various off-axis outer modes are in agreement with results given in \citet{nijboer1997}. However, since this spectrum is calculated using a five times higher resolution, we have much more intricate detail near the Suydam surface, revealing even more off-axis modes (inset on panel $a$). The top two panels on the right side of Figure \ref{fig: suydam} show the sonic and Alfv\'enic Mach numbers (panel $b$), together with the $\bfk \cdot \bfb_0 = B_{02}k_2/r + B_{03}k_3$ and $\bfk \cdot \bfv_0 = B_{02}v_{02}/r + B_{03}v_{03}$ profiles as a function of radius (panel $c$), respectively. The location of the Suydam surface at $r \approx 0.483$ is denoted with a red cross. The bottom row of panels ($d - g$) show the real part of the $\rho$ and $rv_r$ eigenfunctions, for the four modes annotated on panel $a$ with the location of the Suydam surface annotated with a vertical red line. $\omega_1$ and $\omega_3$ correspond to modes on the left side of the Suydam surface, the other two correspond to modes on the right side. All eigenfunctions shown here show the specific variation associated with their location relative to the Suydam surface: $\omega_1$ and $\omega_3$ have their localised behaviour on the left side of the red dashed line, while it is vice-versa for the other two modes. For visual purposes the horizontal axis in panels $e$ and $f$ is different from the one in panels $c$ and $d$, since the former correspond to the next modes in the Suydam sequence, showing stronger radial variation. Note that the original Suydam criterion \citep{book_MHD} is related to static equilibria, the generalisation of these Suydam cluster criteria is presented in \citet{wang2004}.

\subsection{Resistive, Cartesian cases}		\label{subsect: resistive_results}
All cases discussed up to now handled an adiabatic equilibrium configuration with or without the inclusion of flow. The up-down symmetry of all the MHD spectra shown so far is perfectly maintained, related to the fact that these cases are in essence time-reversible. Every instability (or overstability in the case with flow) has a damped counterpart. We now move on to include additional effects. Hence, we now compute spectra for time-irreversible cases, where either resistivity or other non-adiabatic effects enter, which will break the up-down symmetry. Here we first focus on the inclusion of resistivity.

\subsubsection{Resistive homogeneous plasma}
\begin{figure}[t]
	\centering
	\includegraphics[width=0.85\textwidth]{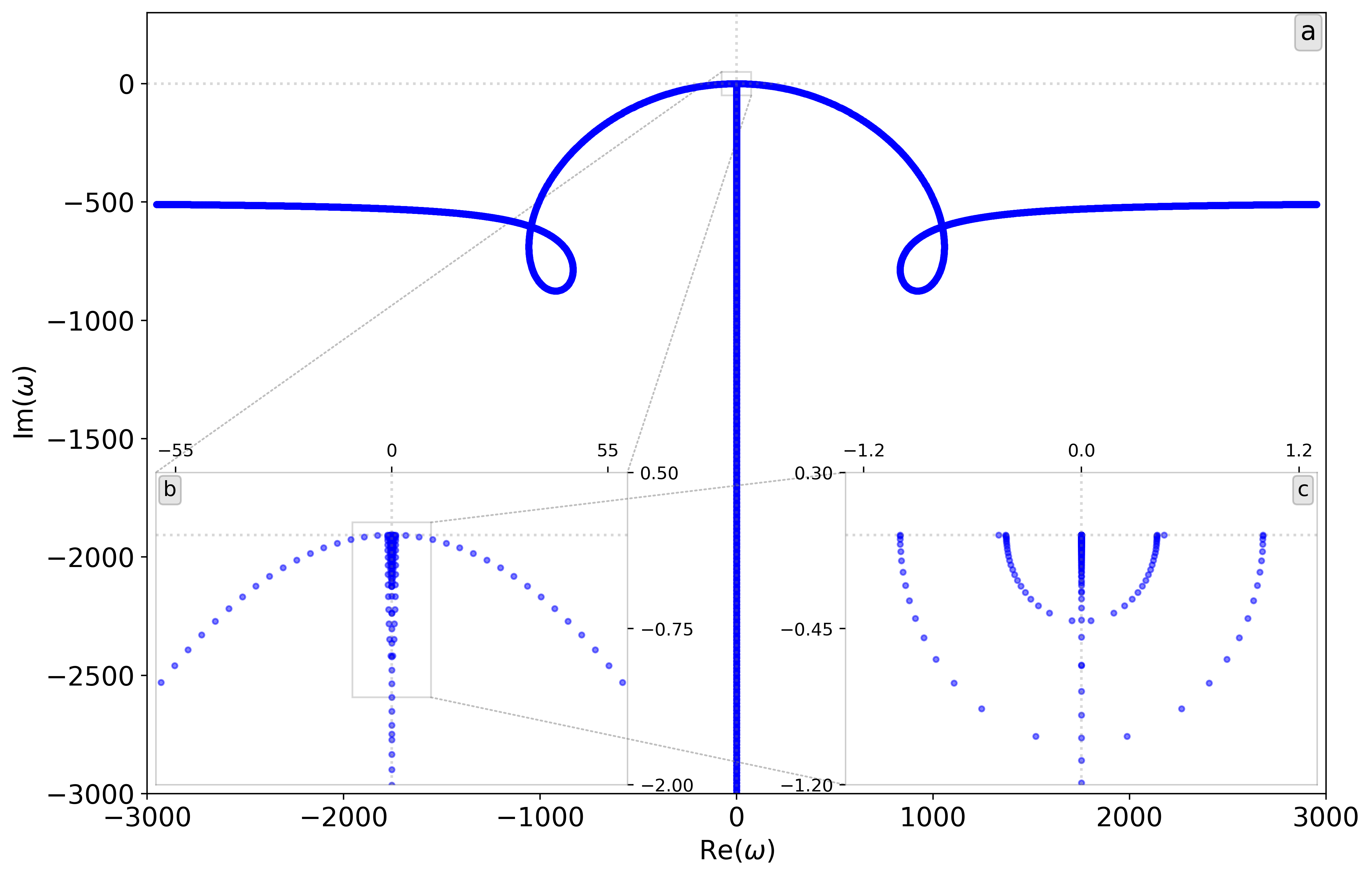}
	\caption{Panel $a$: spectrum for a homogeneous medium with constant resistivity. Panel $b$ zooms in near the origin, showing the start of the fast mode sequence. Panel $c$ zooms in further, revealing the semi-circles traced out by the Alfv\'en and slow modes
			 (outer and inner semi-circles, respectively).}
	\label{fig: resistive_homo}
\end{figure}
First we look at the most simple configuration, that is, a homogeneous plasma in a Cartesian geometry with resistivity included. The uniform equilibrium is given by
\begin{equation}	\label{eq: resistive_homo}
	\begin{aligned}
		\rho_0 &= 1, 	\qquad\qquad& 	B_{02} &= 0, 	\qquad\qquad&		B_{03} &= 1, 	\qquad\qquad&		T_0 &= \dfrac{1}{2}\beta B_0^2,
	\end{aligned}
\end{equation}
where we take a plasma beta of $\beta = 0.25$, $k_2 = k_y = 0$, $k_3 = k_z = 1$ and $x \in [0, 1]$. The value for the resistivity is assumed to be constant and given by $\eta = 10^{-3}$, as described in \citet{book_MHD}.
This spectrum is calculated using 1001 gridpoints, the result is shown in Figure \ref{fig: resistive_homo}.
In ideal MHD the fast modes form a Sturmian sequence (that is, the oscillation of the eigenfrequencies increases when the number of modes increases, which in turn implies a larger real part of $\omega$) of stable fast magneto-acoustic waves with frequencies accumulating to infinite frequency (related to the $p$-modes in our stratified example). The slow modes have an anti-Sturmian sequence towards their accumulation point (in essence the collapsed slow continuum) and the Alfv\'en modes are degenerate \citep{book_MHD}. When resistivity is included the fast modes become damped and the Alfv\'en and slow modes trace out semi-circles in the bottom-half of the complex plane. The semi-circles and initial fast mode sequence shown in Figure \ref{fig: resistive_homo} are in perfect agreement with the spectrum depicted in \citet[fig. 14.6]{book_MHD}. These semi-circles can be quantified analytically, their radius does not depend on the resistivity. The magnetic Reynolds number, calculated as $R_m = (x_1 - x_0)c_A / \eta$, is equal to 1000.

Due to the rather extreme resolution employed here we trace much further into the fast mode sequence, where we see something interesting: the fast modes appear on curves that loop around, breaking the purely Sturmian behaviour for a moment, after which they continue again towards infinity. This implies that initially the fast modes become more damped at higher mode frequencies up to a certain turning point at which they achieve maximal damping (the bottom of the loop). After passing this turning point the oscillation frequency of the modes increases again and the damping frequency seems to converge towards one single value.

Of course, the strong damping for the fast modes as we go further into the fast mode sequence must have consequences for the original uniform (ideal) equilibrium state. In what follows, we will adopt the common practice to compute resistive MHD spectra about an ideal MHD state, which itself will evolve when resistivity is acting.

\subsubsection{Quasi-modes in resistive MHD}		\label{subsect: quasimodes}
We now turn to a non-adiabatic case, where resistivity is important. We will compute the resistive MHD spectrum for a case where the equilibrium varies across an interface which gives rise to so-called quasi-modes. These are essentially surface waves undergoing damping, due to the fact that the global quasi-mode overlaps in frequency with the continuum range, causing resonant absorption. Quasi-modes are quite important in solar physics, since they can be indirectly related to the coronal heating problem as discussed in for example \citet{poedts1989, poedts1991}. A detailed analytical treatment of quasi-modes including theoretical growth rates is given in \citet{book_priest}, where they start from an inhomogeneous layer of width $l$ connecting two regions of uniform plasma. This can be reproduced by introducing a linear density profile between two homogeneous regions in a Cartesian geometry, however, this would mean that the density derivative shows rather strong discontinuities near the edges of the transition layer.
We therefore opt for a smooth profile by introducing a sine dependence such that
\begin{equation}	\label{eq: rho_quasimodes}
	\rho_0(x) =
	\begin{cases}
		\rho_1 			&\qquad \text{if}~ x_0 \leq x < s - \dfrac{1}{2}l,	\\
		\dfrac{1}{2}\rho_1\left[1 + \dfrac{\rho_2}{\rho_1} - \left(1 - \dfrac{\rho_2}{\rho_1}\right)\sin\left(\dfrac{\pi(x - s)}{l}\right)\right]		&\qquad \text{if}~ s - \dfrac{1}{2}l \leq x \leq s + \dfrac{1}{2}l,	\\
		\rho_2 			&\qquad \text{if}~ s + \dfrac{1}{2}l < x < x_1,
	\end{cases}
\end{equation}
where $s$ denotes the midpoint between $x_0$ and $x_1$, representing the left and right edges of the Cartesian grid, respectively, and equal to $0$ and $1$ such that $s = 0.5$.
\begin{figure}[b]
	\centering
	\includegraphics[width=\textwidth]{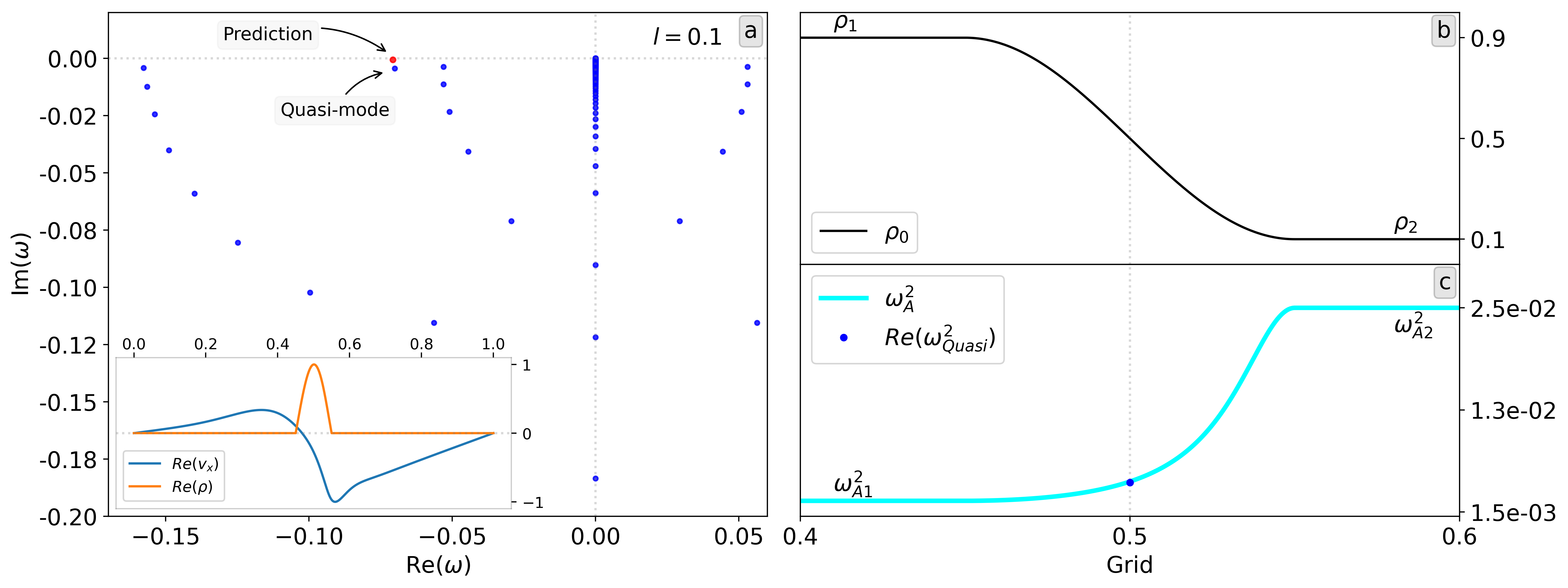}
	\caption{Panel $a$: part of the MHD spectrum containing the quasi-mode, the red dot denotes the theoretical prediction of Eq. \eqref{eq: quasimode_loc}.
		Panels $b$ and $c$ show plots of the sinusoidal density profile and Alfv\'en frequency as a function of the grid, respectively.
		Note that these two panels only show part of the grid, near the transition region. The inset shows the real part of the $v_x$ (orange) and $\rho$ (blue) normalised eigenfunctions associated with the quasi-mode.
		The blue dot in panel $c$ denotes the squared quasi-mode frequency, matching the squared Alfv\'en frequency at $x = 0.5$. The vertical dotted line denotes the middle of the grid.}
	\label{fig: quasi-modes}
\end{figure}
The width of the transition region $l = 0.1$ is taken to be small as to reduce the influence of the walls. If $l \rightarrow 0$ the inhomogeneous region disappears and we simply have a discontinuous jump in the plasma.
Furthermore we take $\rho_1 = 0.9$, $\rho_2 = 0.1$, $B_{02} = 0$, $B_{03} = 1$ and $T_0 = 0$. The magnetic field is unidirectional along the $z$-axis, and setting the pressure (and thus temperature) to zero provides an additional test on the handling of zero rows in the matrix. This zero-temperature case is frequently encountered in fully nonlinear MHD simulations which artificially adopt a zero plasma beta. It has as an important consequence that the slow continuum collapses to marginal frequency, eliminating many interesting modes from the spectrum. Additionally we adopt wavenumbers $k_3 = 0.05$ and $k_2 = 1.0$, such that $k_3 \ll k_2$. As discussed in \citet{book_priest} the quasi-modes are damped,
such that they move away from the real eigenfrequency axis with complex eigenvalues $\omega = \omega_R + i\omega_I$ given by
\begin{equation}	\label{eq: quasimode_loc}
	\omega_R = \pm \sqrt{\frac{\rho_1 \omega_{A1}^2 + \rho_2 \omega_{A2}^2}{\rho_1 + \rho_2}},		\qquad\qquad \omega_I = -\frac{\pi k_z l \left(\omega_{A2}^2 - \omega_{A1}^2\right)}{8 \omega_R},
\end{equation}
with $\omega_{A}^2 = k_z^2v_A^2$. This analytic result originates from a complex analysis of the linearised initial value problem. However, there is one caveat: it is impossible to obtain complex eigenvalues away from the real or imaginary axes in an ideal plasma, due to the matrix operator being hermitian in ideal MHD. Since the (ideal) quasi-mode is damped, we therefore include a (small) value for the resistivity, $\eta = 10^{-4}$, which is sufficient to make the matrix operator non-Hermitian and allows for complex eigenvalues away from the horizontal axis.

The magnetic Reynolds number in this case varies between $R_m \approx 10^4$ and $R_m \approx 3 \times 10^4$. Figure \ref{fig: quasi-modes} shows the MHD spectrum on panel $a$, for 501 gridpoints, with the theoretical prediction for the global quasi-mode location annotated with a red dot. The actual quasi-mode is also denoted on the figure, and is slightly more damped than its theoretical counterpart. This is to be expected, since the inclusion of resistivity imposes additional damping on the eigenvalues, shifting them downwards. It should be noted that increasing the width $l$ of the transition layer moves the quasi-mode further down and to the right on the figure, such that it eventually merges with the damped modes found on the branch immediately to its right. This phenomenon, where the quasi-modes merge with the resistive branches is discussed in more detail in \cite{vandoorsselaere2007}. The inset on panel $a$ depicts the $\rho$ (blue) and $v_x$ (orange) eigenfunctions associated with the quasi-mode, showing localised variation near the transition region as expected. Note that the spectrum as shown in Figure \ref{fig: quasi-modes}, left panel $a$, is still left-right symmetric, but that resistivity has broken the up-down symmetry, with all modes here found in the stable (damped) half plane.

\subsubsection{Resistive tearing modes}
\begin{figure}[t]
	\centering
	\includegraphics[width=\textwidth]{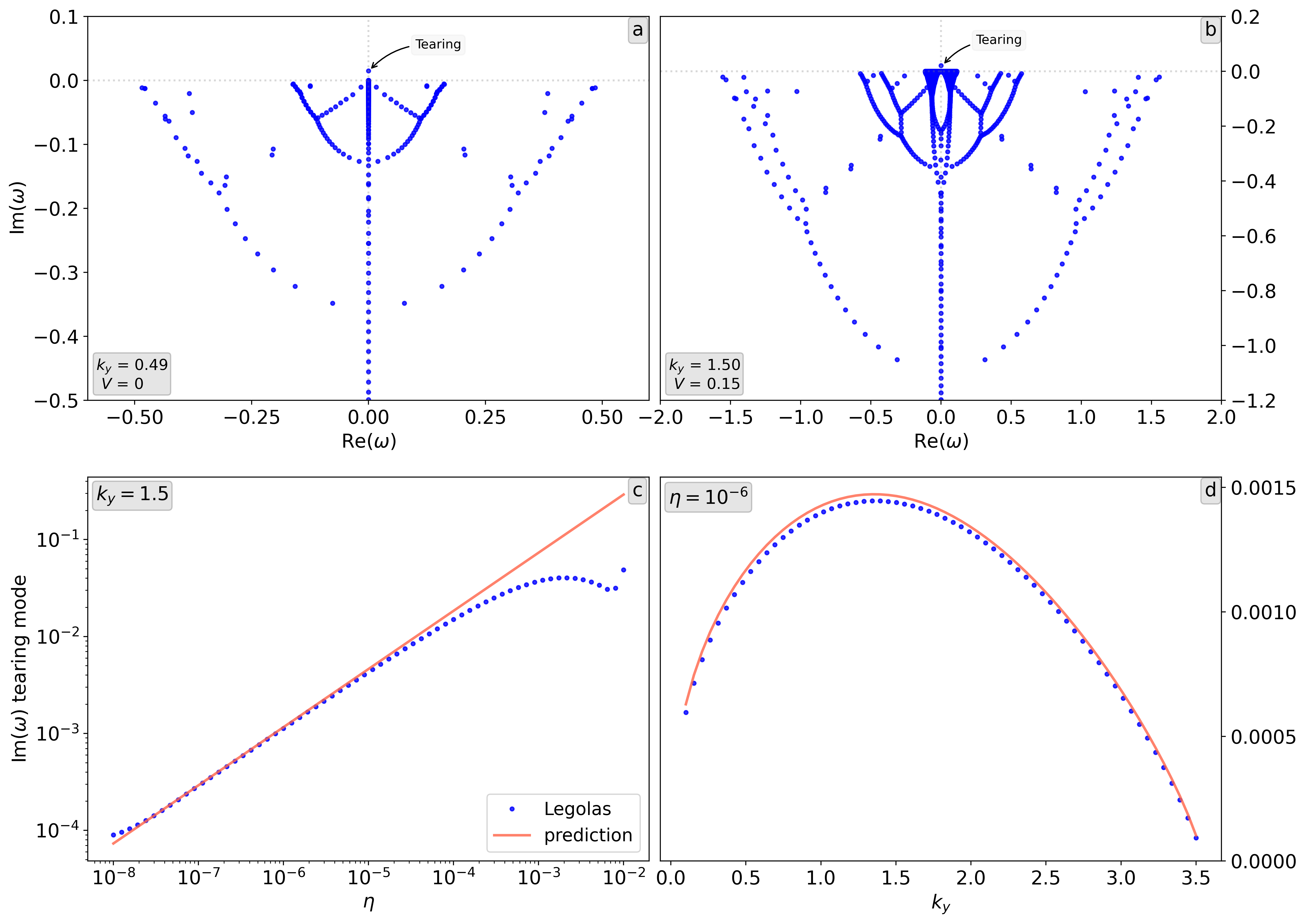}
	\caption{Resistive spectra of the equilibrium given in \eqref{eq: resistive_tearing}, unstable to the tearing mode (annotated with an arrow). Panels $a$ and $b$ show the spectrum without and with a linear flow profile, respectively,
		zoomed in to reveal the complex structure of the slow and Alfv\'en modes (fast modes are not shown). Panels $c$ and $d$: growth rate of the tearing mode versus fixed $k_y$, varying $\eta$ and fixed $\eta$, varying $k_y$, respectively.
		Both bottom panels are cases without flow, with 64 runs of 351 gridpoints each, per panel. The red line in panels $c$ and $d$ annotates the theoretical growth rate prediction given in Eq. \eqref{eq: tearing_theory}.}
	\label{fig: resistive_tearing}
\end{figure}

Next we move on to an inhomogeneous medium with the inclusion of resistivity and an optional linear flow profile, discussed in \citet{book_MHD}. The geometry is Cartesian, with $x \in [-0.5, 0.5]$ and an equilibrium configuration given by
\begin{equation}	\label{eq: resistive_tearing}
	\begin{aligned}
		\rho_0 &= 1, 						\qquad\qquad&		B_{02} &= \sin(\alpha x), 	\qquad\qquad&		B_{03} &= \cos(\alpha x),	\\
		T_0 &= \dfrac{1}{2}\beta B_0^2,		\qquad\qquad& 		v_{02} &= Vx, 				\qquad\qquad& 		v_{03} &= 0,
	\end{aligned}
\end{equation}
where $k_3 = k_z = 0$, $\beta = 0.15$ and $\alpha = 4.73884$, with a constant resistivity value of $\eta = 10^{-4}$.
The magnetic configuration chosen here is a linear force-free field with shear, with $\alpha$ the proportionality constant between the current and magnetic field, that is, $\boldsymbol{j} = \alpha\bfb_0$. This specific choice of parameters violates the tearing mode stability criterion \citep{book_MHD}, which results in an isolated, unstable tearing mode. Spectra are calculated both for $V = 0$ (no flow) and $V = 0.15$, the inclusion of the linear velocity profile in the latter introduces a Doppler shift in the slow and Alfv\'en continuum, the results are shown for 501 gridpoints in Figure \ref{fig: resistive_tearing}. Panels $a$ and $b$ show spectra for $V = 0$, $k_y = 0.49$ (no flow) and $V = 0.15$, $k_y = 1.5$ (linear flow profile), respectively, revealing the intricate behaviour of the damped slow and Alfv\'en sequences. The purely imaginary unstable tearing mode is annotated with an arrow. The resistivity $\eta$ is equal to $10^{-4}$ for both cases, yielding a magnetic Reynolds number of $\approx 10^4$. These results are in perfect agreement with the original spectra depicted in \citet[fig. 14.7-14.9]{book_MHD}. Note that these cases still have the left-right symmetry maintained, despite the presence of equilibrium flow. However, this is purely because the flow profile and domain happens to be chosen in a symmetric way, such that forward and backward modes behave symmetrically.

For an excellent discussion on tearing modes we refer to \citet{book_MHD}, where they perform a detailed analysis on the resistive MHD equations to derive an analytical expression for the growth rate of the tearing mode, given by
\begin{equation}	\label{eq: tearing_theory}
	Im(\omega) = R_m^{-3/5}\left(KH\right)^{2/5}\left(\frac{\Delta'}{C}\right)^{4/5}v_A/a, \qquad \text{in which} \qquad \Delta' = -2\sqrt{H^2 - K^2}\cot\left(\frac{1}{2}\sqrt{H^2 - K^2}\right),
\end{equation}
with $R_m = av_A/\eta$ the magnetic Reynolds number, $a$ the width of the slab and $v_A$ the Alfv\'en velocity. The other parameters in this equation are variables introduced during the analysis,
given by $H = \alpha a$, $K = k_0a$ and $C = 2\pi\Gamma(3/4)/\Gamma(1/4) \approx 2.1236$. Panels $c$ and $d$ of Figure \ref{fig: resistive_tearing} show a comparison of the tearing mode growth rate between \texttt{Legolas} results (using the no-flow case) and Eq. \eqref{eq: tearing_theory}. Panel $c$ holds $k_y = 1.5$ constant but varies the resistivity, reaching Reynolds numbers of $\approx 10^8$, a feat that is nearly impossible to achieve if fully nonlinear codes would be used instead. The correspondence between the theoretical and numerical growth rates is nearly one-to-one, except for large resistivity values since the theoretical approximation starts to break down in those regimes.
In panel $d$ we keep $\eta = 10^{-6}$ (magnetic Reynolds number of $\approx 10^6$) constant and vary $k_y$ between 0.1 and 3.5, which again yields an excellent agreement between theory and numerical results. For both panels $c$ and $d$ we performed 64 runs of 351 gridpoints each.

\subsubsection{Resistive rippling modes}
\textit{This subsection is accompanied by an Erratum, see Appendix \ref{erratum}. The original figure and text in this subsection are kept to match the original publication in ApJS 2020, \textbf{251}, 25.}

\begin{figure}[t]
	\centering
	\includegraphics[width=\textwidth]{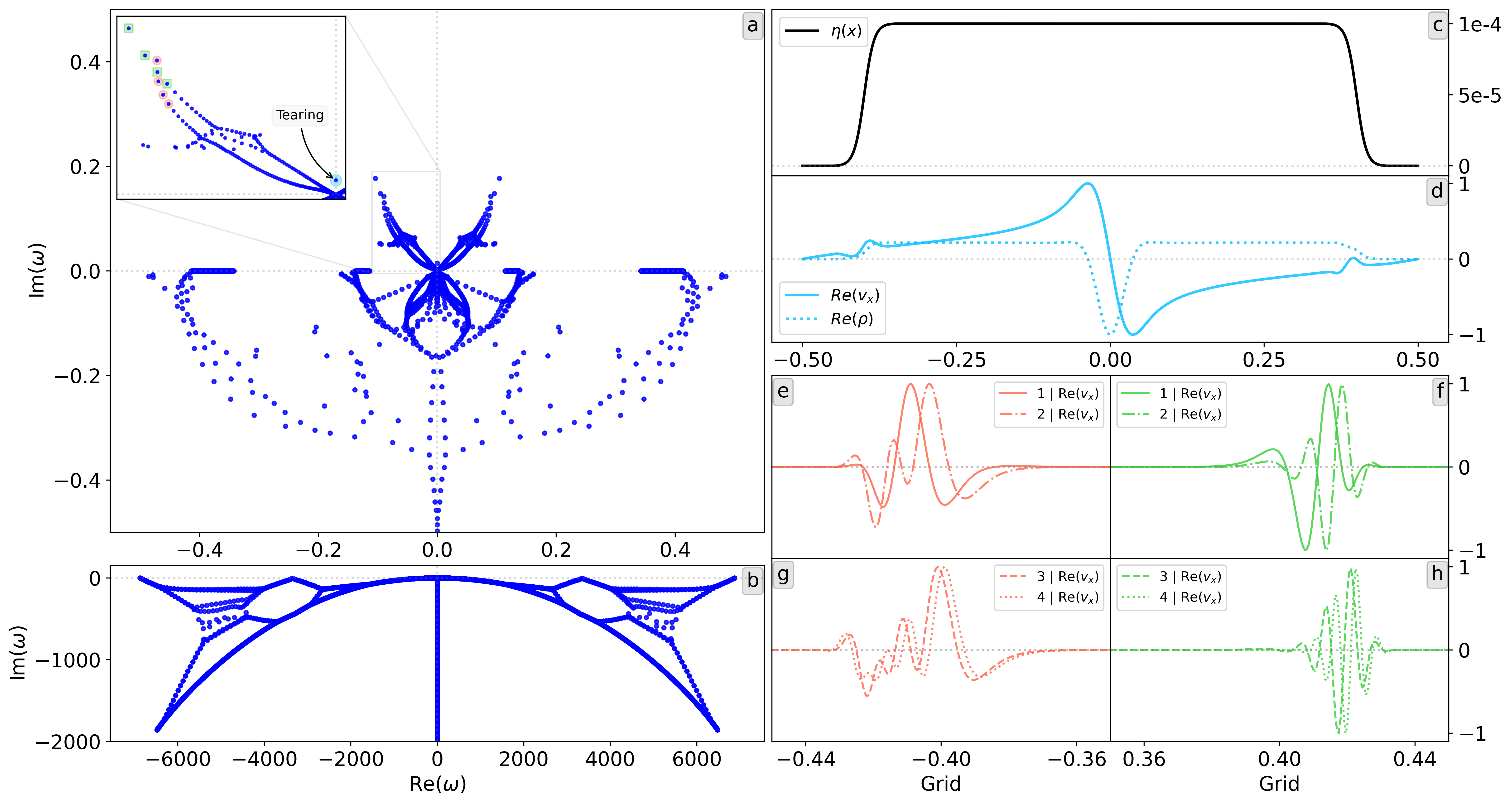}
	\caption{Resistive spectrum using the same parameters as panel $a$ in Figure \ref{fig: resistive_tearing}, but with a spatially varying $\eta(x)$-profile (panel $c$). Panel $b$: full spectrum with modified fast mode sequences. Panel $a$: zoom in on the
			 slow and Alfv\'en modes, the inset zooms in on the tearing and unstable rippling modes. Panel $d$: $v_x$ (solid) and $\rho$ (dotted) eigenfunctions of the tearing mode (inset, blue). Panels $e$ and $g$: $v_x$ eigenfunctions of the first four modes
		 	 associated with the left part of the $\eta(x)$ profile (inset, red circles). Panels $f$ and $h$: $v_x$ eigenfunctions of the first four modes associated with the right part of the $\eta(x)$ profile (inset, green squares).}
	\label{fig: resistive_ripple}
\end{figure}

As an extension to the tearing modes described above we take a look at resistive rippling modes, which originate whenever there is a spatially varying resistivity profile as discussed in for example \citet{book_priest}.
To that extent we take the same equilibrium and parameters as Eq. \eqref{eq: resistive_tearing} (with $V = 0$, so without flow and $k_y = 0.49$), but impose a hyperbolic tangent profile on the resistivity to let it smoothly drop down to zero near the edges of the domain. This spatial variation in resistivity will excite (unstable) rippling modes, in a current-carrying equilibrium that is also unstable to the tearing mode. The resulting spectrum is shown in Figure \ref{fig: resistive_ripple} for 1001 gridpoints, which is again left-right symmetric. The adopted $\eta(x)$ profile is given explicitly in Eq. \eqref{eq: eta_profile} and depicted in Fig. \ref{fig: resistive_ripple}, panel $c$. Here $\eta_0 = 10^{-4}$ denotes the constant resistivity value, $s_L = -0.4$ and $s_R = 0.4$ represent the centre of the left and right transition region, respectively, having a width of $l = 0.1$.

\begin{equation}	\label{eq: eta_profile}
	\eta(x) =
	\begin{cases}
		\dfrac{\eta_0}{2} + \dfrac{\eta_0}{2\tanh{\pi}}\tanh\left(\dfrac{2\pi(x - s_L)}{l}\right)		&\qquad \text{if}~ s_L - \frac{l}{2} \leq x \leq s_L + \frac{l}{2},		\\
		\eta_0 																							&\qquad \text{if}~ s_L + \frac{l}{2} < x < s_R - \frac{l}{2},			\\
		\dfrac{\eta_0}{2} + \dfrac{\eta_0}{2\tanh{\pi}}\tanh\left(\dfrac{2\pi(s_R - x)}{l}\right)		&\qquad \text{if}~ s_R - \frac{l}{2} \leq x \leq s_R + \frac{l}{2},		\\
		0																								&\qquad \text{elsewhere}
	\end{cases}
\end{equation}

The inclusion of this relatively simple $\eta(x)$ profile has a major influence on the resulting spectrum: the fast mode sequences trace out intricate patterns in the complex eigenvalue plane.
The semi-circles traced out by the slow and Alfv\'en modes are still vaguely present, but are much more scattered than their constant-$\eta$ counterparts. The inset on panel $a$ of Figure \ref{fig: resistive_ripple} zooms in near the origin,
revealing that the tearing mode is still present, annotated with an arrow and a large blue dot. Panel $d$ shows the $v_x$ and $\rho$ eigenfunctions of the tearing mode, with relatively sharp transitions near the $x = 0$ point and a (minor)
influence from the wings of the $\eta(x)$ profile. The rippling modes visible on the inset come in two branches, annotated with red circles and green squares, which intersect near the top of the branches.
The red circles correspond to the influence of the left wing of the $\eta(x)$ profile, visible in the eigenfunctions which are shown in panels $e$ (first and second dot) and $g$ (third and fourth dot), and are very localised near the $\eta$ transition region as can be seen by the grid indication on the horizontal axis. Analogously, the branch annotated by green squares corresponds to the right wing of the $\eta(x)$ profile, with $v_x$ eigenfunctions shown in panels $f$ and $h$.

The fact that the rippling modes appear more unstable than the tearing mode indicates that their importance should not be underestimated, and that they will most likely be prominently present when fully realistic $\eta$ profiles are used. Indeed, in actual plasmas the temperature variation in the equilibrium can itself already cause a spatially varying resistivity profile, and this is accounted for in \texttt{Legolas}. This allows for future systematic studies of rippling versus tearing mode dominance in slabs or loop-like settings.

\subsection{Non-adiabatic, cylindrical cases}	\label{subsect: nonadiabatic_results}
Since \texttt{Legolas} is the first 1D code to simultaneously include non-adiabatic effects, resistivity and flow, there are no known previously calculated spectra where all these physical effects are included. Nevertheless, there exist some spectra in the literature where solely non-adiabatic effects are included (and where the equilibrium is static and no resistivity is incorporated), so we will use these as a base comparison for testing the non-adiabatic terms in the implementation.

\subsubsection{Non-adiabatic discrete Alfv\'en waves}
The first spectrum we will look at is that of non-adiabatic discrete Alfv\'en waves, described in \citet{keppens1993}. The basic cylindrical equilibrium represents a solar coronal loop, and non-adiabatic effects included were optically thin radiative losses and parallel thermal conduction. It was pointed out how a cluster sequence of discrete Alfv\'en waves may become unstable, and could lead to disruptions or oscillatory behaviour in loops of prominences.

This particular equilibrium uses an axial current profile in a cylindrical geometry, yielding the same magnetic configuration as given in Eq. \eqref{eq: Bnu_profile} although this time with $\nu = 2$ such that a current distribution is present throughout the loop.
The pressure profile can be obtained through integration of the equation for magnetostatic equilibrium \eqref{eq: eq_condition_1} without flow.
As a boundary condition we impose that the pressure vanishes at the plasma boundary $r = R$, which can be used to constrain the integration constant. Furthermore a parabolic density profile is used, yielding an equilibrium configuration given by
\begin{equation}	\label{eq: discrete_alfven}
	\begin{aligned}
		\rho_0 &= 1 - \left(1 - d\right)\left(\dfrac{r}{R}\right)^2, 	\qquad\qquad& 	B_{02} &= \dfrac{1}{6}j_0 r \left(r^4 - 3r^2 + 3\right), 		\qquad\qquad& 	B_{03} &= 1,	\\
		p_0 &= \mathrlap{\dfrac{j_0^2}{720}\Bigl[12\left(R^{10} - r^{10}\right) - 75\left(R^8 - r^8\right) + 200\left(R^6 - r^6\right) - 270\left(R^4 - r^4\right) + 180\left(R^2 - r^2\right)\Bigr],}
	\end{aligned}
\end{equation}
in which $d = 0.2$ and $j_0 = 0.125$. The cylinder wall is taken at $R = 1$. The equilibrium temperature profile can be derived using $T_0 = p_0/\rho_0$ and $T_0' = \left(p_0'\rho_0 - \rho_0'p_0\right) / \rho_0^2$, where the prime denotes the derivative with
respect to $r$. Only conduction parallel to the magnetic field lines is taken into account, since cross-field thermal conduction acts on too long a timescale in this case \citep{keppens1993}.

\begin{figure}[t]
	\centering
	\includegraphics[width=\textwidth]{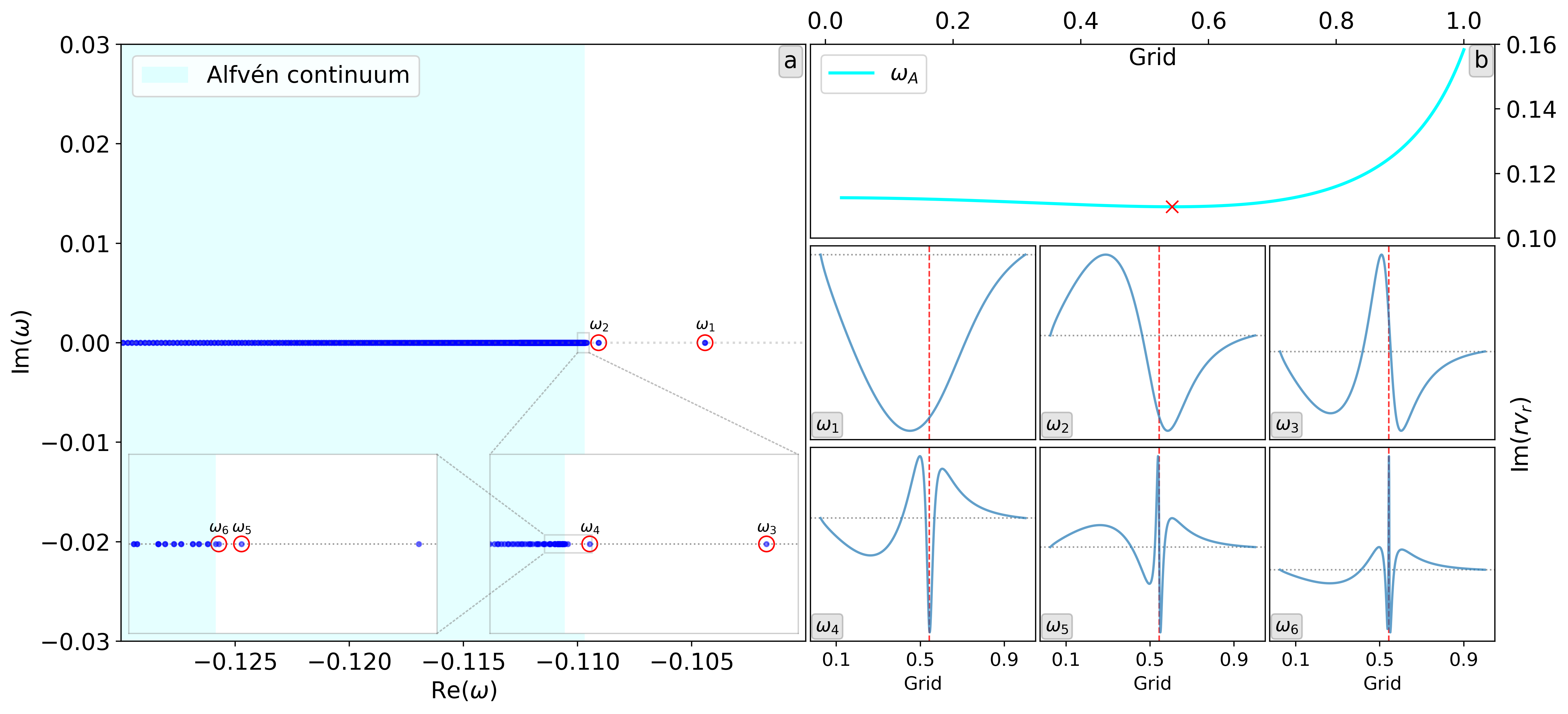}
	\caption{Panel $a$: discrete Alfv\'en spectrum, the Alfv\'en continuum is shaded in cyan. Six discrete modes were found, annotated $\omega_1$ through $\omega_6$. The insets depict zoom-ins, the regions are annotated on the figure.
			 Panel $b$ shows a plot of the Alfv\'en continuum versus radius, the minimum is denoted by a red cross. The six panels in the bottom-right corner show the imaginary part of the $rv_r$ eigenfunctions for modes $\omega_1$ through $\omega_6$,
			 the red dotted line represents the location of the minimum of the Alfv\'en continuum.}
	\label{fig: discrete_alfven}
\end{figure}

The wave numbers $k_2 = m$ and $k_3 = k$ are taken to be $1$ and $0.05$, respectively. Optically thin radiative losses are included as described in Eq. \eqref{eq: radiative_cooling} in which we assume that the heating is such that it exactly balances out the cooling terms in the equilibrium state. We use the cooling curve introduced by \citet{rosner1978}, which represents a piecewise cooling law with predetermined coefficients. Since the inclusion of non-adiabatic effects requires us to specify unit normalisations in order to look up the dimensional values in the cooling tables, we take reference values of $50$ Gauss for the magnetic field, $1.5 \times 10^{-15}$ g cm$^{-3}$ for the density and $10^{10}$ cm as a length scale. This automatically constrains all other normalisations through the ideal gas law, assuming a fully ionised plasma.

The spectrum is calculated using 501 gridpoints, Figure \ref{fig: discrete_alfven} shows the discrete Alfv\'en spectrum (panel $a$) along with a plot of the Alfv\'en continuum (panel $b$). In total six discrete modes (that is, modes that fall outside of the Alfv\'en continuum) were found in contrast with the five discrete modes in the original paper \citep{keppens1993}. The discrete modes are annotated according to their overtones, where $\omega_1$ represents the fundamental mode (FM) and $\omega_6$ the fifth overtone (OT). This last overtone does not show up for resolutions below $\sim 200$ gridpoints, which explains why it is not present in the original work.
The imaginary part of the $rv_r$ eigenfunction corresponding to each of these six modes is shown in the bottom-right panels of Figure \ref{fig: discrete_alfven}, where the location of the minimum in the Alfv\'en continuum is indicated with a red dotted line.
Note that the number of eigenfunction nodes increases when considering modes further in the Alfv\'en sequence: the eigenfunction corresponding to $\omega_1$ has no nodes, $\omega_2$ through $\omega_6$ have one, two, three, four and five nodes, respectively.
Table \ref{tab: discrete_alfven} shows a detailed comparison between the discrete modes found by \texttt{Legolas} and the ones from \citet{keppens1993}. We multiplied these latter by $i$, since they use the convention $\exp(st)$ rather than $\exp(-i\omega t)$ in the Fourier analysis (thus $\omega = is$).
\begin{table}[t]
	\centering
	\caption{Comparison of discrete Alfv\'en eigenvalues between \texttt{Legolas} and the original work.}
	\begin{tabular}{c c c}
		\hline
		\T\B 	\textbf{Eigenvalue}		&		\texttt{Legolas}									&				\citet{keppens1993}		\\
		\hline
		\T 		FM $\omega_1$			&		$-0.1044165837 + 1.7260866\times 10^{-07}i$		&			$-0.10436875 + 1.7\times 10^{-07}i$		\\
				1\ts{st} OT $\omega_2$	&		$-0.1090793809 + 6.5883040\times 10^{-08}i$		&			$-0.10907061 + 6.7\times 10^{-08}i$		\\
				2\ts{nd} OT $\omega_3$	&		$-0.1096033464 + 8.3813211\times 10^{-09}i$		&			$-0.10960184 + 8.4\times 10^{-09}i$		\\
				3\ts{rd} OT $\omega_4$ 	&		$-0.1096779374 + 2.5325093\times 10^{-09}i$		&			$-0.10967774 + 2.5\times 10^{-09}i$		\\
				4\ts{th} OT $\omega_5$	&		$-0.1096871498 + 3.7770023\times 10^{-10}i$		&			$-0.10968708 + 4.0\times 10^{-10}i$		\\
				5\ts{th} OT $\omega_6$	&		$-0.1096883410 + 7.4335660\times 10^{-11}i$		&						---
	\end{tabular}
	\label{tab: discrete_alfven}
\end{table}

In reality, the mode sequence is expected to be an infinite sequence accumulating to the local minimum in the Alfv\'en continuum, as indicated in panel $b$ on Figure \ref{fig: discrete_alfven}.
Both the real and imaginary parts of the eigenvalues are in excellent agreement. It should be noted that when the non-adiabatic effects are omitted the imaginary parts of these discrete modes become zero, such that they lie on the real axis, representing stable waves. The inclusion of non-adiabatic effects hence has almost no influence on their oscillation frequency, and solely pushes these modes into the unstable part of the imaginary plane.
In \citet{keppens1993}, it was pointed out how these discrete Alfv\'en mode sequences can be studied by means of a WKB analysis. However, this only correctly predicted the damped or overstable nature of the higher order modes: to determine whether the most global modes of the sequence are overstable or damped requires a full numerical computation, possible with general tools like \texttt{Legolas}.

\subsubsection{Magnetothermal instabilities}
As a final test we look at magnetothermal instabilities, originally depicted in \citet{vanderlinden1992}. The geometry is cylindrical with $r \in [0, 1]$ and an isothermal equilibrium profile given by
\begin{equation}
	T_0 = 1,		\qquad\qquad		B_{02} = \dfrac{r}{1 + r^2},		\qquad\qquad		B_{03} = 0,	\qquad\qquad	 p_0 = \dfrac{1}{2\left(1 + r^2\right)^2},		\qquad\qquad 		\rho_0 = \dfrac{p_0}{T_0},
\end{equation}
with $k_2 = m = 0$ and $k_3 = k = 1$. There is no $B_z$, such that this configuration actually represents a z-pinch (which is very unstable), and we are looking at axisymmetric (sausage) $m=0$ modes here.
Field-aligned thermal conduction is included, cross-field thermal conduction is omitted. Optically thin radiative losses are accounted for, and we use the same cooling curve and heating assumptions as for the non-adiabatic discrete Alfv\'en waves in \eqref{eq: discrete_alfven}. Reference values are taken to be $2.6$ MK for the temperature, $10$ Gauss for the magnetic field and $10^8$ cm for the length scale. As before, this automatically constrains all other normalisations as well. These parameters are representative for solar coronal loops and arcades.

\begin{figure}[t]
	\centering
	\includegraphics[width=\textwidth]{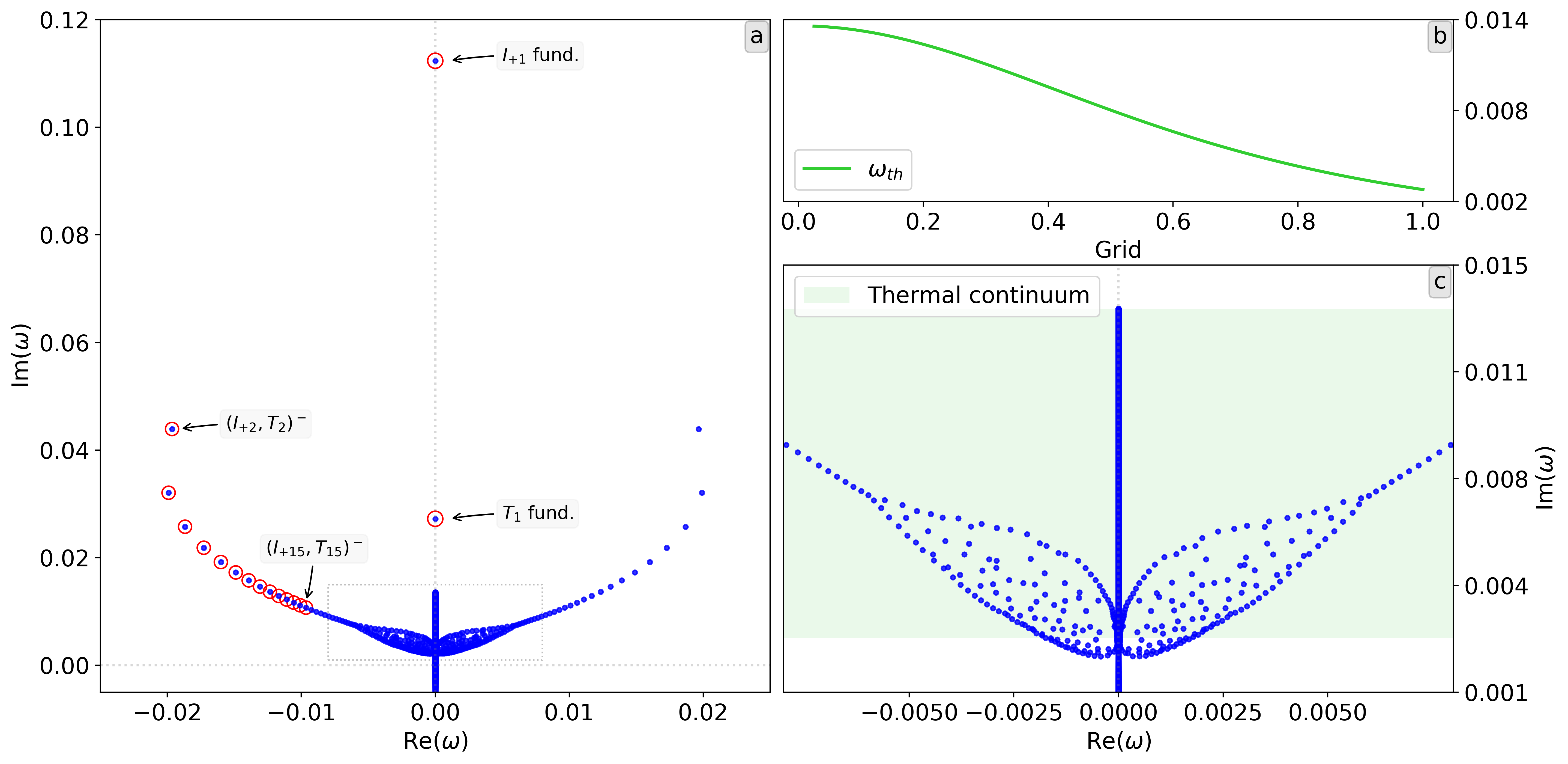}
	\caption{Magnetothermal instabilities for $m=0$ and $k=1$. The fundamental modes and 14 overtones discussed in the original work are encircled by red and denoted on the left panel ($a$). Panel $b$ shows the thermal continuum as a function of radius.
			 Panel $c$ zooms into the region denoted with dashed lines on panel $a$, revealing various other modes forming intricate structures.}
	\label{fig: magnetothermal}
\end{figure}

Both the Alfv\'en and the slow continuum collapse into marginal (zero) frequency. However, including finite parallel thermal conduction and radiative losses introduces the thermal continuum. Together with the marginal slow-Alfv\'en frequencies, this organises the modes such that thermal instabilities merge with magnetic modes, introducing magnetothermal branches. Figure \ref{fig: magnetothermal} shows the MHD spectrum for 1001 gridpoints in the region of the magnetothermal branches, where the bottom-right panel ($c$) zooms in further near the origin. The modes denoted by $I_{+1}$ and $T_1$ represent fundamental magnetic and thermal modes, respectively, meaning they are not coalesced. The overtones of these modes on the other hand are coalesced, these are denoted by $(I_{+2}, T_2)^-$ and $(I_{+15}, T_{15})^-$ for the first and $14^{\text{th}}$ overtones, respectively. The minus sign in superscript means that these are located on the negative part of the real axis. There are corresponding modes on the positive part of the real axis, since without flow, we maintain the left-right symmetry of the spectrum.
The first 14 overtones are in excellent agreement with those described in \citet{vanderlinden1992}, and are encircled in red on the left panel ($a$). However, the original work only displays solutions up to the $14^{\text{th}}$ overtone.
The resolution used here is much higher than the originally published results, allowing us to probe the region near the origin as well. Panel $c$ zooms in near the origin, denoted by the dotted rectangle on panel $a$, revealing a complex mixture of different branches and scattered modes. It seems that the original magnetothermal branches split and a dense region covered with many magnetothermal modes appears. The analytically known thermal continuum is shaded in green on the figure, and is represented by a dense, continuous range of discrete eigenvalues on the imaginary axis. Since this is the first time that high resolution is possible for computing the magnetothermal modes, it is left to future work to clarify how the MHD spectrum allows for such complex mode interactions, revealing the possible presence of areas covered by modes in the complex eigenfrequency plane.

\section{Convergence}	\label{sect: convergence}
As discussed in this paper, increasing the resolution can have a major influence on the spectrum, depending on whether or not all modes are sufficiently resolved. Generally speaking, the further one goes in a specific sequence, that is, looking at larger mode numbers which represents overtones having more and more nodes in their eigenfunctions, the higher the resolution that is required in order to resolve the mode completely. For eigenfunctions it is usually immediately clear if a mode is resolved or not, since higher mode numbers translate into more oscillations in the eigenfunction. Hence, once the number of oscillations approaches the amount of gridpoints, the eigenfunction is no longer resolved.

For the eigenfrequencies it is not a priori clear when a specific $\omega$ is resolved. One way to constrain an eigenvalue is to do multiple runs, each time increasing the resolution. Once an eigenvalue no longer shifts in the complex plane it can be considered resolved, and increasing the resolution even further will not have (much) effect on its value. Now, the question naturally arises how many gridpoints are typically needed to be able to speak of a ``resolved" spectrum. This will strongly depend on the type of equilibrium considered: for smooth equilibria without sharp transitions one can get away with a few dozen gridpoints and already reach an acceptable accuracy for most modes. However, in the case of equilibria with large gradients, or even with localised discontinuities (interfaces), one has to make sure that a sufficient number of gridpoints are taken in order to sufficiently resolve that jump.

\begin{figure}[t]
	\centering
	\includegraphics[width=\textwidth]{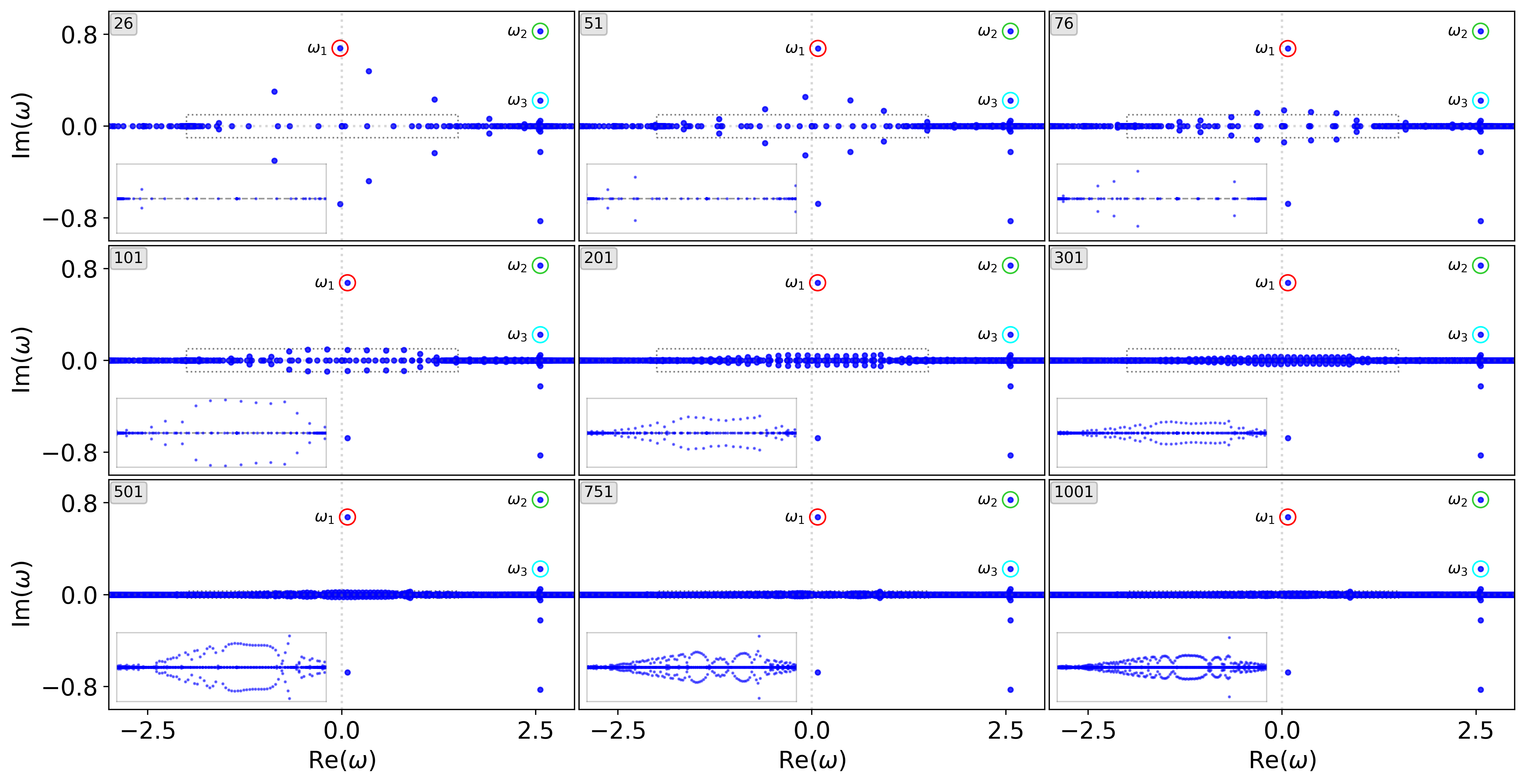}
	\caption{Spectrum of the Kelvin-Helmholtz and current-driven equilibrium as shown earlier in Figure \ref{fig: kh_cd}, using the same parameters as in Section \ref{subsect: kh_cd}, at various resolutions (indicated on the top-left of each panel).
			 The inset on the top two rows of panels zooms near the horizontal axis, for vertical axis values between $-0.1$ and $0.1$. The insets on the bottom row of panels have slightly different vertical bounds, from $-0.03$ to $0.03$.}
	\label{fig: convergence}
\end{figure}

Furthermore, the amount of eigenvalues in the spectrum increases with resolution, since there are as many eigenfrequencies as the dimension of the matrices (which is 16 times the number of gridpoints). It is therefore entirely possible that one starts probing parts of the spectrum that were initially not visible at lower resolutions, simply because there were no eigenvalues in that region for that amount of gridpoints. An example is given here, where we look back at the Kelvin-Helmholtz equilibrium discussed in Section \ref{subsect: kh_cd}. Figure \ref{fig: convergence} shows this equilibrium for the same values as employed earlier, but every panel shows the spectrum for a different resolution, with the amount of gridpoints given in the top-left corner.
The first three panels start out at low resolution, increasing from 26 to 76 gridpoints. Three modes are annotated on the figure, corresponding to the Kelvin-Helmholtz (KH) and current-driven (CD) modes discussed earlier. These modes are already decently resolved even at the lowest resolutions, and can be considered completely resolved at around 100 gridpoints. However, the region near the horizontal axis is another matter entirely. At low resolutions (first three panels) we see that most of the modes here are almost randomly scattered in the complex plane. At around 100 gridpoints they start lining up and form a more intricate ``elliptic" pattern. This is the point where the original paper by \citet{baty2002} stopped, since it was not really feasible to run these codes at higher resolutions, and they were only interested in the most unstable KH and CD modes that were resolved and that determined the early evolution of a full non-linear MHD run that they performed.

When the resolution is increased even further, we see that the near-axis modes shift even closer to the real eigenfrequency axis, and as visible on the insets on each panel these modes start to form intriguing patterns at high resolutions.
The amount of ``scattered" modes decreases considerably, and a helix-like structure is formed at very high resolutions (1001 gridpoints, lower right panel). These complex and almost geometric patterns may arise in various different equilibria as well, and call for extensive research in this topic. A complementary means to identify which modes are actually resolved is to exploit the spectral web, see \citet{goedbloed2018web1, goedbloed2018web2} and \citet[chapter~12-13]{book_MHD}, which locates eigenmodes on specific curves and their intersections. These curves relate to the two self-adjoint operators at play in stationary, adiabatic MHD. Only a combined approach using high resolution \texttt{Legolas} runs, modern linear algebra solvers and physical insight in MHD spectroscopy, will in time reveal the true importance of these modes.

Based on the conclusions drawn here, it is clear that the resolution required depends on the part of the spectrum that is to be investigated. For isolated modes about 100 gridpoints is more than sufficient to resolve most of them, although this depends on the specified equilibrium. For large-scale surveys of the spectrum large resolutions should be employed, in order to reveal possible regions of interest.

\section{Conclusion and outlook}
In this paper we introduced the novel 1D finite element code \texttt{Legolas}, meant to tackle the complete MHD spectrum including all kinds of physical effects in order to be able to look at various realistic equilibrium configurations. We performed a Fourier mode analysis on the linearised MHD equations, after which the resulting matrix eigenvalue problem is solved using a finite element approach. A general formalism was introduced to handle both Cartesian and cylindrical geometries.
\texttt{Legolas} is the first spectral code to combine the effects of flow, resistivity, gravity, radiative cooling and anisotropic thermal conduction. This opens the door to novel, in-depth studies of the complete MHD spectrum, which was up to now impossible with existing numerical tools.

We tested \texttt{Legolas} against various previously established results from the literature, looking at the comparison to analytical results as well as to spectra previously obtained by similar numerical codes. Correspondence with existing results is in most cases one-to-one, greatly increasing the confidence in this new tool. Some cases were run in higher resolutions than their original counterparts, revealing interesting features and additional structure in the spectra.
The resistive homogeneous case in Figure \ref{fig: resistive_homo} is in perfect agreement with results from the literature near the origin, revealing the semi-circles traced out by Alfv\'en and slow modes. However, far along the fast mode sequence an intriguing loop appears, locally breaking the Sturmian behaviour of the fast modes. Even this case calls for further investigation, since this phenomenon has not been described before as no extremely high resolution studies of that part of the spectrum have been done to date.

Something similar can be seen when looking at the instabilities of a cylindrical magnetised jet flow in Section \ref{subsect: kh_cd}, where the outer KH and CD modes are in perfect correspondence with the original results. However, the high resolution revealed complex structures which were not probed before. This becomes even more clear in the convergence study of Section \ref{sect: convergence}, where we used the same setup but increased the resolution even further. It is clear that the spectrum evolves drastically at higher resolutions, when more and more modes are becoming properly resolved. Since a complete knowledge of the MHD spectrum of flowing equilibria is lacking, tools like \texttt{Legolas} and careful converged studies will be essential to unravel the role of as yet unresolved spectral structure. We speculate that these spectral structures have a physical meaning, which can be corroborated in adiabatic flowing cases with a complementary spectral web approach. Our speculation extends to cases that address MHD modes in cylindrical accretion disks \citep{keppens2002}, where it is becoming clear that many more modes than the celebrated MRI instability enrich the spectral structure.

Also in the non-adiabatic cases we found something interesting: the outer modes of the magnetothermal sequence are again in perfect correspondence with modes found in the original work.
However, near the origin - a region that has not been investigated until now - there are indications of a splitting of the magnetothermal sequence in its thermal and magnetic counterparts with various modes scattered in between.
An in-depth study of magnetothermal instabilities is called for, a topic of research that essentially stopped progressing for the last two decades.
Now that high resolution MHD simulations start to reveal the complexity of thermal instability driven evolutions \citep{xia2016, claes2020}, a revival of this topic is urgently needed.

As a side note it can be argued that a standard Fourier decomposition as done in all cases discussed in this paper misses out on non-modal growth, that is, so-called transient modes, which have been shown to be of significant importance in resistive systems
\citep{mactaggart2018}. However, it is the authors' opinion that a calculation of the complete spectrum is all that is needed. Transient growth may well be a consequence of solving an actual initial value problem using a Laplace transformation, while taking all discrete and continuous modes into account. A detailed proof is far from evident, however, in adiabatic cases we have self-adjoint operators (one for a static case, two if flow is included), such that it seems plausible that a full decomposition in terms of a complete basis of eigenfunctions is possible. The inclusion of flow makes these eigenfunctions non-orthogonal, while in ideal, static plasmas the eigenfunctions are orthogonal \citep[chapter~12]{book_MHD}. For non-ideal, stationary plasmas non-orthogonality can again be expected. How this decomposition would be affected by the inclusion of non-adiabatic effects such as resistivity or viscosity is not a priori clear. The discussion of transient growth is usually done in the context of non-modal analysis and uses the concept of pseudo-spectra, that is, also allowing for modes that are nearly eigenvalues.

All of the above is a clear indication that a thorough investigation of the entire spectrum is in order to yield new insights in (M)HD instabilities. For solar applications alone the effect of thermal conduction on the linear modes as indicated by \cite{vanderlinden1991} has never been investigated in fully realistic setups, possibly allowing for a deeper understanding of fine structure in solar prominences. However, since \texttt{Legolas} is such a versatile code the possibilities are endless,
from various hydrodynamic configurations to more magnetically-oriented astronomical cases like accretion disks or astrophysical jets. The discussion of the more ``advanced" cases in this paper alone reveals how much of the MHD spectrum is still not thoroughly investigated. We plan to tackle various new cases in future work, combining linear results from this code with state of the art nonlinear numerical simulations.
Including additional physical effects is also planned, with an extension to viscous (M)HD (Navier-Stokes), Hall MHD and ambipolar terms. Additionally, we plan to include ``real" atmospheric models, that is, numerically generated hydrostatic equilibria based on a given temperature profile. The possibilities are endless, and modern MHD spectroscopy will no doubt shed new light on various physical phenomena.

\vfill\noindent
\footnotesize{\emph{Acknowledgements.} The authors would like to thank Hans Goedbloed for useful discussions and suggestions. This work is supported by funding from the European Research Council (ERC) under the European Unions Horizon 2020 research
					and innovation programme, Grant agreement No. 833251 PROMINENT ERC-ADG 2018;
			  		by the VSC (Flemish Supercomputer Center), funded by the Research Foundation – Flanders (FWO) and the Flemish Government – department EWI; and by internal funds KU Leuven, project C14/19/089 TRACESpace.}

\normalsize\newpage

\appendix
\section{Numerical approach}	\label{sect: numerical_approach}
\subsection{Finite element analysis}
\begin{figure}[t]
	\centering
	\includegraphics[width=\textwidth]{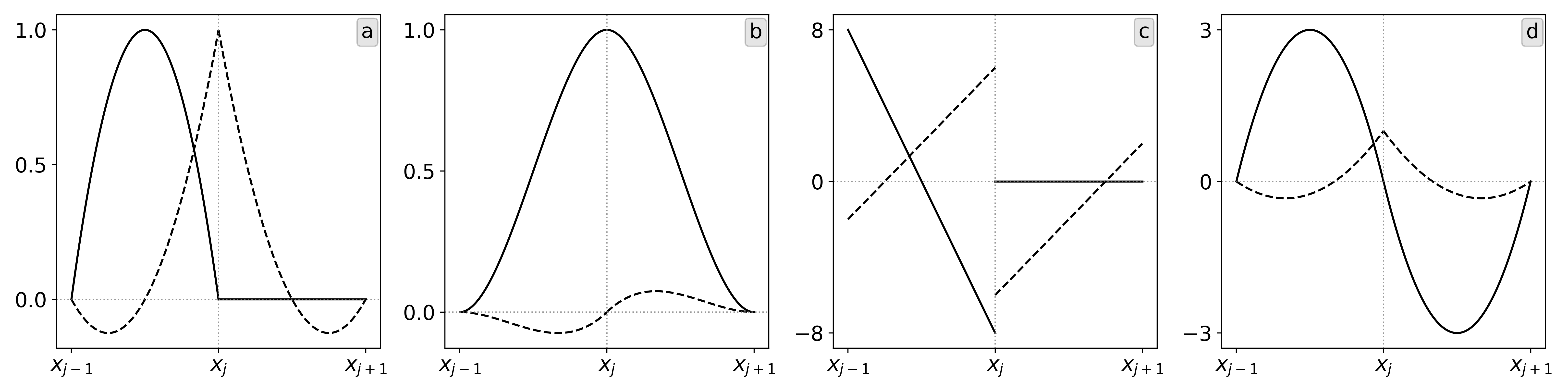}
	\caption{The quadratic (a) and cubic (b) basis functions for the interval $[x_{j-1}, x_{j+1}]$ along with their derivatives (c) and (d), respectively. The solid lines denote $H^1$, the dashed lines $H^2$.}
	\label{fig: basis_functions}
\end{figure}

The system of equations \eqref{eq: linearised_rho_trans}-\eqref{eq: linearised_a3_trans} actually defines a generalised eigenvalue problem in $\omega$ of the form $\mathcal{A}\bfX = \omega\mathcal{B}\bfX$ in which the state vector $\bfX$ is given by
\begin{equation}	\label{eq: state_vector}
	\bfX = (\rho_1, v_1, v_2, v_3, T_1, a_1, a_2, a_3) = (x^1, x^2, x^3, x^4, x^5, x^6, x^7, x^8),
\end{equation}
where each superscript on the right hand side denotes the index of the unknown variable in the state vector. The matrices $\mathcal{A}$ and $\mathcal{B}$ contain the various equilibrium quantities and differential operators with respect to $u_1$.
The eigenvalue problem is solved by using the Finite Element Method (FEM). The basic idea is to discretise the domain in (not necessarily equally spaced) subdomains that are separated by nodes, that is, prechosen fixed points $x_j$ with $j$ the node number ($0$ to $N$) on the interval $x$ (in the Cartesian case), after which a linear combination of basis functions is used to approximate the unknown variable $x^i$. These basis functions are represented by local piecewise polynomials on every subdomain, and vanish outside this subdomain. As shown in for example \citet{rappaz1977}, using the same basis functions for all variables leads to spectral pollution, while using a combination of linear and constant finite elements can yield inaccurate eigenvalues \citep{kerner1985}. Hence, \texttt{Legolas} uses a combination of higher-order finite elements, namely quadratic basis functions for the variables $\rho_1, v_2, v_3, T_1, a_1$ and cubic hermite basis functions for the variables $v_1, a_2, a_3$. This mixture of high-order finite elements ensures that the eigenfunctions can represent certain physical properties of specific eigenfunctions exactly, and can for example allow for truly incompressible modes that have $\nabla \cdot v_1 = 0$ everywhere on the domain. The trade-off in using higher-order basis functions is that this increases the accuracy of the spectrum, but also increases the size of the final matrices by a factor two compared to linear finite elements. Hence, the approximation of a state variable $x^i$ for a domain divided into $N$ subdomains (nodes) is given by
\begin{equation}	\label{eq: FE_sum}
	x^i(u_1) \approx \sum_{j=0}^{N}\biggl(H^1_j(u_1)x_{j1}^i + H^2_j(u_1)x_{j2}^i\biggr),
\end{equation}
where $H_j^1(u_1)$ and $H_j^2(u_1)$ denote the quadratic or cubic elements at node $j$. The hermite ($C_j$) and quadratic ($Q_j$) basis functions themselves are given by
\begin{equation}	\label{eq: basis_functions}
	\begin{alignedat}{2}
		&Q^1_j(x) =
		\begin{cases}
			\dfrac{4\left(x - x_{j-1}\right)\left(x_j - x\right)}{\left(x_j - x_{j-1}\right)^2},	\\
			0, 	\\
			0,
		\end{cases}
		&C^1_j(x) &=
		\begin{cases}
			\left(\dfrac{x - x_{j-1}}{x_j - x_{j-1}}\right)^2\left(3 - 2\dfrac{x - x_{j-1}}{x_j - x_{j-1}}\right) 	\qquad &\text{for}~ x_{j-1} \leq x \leq x_j,	\\
			\left(\dfrac{x_{j+1} - x}{x_{j+1} - x_j}\right)^2\left(3 - 2\dfrac{x_{j+1} - x}{x_{j+1} - x_j}\right) 	\qquad &\text{for}~ x_j \leq x \leq x_{j+1},	\\
			0	\qquad &\text{elsewhere},
		\end{cases}	\\
		&Q^2_j(x) =
		\begin{cases}
			\dfrac{\left(2x - x_j - x_{j-1}\right)\left(x - x_{j-1}\right)}{\left(x_j - x_{j-1}\right)^2},	\\
			\dfrac{\left(2x - x_{j+1} - x_j\right)\left(x - x_{j+1}\right)}{\left(x_{j+1} - x_j\right)^2}, 	\\
			0,
		\end{cases}
		&C^2_j(x) &=
		\begin{cases}
			\left(x - x_j\right)\left(\dfrac{x - x_{j-1}}{x_j - x_{j-1}}\right)^2	\qquad &\text{for}~ x_{j-1} \leq x \leq x_j,	\\
			\left(x - x_j\right)\left(\dfrac{x_{j+1} - x}{x_{j+1} - x_j}\right)^2	\qquad &\text{for}~ x_j \leq x \leq x_{j+1},	\\
			0	\qquad &\text{elsewhere},
		\end{cases}
	\end{alignedat}
\end{equation}
as in for example \citet{kerner1998, book_MHD}, and are shown graphically along with their derivatives in Figure \ref{fig: basis_functions}. Since we have two basis functions per subdomain, this allows for the approximation of both the original variable and its derivative by differentiating Eq. \eqref{eq: FE_sum}.
To turn the problem algebraic in nature, we use the Galerkin method such that the eigenvalue problem is written as a set of integral equations by multiplying each of the eight equations by an appropriate element of the chosen basis, denoted by $h_j$, and integrate over the relevant domain. Mathematically, this can be written as
\begin{equation}	\label{eq: galerkin_form}
	\int_\Omega \left(\mathcal{A}\bfX - \omega\mathcal{B}\bfX\right)h_jdu_1 = 0.
\end{equation}
However, the matrix $\mathcal{A}$ contains second-order derivatives with respect to $u_1$. In order to reduce these to derivatives of first order, hence simplifying the integrals, we make use of the Galerkin weak formulation.
This is achieved by performing integration by parts, which introduces additional surface terms that have to be evaluated at the boundaries.
These surface terms can in turn be exploited to enforce boundary conditions, which will be discussed in Subsection \ref{subsect: implement_boundaries}.
Since there are eight unknowns in our eigenvalue system, together with two basis functions and $N$ subdomains, this implies that the final matrix eigenvalue problem will have $16N$ equations, resulting in a $16N\times 16N$ size matrix. By extension, it becomes clear that using basis functions of even higher order will increase the size of this eigensystem considerably.

\subsection{Matrix assembly}	\label{subsect: matrix_assembly}
The actual expression for the matrix elements can be obtained by applying Eq. \eqref{eq: galerkin_form} to the system of differential equations \eqref{eq: linearised_rho_trans}-\eqref{eq: linearised_a3_trans}. As an example we will look at the $v_1$ component of the linearised continuity equation \eqref{eq: linearised_rho_trans}, corresponding to element $(1, 2)$ in the matrices. This ``number" links to the indices of the state vector $\bfX$, since the continuity equation is associated with $\rho_1$, or index 1 in the state vector \eqref{eq: state_vector}, and the $v_1$ component is associated with index 2. How these indices translate to the actual position in each matrix will be discussed further in this section.

In the finite element representation adopted here, the $(\rho_1, v_1)$ contribution can be expanded as
\begin{equation}
	\sum_{j=0}^{N}\omega\int \frac{1}{\eps}h_j^1h_k^1du_1 = -\sum_{j=0}^{N}\int \frac{1}{\eps}\rho_0'h_j^1h_k^2du_1 - \sum_{j=0}^{N}\int \frac{1}{\eps}\rho_0 h_j^1\frac{dh_k^2}{du_1}du_1,
\end{equation}
where the $u_1$-dependence of the equilibrium density $\rho_0$ and the basis functions $h^1$ and $h^2$ is implied. In this case $h^1$ is quadratic, since $\rho_1$ is associated with a quadratic basis function; analogously $h^2$ is cubic. The matrix elements for this particular contribution are hence given by
\begin{equation}	\label{eq: example_matrix_elements}
	\mathcal{B}_{jk}(1, 1) = \int\frac{1}{\eps}h_j^1h_k^1du_1,	\qquad\qquad 	\mathcal{A}_{jk}(1, 2) = -\int\left(\frac{1}{\eps}\rho_0'\right)h_j^1h_k^2du_1 - \int\left(\frac{1}{\eps}\rho_0\right)h_j^1\frac{dh_k^2}{du_1}du_1.
\end{equation}
If this reasoning is applied to all equations in the linear system, it follows that $\mathcal{B}$ will only have equal-number elements, implying that $\mathcal{B}$ is fully symmetric and real. $\mathcal{A}$ on the other hand will have cross-term elements such that it is, in general, not symmetric. Furthermore, it might be complex, depending on the included physical effects. Terms that contain derivatives of the state vector components, as for example $(v_1, \rho_1)$ which corresponds to element (2, 1), will be integrated by parts. It is exactly this integration that gives rise to the surface terms, that is, terms that do not contain an integral which translate into the natural boundary conditions. These are discussed in the next subsection.

The actual assembly of both matrices $\mathcal{A}$ and $\mathcal{B}$ in \texttt{Legolas} is done by sequential iteration over the gridpoints. From \eqref{eq: basis_functions} we see that finite elements have a localised nature around every gridpoint $j$, meaning that only the elements associated with a certain region (that is, the gridpoint $j$ itself and its neighbours $j-1$ and $j+1$) yield a non-zero contribution. However, it is actually easier implementation-wise to loop over the elements in the interval $[x_{j-1}, x_j]$ instead of over those in $[x_{j-1}, x_{j+1}]$, since then the actual integration of the matrix elements can be done in the same way, independent of whether the basis functions are cubic or quadratic. If only the interval $[x_{j-1}, x_j]$ is considered we end up with 16 possible combinations of the basis functions for every gridpoint. This translates into a $4 \times 4$ sub-matrix for every variable, where every one of the 16 elements corresponds to one specific combination of the shape functions. As there are eight variables in the state vector this implies a $32 \times 32$ matrix block for every gridpoint, hereafter dubbed a ``quadblock" since it consists of four $16 \times 16$ blocks (hereafter called ``subblocks"). Every one of those subblocks inside a quadblock corresponds to a quarter section of the aforementioned sub-matrix, which is a $2 \times 2$ block in every subblock. Hence, to recap, a $2 \times 2$ block times eight variables represents a $16 \times 16$ subblock, of which four combined form a $32 \times 32$ quadblock for every gridpoint.

Of course, every matrix element contains one or more integrals that still have to be calculated. Since the coefficient functions are in general complicated expressions depending on $u_1$, this is done numerically using a 4-point Gaussian quadrature for which an integral in the interval $[x_{j-1}, x_j]$ can be expressed as
\begin{equation}
	\int_{x_{j-1}}^{x_j} f(u_1)du_1 \approx \frac{1}{2}\left(x_j - x_{j-1}\right) \sum_{i=1}^{4}w_i f\left(\frac{1}{2}(x_j - x_{j-1})\xi_i + \frac{1}{2}(x_{j-1} + x_j)\right),
\end{equation}
where $\xi_i$ and $w_i$ are the evaluation points and weights of the Gaussian integration. These values can be found in various textbooks, as for example given in \citet{book_MHD}. The function $f$ denotes the integral coefficients, which are essentially the equilibrium quantities and basis functions evaluated in the various evaluation points. This actually implies that every grid interval $[x_{j-1}, x_j]$ is subdivided into four points, meaning that the equilibrium expressions are probed using $4(N-1)$ points rather than $N$ points. The way the matrices are then assembled is thus done on a double-loop basis, where the outer loop iterates over the intervals $[x_{j-1}, x_j]$ and the inner loop iterates over the four Gaussian points. This inner loop will calculate the basis functions and matrix elements at every point, then multiply the coefficients with the Gaussian weights and finally add them all together in a consistent manner.

\begin{figure}[t]
	\centering
	\includegraphics[width=\textwidth]{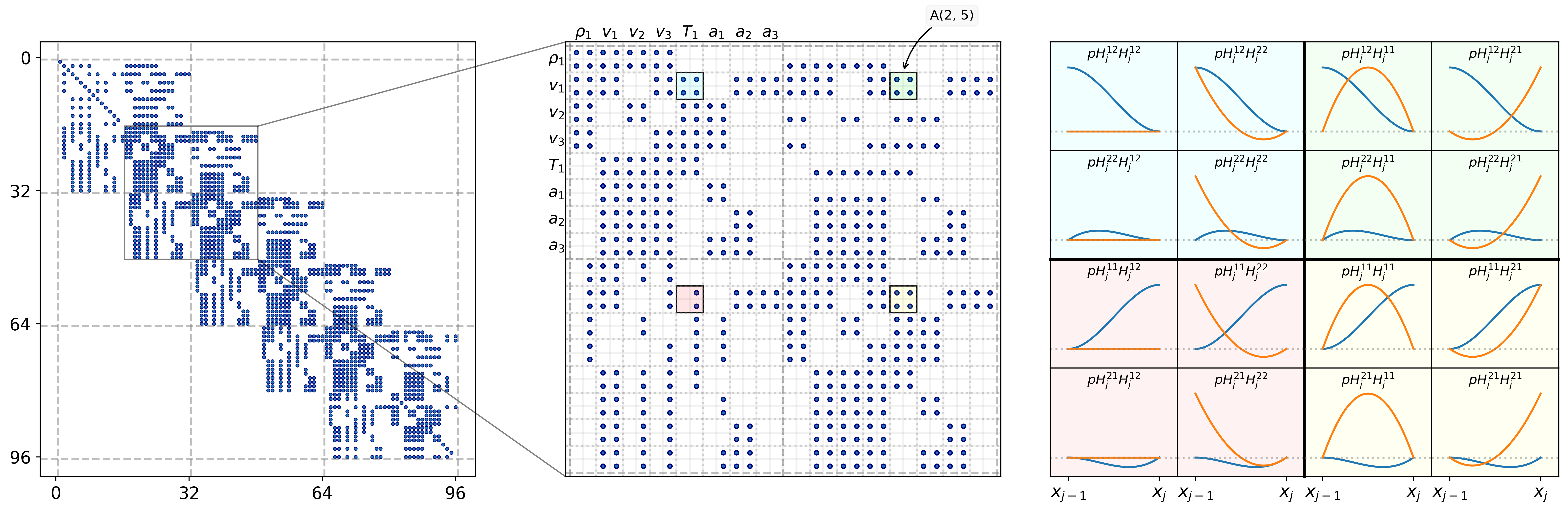}
	\caption{General assembly and structure of the finite element (complex) $\mathcal{A}$ matrix. Left: Example of a full matrix for six gridpoints, showing the block-tridiagonal structure where dots represent non-zero values.
		The middle panel zooms in on one quadblock, showing the dependence of different subblock positions with respect to the variables. The $2 \times 2$ blocks corresponding to $\mathcal{A}(2, 5)$ (cubic, quadratic) are highlighted.
		Right panel: $2 \times 2$ block assembly for a general $\mathcal{A}$ (cubic, quadratic) matrix element. Cubic elements are shown in blue, quadratic ones in orange. The dotted grey line denotes zero.}
	\label{fig: matrix_assembly}
\end{figure}

Since every quadblock corresponds to the interval $[x_{j-1}, x_j]$ we still have to account for the contribution of the $x_{j+1}$ gridpoint. This is done by partially overlapping the quadblock in the next gridpoint with the one from the previous gridpoint. Figure \ref{fig: matrix_assembly} shows a visual representation of the structure and assembly process for the $\mathcal{A}$ matrix, using the Kelvin-Helmholtz and current-driven equilibrium discussed in Section \ref{subsect: kh_cd} with only six gridpoints here for the purpose of illustration. The left panel shows the general matrix, highlighting the block-tridiagonal structure. The dashed grey lines denote the $32 \times 32$ quadblocks of the matrix, and every dot represents a non-zero value. In total five quadblocks can be distinguished for six gridpoints, one for every grid interval. The middle panel shows a zoom-in of the quadblock corresponding to the second grid interval as annotated on the left panel, where it can be seen that the top-left corner of this quadblock overlaps with the bottom-right corner of the previous quadblock corresponding to the first grid interval.

A single quadblock is further divided into four subblocks, with the location of the different state variables annotated on the middle panel of Figure \ref{fig: matrix_assembly}. The matrix element $\mathcal{A}(2, 5)$ is highlighted in every subblock which corresponds to the $(v_1, T_1)$ contribution, which are cubic ($v_1$) and quadratic ($T_1$) variables. The $2 \times 2$ block in the top-left corner of the quadblock corresponds to the top-left $2 \times 2$ corner of the right panel, as indicated by the background colours. The right panel shows the various combinations of the regular basis functions for the $\mathcal{A}(2, 5)$ contribution, where the blue curve corresponds to the cubic basis functions and the orange curves to the quadratic ones. It should be noted that the right panel shows one specific case, that is, the regular $h_j^2 h_k^5$ terms of the matrix element. If for example the $h_j^2h_k^7$ element is calculated, corresponding to the $a_2$ term in \eqref{eq: linearised_v1_trans}, the quadratic basis functions have to be replaced by their cubic counterparts, since $a_2$ is also a cubic variable. A similar reasoning can be made for the other matrix elements.

The term $pH_j^{\alpha\beta}H_j^{\alpha\beta}$ at the top of every sub-panel on the right of Figure \ref{fig: matrix_assembly} denotes which combination of the basis functions should be used, where $p$ stands for the integral coefficients. The exponent $\alpha\beta$ refers to the expressions for the basis functions in \eqref{eq: basis_functions}, where $\alpha = 1$ stands for $Q_j^1$ or $C_j^1$ and $\alpha = 2$ stands for $Q_j^2$ or $C_j^2$, depending on the variable under consideration.
$\beta$ stands for which part of the basis function that should be taken, $\beta = 1$ means the first equation in cases, while $\beta = 2$ means the second one. As an example we can look at $H_j^{12}H_j^{21}$: since $\mathcal{A}(2, 5)$ represents a cubic and quadratic variable, this translates into $C_j^{12}Q_j^{21}$. For the cubic part we therefore take the second equation of $C_j^1$, corresponding to $x_j \leq x \leq x_{j+1}$. The quadratic part on the other hand is given by $Q_j^{21}$, which means that we take the first equation of $Q_j^2$, corresponding to $x_{j-1} \leq x \leq x_j$. Boundary conditions are imposed after matrix assembly is completed.

\subsection{Implementation of boundary conditions}	\label{subsect: implement_boundaries}
Integration by parts on \eqref{eq: galerkin_form} gives rise to additional surface terms, which should be evaluated at the boundaries. These kind of conditions are called natural boundary conditions, since they emerge in a natural way by rewriting the eigenvalue problem. The regularity and fixed wall conditions considered earlier on the other hand are called essential boundary conditions, and have to be handled explicitly. Since the additional surface terms originate from reducing second-order derivatives to first order derivatives, we only have these terms for the variables $v_1, T_1, a_2$ and $a_3$, as these are the only equations that contain derivatives of higher order.
Hence, for the momentum equation, the additional surface terms can be written as
\begin{equation}	\label{eq: v1_boundary}
	\begin{aligned}
		S_{v_1} &= \left[\frac{T_0}{\eps}h_j^2h_k^1 + \frac{\rho_0}{\eps}h_j^2h_k^5 + \left(B_{02}k_3 - \frac{1}{\eps}B_{03}k_2\right)h_j^2h_k^6 + \frac{B_{03}}{\eps}h_j^2\frac{dh_j^7}{du_1} - B_{02}h_j^2\frac{dh_k^8}{du_1}\right]_{\partial \Omega} \\
				&= v_1\left[\frac{T_0}{\eps}\rho_1 + \frac{\rho_0}{\eps}T_1 + \left(B_{02}k_3 - \frac{1}{\eps}B_{03}k_2\right)a_1 + \frac{B_{03}}{\eps}a_2' - B_{02}a_3'\right]_{\partial \Omega},
	\end{aligned}
\end{equation}
where the number in superscript on $h_{j/k}$ denotes the index of the variable in the state vector $\bfX$. The subscript $\partial \Omega$ means that these terms should be evaluated at the left- or inner edge,
as well as at the right- or outer edge. In a similar manner are the surface terms for the energy and induction equations given by
\begin{align}
	S_{T_1} &= \frac{i(\gamma - 1)}{\eps}T_1\Biggl[\Bigr.
	\begin{alignedat}[t]{1}
		&T_0'\frac{\partial \kappa_\bot}{\partial \rho}\rho_1 + \left(T_0'\frac{\partial\kappa_\bot}{\partial T} - \frac{\eps'}{\eps}\kappa_\bot\right)T_1 + \kappa_\bot T_1' \\
		&+2\left(T_0'\bigl(\eps B_{02}k_3 - B_{03}k_2\bigr)\frac{\partial \kappa_\bot}{\partial (B^2)} - \eta B_{03}'k_2 + \eta k_3(\eps B_{02})'\right)a_1	\\
		&+2\left(T_0'B_{03}\frac{\partial\kappa_\bot}{\partial(B^2)} + \eta B_{03}'\right)a_2' - 2\left(\eps T_0'B_{02}\frac{\partial\kappa_\bot}{\partial(B^2)} + \eta(\eps B_{02})'\right)a_3'\Biggl.\Biggr],	\label{eq: T1_boundary}
	\end{alignedat}	\\
	S_{a_2} &= \dfrac{i\eta}{\eps}a_2\Bigl(-k_2a_1 + a_2'\Bigr),	\label{eq: a2_boundary}	\\
	S_{a_3} &= i\eta a_3\Bigl(-k_3a_1 + a_3'\Bigr),	\label{eq: a3_boundary}
\end{align}
which should all be evaluated at the boundaries. For the case of a solid wall we see that the natural boundaries simplify considerably, since if $v_1, a_2, a_3$ are all zero at the wall, $S_{v_1}, S_{a_2}$ and $S_{a_3}$ are also zero and have to be omitted. The natural boundary condition on $T_1$ is only relevant if resistivity or perpendicular thermal conduction is included. However, in the case of the latter, the additional essential boundary condition requires that $T_1 = 0$, in which case $S_{T_1}$ drops out as well. The only combination in which the surface terms for the energy equation are non-zero is when resistivity is included, but perpendicular thermal conduction is omitted. In that case the resistive heating terms should be included in the calculation, which is done by adding the appropriate terms to the matrix elements $A(5, 6), A(5, 7)$ and $A(5, 8)$.

Currently only fixed wall boundary conditions are implemented in \texttt{Legolas}. However, the surface terms described here can be used to impose other types of boundary conditions as well. In the case of a plasma-vacuum-wall transition we have for example Bessel functions at the outer boundary of a cylindrical geometry \citep{book_roberts}, which encode the analytic vacuum solution for the electromagnetic field in the outer vacuum region. These expressions can then be used to rewrite the surface terms \eqref{eq: v1_boundary}-\eqref{eq: a3_boundary} in an appropriate way such that they can be added to their respective subblock positions in the matrix. This is a planned extension to be included in future versions of \texttt{Legolas}. This functionality was previously available in some LEDA versions \citep{vanderlinden1992}.

The essential boundary conditions as described in Section \ref{subsect: boundary_conditions} have to be implemented explicitly. This is done by omitting the relevant basis functions that do not satisfy the boundary conditions on the edges. Consider as an example the variable $v_1$, which is associated with a cubic basis function. From Figure \ref{fig: basis_functions} we see that the only cubic element which is non-zero at the left boundary is $C_j^{12}$, which implies that the matrix elements where it appears should be zeroed out. Looking back at how the quadblock is composed in Figure \ref{fig: matrix_assembly}, the boundary condition $v_1 = 0$ corresponds to forcing the odd rows and columns of the $v_1$ contribution to zero, for subblocks 1, 2 and 3 on the left boundary (that is, the first node $j$) since these correspond to the $2 \times 2$ blocks in blue, green and red on the right panel. Similarly, on the right side (viz. the last node $j$) only $C_j^{11}$ is non-zero, which implies that the odd rows and columns of subblocks 2, 3 and 4 have to be zeroed out (corresponding to green, red and yellow). Extending this reasoning to the essential boundary condition on $T_1$, we see that in this case the even rows and columns should be handled since $T_1$ is associated with a quadratic element. This is done for both matrices.

Of course, ``just" zeroing out rows and columns in a matrix has the unpleasant side-effect that the matrices become singular. For the $\mathcal{A}$ matrix this is not necessarily a problem, however, the $\mathcal{B}$ matrix can never be singular since it is inverted when solving the general eigenvalue problem using the QR algorithm. Therefore, we introduce a one on $\mathcal{B}$'s diagonal at the location that was zeroed out, and an element $\delta$ on $\mathcal{A}$'s diagonal. The effect of this is that one essentially ``forces" the boundary condition, since this implies that $\delta x^i_j = \omega x^i_j$. By extension, if $\delta$ is taken to be a large number (we take $\delta = 10^{20}$), this means that $x^i_j = 0$, which corresponds to the essential boundary condition we wanted to impose.
The only side-effect of this approach is that it introduces eigenvalues equal to $\delta$. However, since $\delta$ is taken to be large, these will not influence the spectrum in any way and they can be easily filtered out during post-processing.
This method thus provides a relatively easy and straightforward way to impose Dirichlet boundary conditions at the edges. The imposed boundary conditions are noticeable on the left panel of Figure \ref{fig: matrix_assembly}, especially for the first node. The odd rows and columns that were zeroed out can clearly be seen, together with the large numbers introduced on the main diagonal.

\section{Erratum: ``Legolas: a modern tool for magnetohydrodynamic spectroscopy" \\ (2020, ApJS, 251, 25)} \label{erratum}
\begin{figure}[t]
	\centering
	\includegraphics[width=\textwidth]{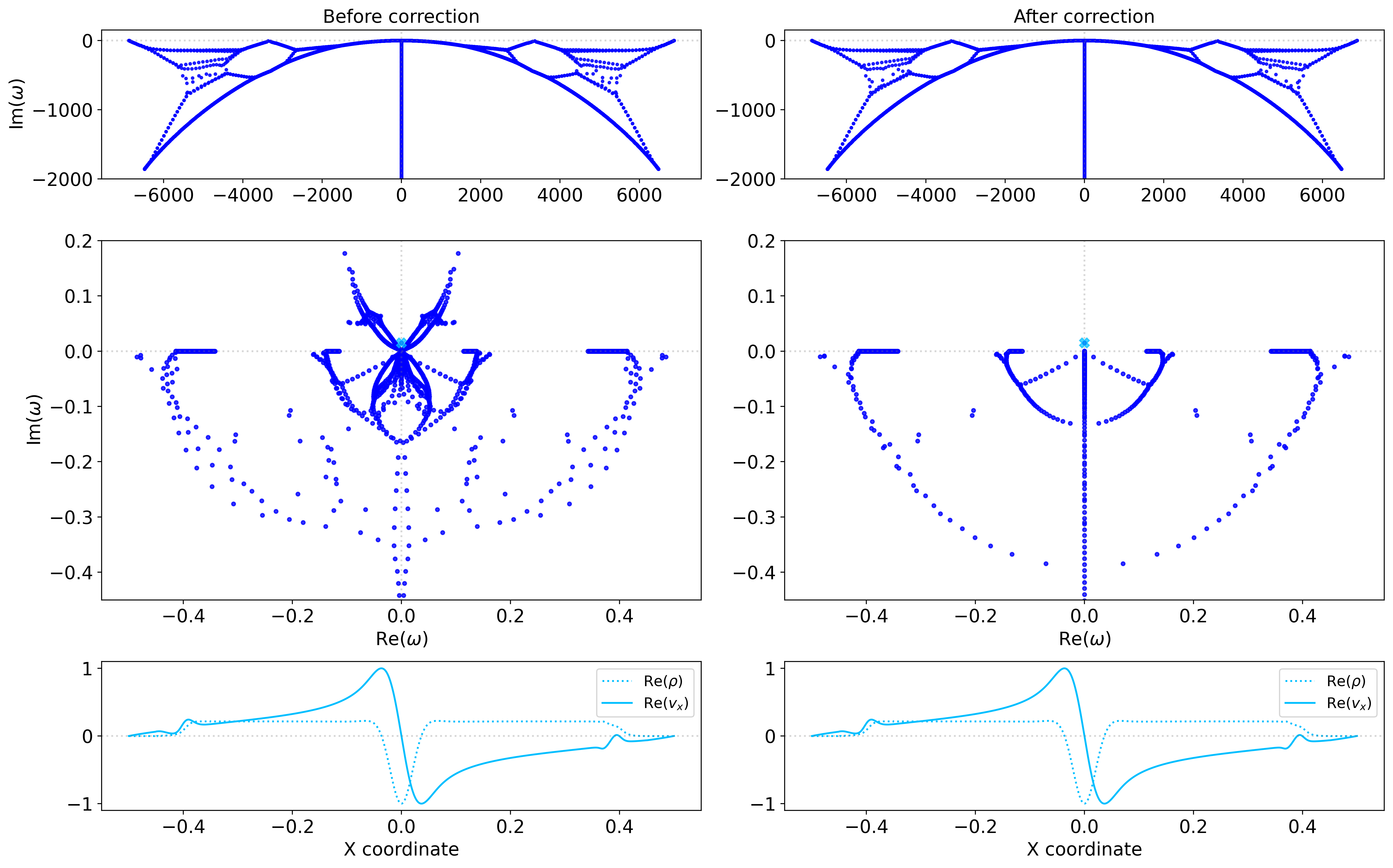}
	\caption{Spectra before (left) and after (right) correction of the erroneous terms. The large-scale spectrum remains mostly unaffected (top panels), as are the tearing mode and its eigenfunctions (bottom panels).
			 The damped slow and Alfv\'en sequences (middle panels) are in fact modified, and there are no longer rippling modes present. The tearing mode is annotated with a cyan cross.}
	\label{fig: spectrum}
\end{figure}

In this work, published in ApJS \textbf{251}, 25 (2020), we reported on the \texttt{Legolas} code, where we gave a detailed overview of the code itself and discussed its application on various equilibria. The case in Section 3.3.4 treated so-called rippling modes, which may arise whenever there is a spatially varying resistivity profile present. Herein we imposed a hyperbolic tangent profile for the resistivity $\eta(x)$, and showed a spectrum that has multiple unstable branches on the left and right side of the imaginary axis as depicted here in Fig. \ref{fig: spectrum}, middle left panel. Based on the very localized nature of the eigenfunctions, we concluded that these were rippling modes alongside an already present tearing mode.
However, during an extension of the code we discovered a bug in one of the terms of the matrix elements $A(7, 5)$ and $A(8, 5)$, which both correspond to the $T_1$ resistive components in the linearized $a_2$ and $a_3$ equations. More specifically, the resistivity derivative with respect to temperature $d\eta/dT$ was erroneously treated as a spatial derivative. This implies that these two elements were nonzero, while they should vanish for the imposed temperature-independent resistivity profile. These nonzero values in turn led to a modification of the spectrum and gave rise to the two unstable branches.

After these two terms were corrected, the rippling modes are no longer present as can be seen in the right-hand side of Figure \ref{fig: spectrum}, while the tearing mode and its associated eigenfunctions remain unaffected. The large-scale spectrum is barely modified, however on smaller scales it can be seen that there is a major effect on the damped slow and Alfv\'en sequences. After correction of the terms these sequences better trace out the semicircles in the complex plane, in much closer resemblance to the original tearing mode spectra in Section 3.3.3 of the original work.

It should be noted that the general conclusions regarding rippling modes remain valid. For the spatially varying resistivity profile that we imposed there are no rippling modes, but for a more general $\eta(x, T(x))$ profile the two matrix elements discussed here will indeed be nonzero. As such it is entirely possible that for some profiles rippling modes may arise, and the question of rippling- versus tearing mode dominance in more realistic resistivity profiles remains relevant.

\bibliography{bibfile}{}

\begin{thebibliography}{}
\expandafter\ifx\csname natexlab\endcsname\relax\def\natexlab#1{#1}\fi
\providecommand{\url}[1]{\href{#1}{#1}}
\providecommand{\dodoi}[1]{doi:~\href{http://doi.org/#1}{\nolinkurl{#1}}}
\providecommand{\doeprint}[1]{\href{http://ascl.net/#1}{\nolinkurl{http://ascl.net/#1}}}
\providecommand{\doarXiv}[1]{\href{https://arxiv.org/abs/#1}{\nolinkurl{https://arxiv.org/abs/#1}}}

\bibitem[{Anderson {et~al.}(1999)Anderson, Bai, Bischof, Blackford, Demmel,
  Dongarra, Du~Croz, Greenbaum, Hammarling, McKenney, \& Sorensen}]{lapack}
Anderson, E., Bai, Z., Bischof, C., {et~al.} 1999, LAPACK Users' Guide, 3rd
  edn. (Philadelphia, PA: Society for Industrial and Applied Mathematics)

\bibitem[{Balbus \& Hawley(1991)}]{balbus1991}
Balbus, S.~A., \& Hawley, J.~F. 1991, The Astrophysical Journal, 376, 214

\bibitem[{Ballester(2006)}]{ballester2006}
Ballester, J.~L. 2006, Space science reviews, 122, 129

\bibitem[{Barbulescu {et~al.}(2019)Barbulescu, Ruderman, Van~Doorsselaere, \&
  Erd{\'e}lyi}]{barbulescu2019}
Barbulescu, M., Ruderman, M.~S., Van~Doorsselaere, T., \& Erd{\'e}lyi, R. 2019,
  The Astrophysical Journal, 870, 108

\bibitem[{Baty \& Keppens(2002)}]{baty2002}
Baty, H., \& Keppens, R. 2002, The Astrophysical Journal, 580, 800

\bibitem[{Beli{\"e}n {et~al.}(2002)Beli{\"e}n, Botchev, Goedbloed, van~der
  Holst, \& Keppens}]{belien2002}
Beli{\"e}n, A., Botchev, M., Goedbloed, J., van~der Holst, B., \& Keppens, R.
  2002, Journal of computational physics, 182, 91

\bibitem[{Berger {et~al.}(2012)Berger, Liu, \& Low}]{berger2012}
Berger, T.~E., Liu, W., \& Low, B. 2012, The Astrophysical Journal Letters,
  758, L37

\bibitem[{Blokland {et~al.}(2007{\natexlab{a}})Blokland, Keppens, \&
  Goedbloed}]{blokland2007}
Blokland, J., Keppens, R., \& Goedbloed, J. 2007{\natexlab{a}}, Astronomy \&
  Astrophysics, 467, 21

\bibitem[{Blokland {et~al.}(2007{\natexlab{b}})Blokland, van~der Holst,
  Keppens, \& Goedbloed}]{blokland2007phoenix}
Blokland, J., van~der Holst, B., Keppens, R., \& Goedbloed, J.
  2007{\natexlab{b}}, Journal of computational physics, 226, 509

\bibitem[{Chandrasekhar(2013)}]{book_chandrasekhar}
Chandrasekhar, S. 2013, Hydrodynamic and hydromagnetic stability (Courier
  Corporation)

\bibitem[{Choudhuri(1998)}]{book_choudhuri}
Choudhuri, A.~R. 1998, The Physics of Fluids and Plasmas: An Introduction for
  Astrophysicists (Cambridge University Press),
  \dodoi{10.1017/CBO9781139171069}

\bibitem[{Claes {et~al.}(2020)Claes, Keppens, \& Xia}]{claes2020}
Claes, N., Keppens, R., \& Xia, C. 2020, Astronomy \& Astrophysics, 636, A112

\bibitem[{Colgan {et~al.}(2008)Colgan, Abdallah~Jr, Sherrill, Foster, Fontes,
  \& Feldman}]{colgan2008}
Colgan, J., Abdallah~Jr, J., Sherrill, M., {et~al.} 2008, The Astrophysical
  Journal, 689, 585

\bibitem[{Dalgarno \& McCray(1972)}]{dalgarno1972}
Dalgarno, A., \& McCray, R. 1972, Annual review of astronomy and astrophysics,
  10, 375

\bibitem[{Demaerel \& Keppens(2016)}]{demaerel2016}
Demaerel, T., \& Keppens, R. 2016, Physics of Plasmas, 23, 122118

\bibitem[{Engvold(1998)}]{engvold1998}
Engvold, O. 1998, in Astronomical Society of the Pacific Conference Series,
  Vol. 150, IAU Colloq. 167: New Perspectives on Solar Prominences, ed. D.~F.
  Webb, B.~Schmieder, \& D.~M. Rust, 23

\bibitem[{Goedbloed {et~al.}(2019)Goedbloed, Keppens, \& Poedts}]{book_MHD}
Goedbloed, H., Keppens, R., \& Poedts, S. 2019, Magnetohydrodynamics of
  Laboratory and Astrophysical Plasmas (Cambridge University Press),
  \dodoi{10.1017/9781316403679}

\bibitem[{Goedbloed(2018{\natexlab{a}})}]{goedbloed2018web1}
Goedbloed, J. 2018{\natexlab{a}}, Physics of Plasmas, 25, 032109

\bibitem[{Goedbloed(2018{\natexlab{b}})}]{goedbloed2018web2}
---. 2018{\natexlab{b}}, Physics of Plasmas, 25, 032110

\bibitem[{Goedbloed {et~al.}(2004)Goedbloed, Beli{\"e}n, van~der Holst, \&
  Keppens}]{goedbloed2004}
Goedbloed, J., Beli{\"e}n, A., van~der Holst, B., \& Keppens, R. 2004, Physics
  of Plasmas, 11, 28

\bibitem[{Goedbloed {et~al.}(1993)Goedbloed, Holties, Poedts, Huysmans, \&
  Kerner}]{goedbloed1993}
Goedbloed, J., Holties, H., Poedts, S., Huysmans, G., \& Kerner, W. 1993,
  Plasma physics and controlled fusion, 35, B277

\bibitem[{Goedbloed(2011)}]{goedbloed2011}
Goedbloed, J.~H. 2011, Plasma Physics and Controlled Fusion, 53, 074001

\bibitem[{Hillier(2018)}]{hillier2018}
Hillier, A. 2018, Reviews of Modern Plasma Physics, 2, 1

\bibitem[{Hillier {et~al.}(2019)Hillier, Barker, Arregui, \&
  Latter}]{hillier2019}
Hillier, A., Barker, A., Arregui, I., \& Latter, H. 2019, Monthly Notices of
  the Royal Astronomical Society, 482, 1143

\bibitem[{Hillier \& Polito(2018)}]{hillier2018_KHI}
Hillier, A., \& Polito, V. 2018, The Astrophysical Journal Letters, 864, L10

\bibitem[{Keppens {et~al.}(2002)Keppens, Casse, \& Goedbloed}]{keppens2002}
Keppens, R., Casse, F., \& Goedbloed, J. 2002, The Astrophysical Journal
  Letters, 569, L121

\bibitem[{Keppens {et~al.}(1993)Keppens, Van Der~Linden, \&
  Goossens}]{keppens1993}
Keppens, R., Van Der~Linden, R.~A., \& Goossens, M. 1993, Solar physics, 144,
  267

\bibitem[{Kerner {et~al.}(1998)Kerner, Goedbloed, Huysmans, Poedts, \&
  Schwarz}]{kerner1998}
Kerner, W., Goedbloed, J., Huysmans, G., Poedts, S., \& Schwarz, E. 1998,
  Journal of computational physics, 142, 271

\bibitem[{Kerner {et~al.}(1985)Kerner, Lerbinger, Gruber, \&
  Tsunematsu}]{kerner1985}
Kerner, W., Lerbinger, K., Gruber, R., \& Tsunematsu, T. 1985, Computer physics
  communications, 36, 225

\bibitem[{Mackay {et~al.}(2010)Mackay, Karpen, Ballester, Schmieder, \&
  Aulanier}]{mackay2010}
Mackay, D., Karpen, J., Ballester, J., Schmieder, B., \& Aulanier, G. 2010,
  Space Science Reviews, 151, 333

\bibitem[{MacTaggart(2018)}]{mactaggart2018}
MacTaggart, D. 2018, Journal of Plasma Physics, 84, 905840501,
  \dodoi{10.1017/S0022377818001009}

\bibitem[{Nakariakov \& Ofman(2001)}]{nakariakov2001}
Nakariakov, V., \& Ofman, L. 2001, Astronomy \& Astrophysics, 372, L53

\bibitem[{Nijboer {et~al.}(1997)Nijboer, Holst, Poedts, \&
  Goedbloed}]{nijboer1997}
Nijboer, R., Holst, B., Poedts, S., \& Goedbloed, J. 1997, Computer physics
  communications, 106, 39

\bibitem[{Parker(1953)}]{parker1953}
Parker, E.~N. 1953, The Astrophysical Journal, 117, 431

\bibitem[{Poedts {et~al.}(1989)Poedts, Goossens, \& Kerner}]{poedts1989}
Poedts, S., Goossens, M., \& Kerner, W. 1989, Solar physics, 123, 83

\bibitem[{Poedts \& Kerner(1991)}]{poedts1991}
Poedts, S., \& Kerner, W. 1991, Physical review letters, 66, 2871

\bibitem[{Priest(2014)}]{book_priest}
Priest, E. 2014, Magnetohydrodynamics of the Sun (Cambridge University Press),
  \dodoi{10.1017/CBO9781139020732}

\bibitem[{Rappaz(1977)}]{rappaz1977}
Rappaz, J. 1977, Numerische Mathematik, 28, 15

\bibitem[{Rempel(2012)}]{rempel2012}
Rempel, M. 2012, The Astrophysical Journal, 750, 62

\bibitem[{Roberts(2019)}]{book_roberts}
Roberts, B. 2019, MHD Waves in the Solar Atmosphere (Cambridge University
  Press), \dodoi{10.1017/9781108613774}

\bibitem[{Rosner {et~al.}(1978)Rosner, Tucker, \& Vaiana}]{rosner1978}
Rosner, R., Tucker, W.~H., \& Vaiana, G. 1978, The Astrophysical Journal, 220,
  643

\bibitem[{Ruan {et~al.}(2019)Ruan, Xia, \& Keppens}]{ruan2019}
Ruan, W., Xia, C., \& Keppens, R. 2019, The Astrophysical Journal Letters, 877,
  L11

\bibitem[{Schure {et~al.}(2009)Schure, Kosenko, Kaastra, Keppens, \&
  Vink}]{spex2009}
Schure, K., Kosenko, D., Kaastra, J., Keppens, R., \& Vink, J. 2009, Astronomy
  \& Astrophysics, 508, 751

\bibitem[{van~der Holst {et~al.}(2014)van~der Holst, Sokolov, Meng, Jin,
  Manchester~IV, Toth, \& Gombosi}]{vanderholst2014}
van~der Holst, B., Sokolov, I.~V., Meng, X., {et~al.} 2014, The Astrophysical
  Journal, 782, 81

\bibitem[{Van~der Linden \& Goossens(1991{\natexlab{a}})}]{vanderlinden1991}
Van~der Linden, R., \& Goossens, M. 1991{\natexlab{a}}, Solar physics, 134, 247

\bibitem[{Van~der Linden \& Goossens(1991{\natexlab{b}})}]{vanderlinden1991TI}
---. 1991{\natexlab{b}}, Solar physics, 131, 79

\bibitem[{Van~der Linden {et~al.}(1992)Van~der Linden, Goossens, \&
  Hood}]{vanderlinden1992}
Van~der Linden, R., Goossens, M., \& Hood, A. 1992, Solar physics, 140, 317

\bibitem[{Van~Doorsselaere \& Poedts(2007)}]{vandoorsselaere2007}
Van~Doorsselaere, T., \& Poedts, S. 2007, Plasma Physics and Controlled Fusion,
  49, 261

\bibitem[{Wang {et~al.}(2004)Wang, Blokland, Keppens, \& Goedbloed}]{wang2004}
Wang, C., Blokland, J., Keppens, R., \& Goedbloed, J. 2004, Journal of plasma
  physics, 70, 651

\bibitem[{Xia \& Keppens(2016)}]{xia2016}
Xia, C., \& Keppens, R. 2016, The Astrophysical Journal, 823, 22

\bibitem[{Xia {et~al.}(2017)Xia, Keppens, \& Fang}]{xia2017}
Xia, C., Keppens, R., \& Fang, X. 2017, Astronomy \& Astrophysics, 603, A42

\end{thebibliography}
\bibliographystyle{aasjournal}

\end{document}